\documentclass[journal]{IEEEtran}

\usepackage{blindtext}
\usepackage{graphicx}
\usepackage{rotating}

\usepackage{cite}
\usepackage[hyphens]{url}
\usepackage[hidelinks]{hyperref} 
\usepackage{algpseudocode}

\usepackage[cmex10]{amsmath}
\usepackage{amssymb,amsfonts,bm,balance,multirow}

\usepackage[usenames,dvipsnames]{color}
\usepackage[dvipsnames]{xcolor}
\usepackage[normalem]{ulem}  
\usepackage{pst-sigsys}
\usepackage{stackengine}

\setstackEOL{\\}
\usepackage{tikz}
\usepackage{scalerel}
\usepackage{float}
\usepackage{algorithm}

\graphicspath{{figures/} {figures_R1/}}

\newcommand*{\slopebox}[1]{%
\framebox{\raisebox{0pt}[0.4\baselineskip][0.01\baselineskip]{%
\hbox to 8mm{\hss#1\hss}}}}

\newcommand{\Frac}[2]{{#1}/{#2}}

\newcommand{\defeq}{\stackrel{\text{\tiny{def}}}{=}}

\newcommand*{\probbox}[1]{%
 \framebox{\raisebox{0pt}[0.3\baselineskip][0.1\baselineskip]{%
 \hbox to 8mm{\hss#1\hss}}}}

\usepackage{cleveref}

\crefname{myThm}{Theorem}{Theorems}
\newtheorem{myDef}{Definition}

\crefname{myLem}{Lemma}{Lemmas}
\newtheorem{myExample}{Example}
\crefname{myExample}{Example}{Examples}

\crefname{myProp}{Proposition}{Propositions}

\DeclareMathOperator*{\argmin}{arg\,min}

\DeclareMathOperator*{\E}{\mathbb{E}}

\def\BetaDist{\mathrm{Beta}}

\def\dB{\mathbf{dB}}
\def\DeltaThresh{\Delta_{\mathrm{min}}}

\def\half{\frac{1}{2}}

\def\pML{\widehat{p}_{\rm ML}}

\def\Rcont{R_{\mathrm{cont}}}
\def\Rstop{R_{\mathrm{stop}}}

\def\trialbudget{\eta} 
\def\allocationgain{\gamma_{\rm alloc}}

\def\phat{\widehat{p}}

\def\phatMMSE{\widehat{p}_{\rm MMSE}}
\newcommand{\var}[1]{\mathrm{var}\!\left({#1}\right)}   
\newcommand{\Rbeta}[2]{\sigma_{#1,#2}^2}                
\newcommand{\Risk}[4]{R_{#3,#4}^{#1,#2}}                

\title{
Beyond Binomial and Negative Binomial: \\
Adaptation in Bernoulli Parameter Estimation \\
\thanks{This material is based upon work supported in part by the US National Science Foundation under Grant No.\ 1422034 and Grant No.\ 1815896, and by the DARPA REVEAL program under Contract No.\ HR0011-16-C-0030.}
\thanks{The authors are with the Department of Electrical and Computer Engineering, Boston University, Boston, MA 02215 USA (e-mail: scmedin@bu.edu; johnmb@bu.edu; dac@bu.edu;
v.goyal@ieee.org).}
}

\author{Safa C. Medin,
        John Murray-Bruce, 
        David Casta\~n\'on, and
        Vivek~K~Goyal}%

\begin{document}

\maketitle

\begin{abstract}
Estimating the parameter of a Bernoulli process arises in many applications,
including photon-efficient active imaging where each illumination period is regarded as a single Bernoulli trial.
Motivated by acquisition efficiency when multiple Bernoulli processes (e.g., multiple pixels) are of interest,
we formulate the allocation of trials under a constraint on the mean as an optimal resource allocation problem.
An oracle-aided trial allocation demonstrates that there can be a significant advantage from varying the allocation for different processes and inspires the introduction of a simple trial allocation gain quantity.
Motivated by achieving this gain without an oracle, we present a trellis-based framework for representing and optimizing stopping rules.
Considering the convenient case of Beta priors,
three implementable stopping rules with similar performances are explored,
and the simplest of these is shown to asymptotically achieve the oracle-aided trial allocation.
These approaches are further extended to estimating functions of a Bernoulli parameter.
In simulations inspired by realistic active imaging scenarios, we demonstrate significant mean-squared error improvements up to 4.36~{dB} for the estimation of $p$ and up to 1.86~{dB} for the estimation of $\log p$.
\end{abstract}


\section{Introduction}
Estimating the parameter of a Bernoulli process is a fundamental problem in statistics and signal processing.
From the binary-valued outcomes of independent and identically distributed (i.i.d.) trials (generically failure (0) or success (1)), we wish to estimate the probability $p$ of success. 
Among myriad applications, our primary interest is raster-scanned active imaging in which a scene patch is periodically illuminated with a pulse, and each illumination period (Bernoulli trial) either has a photon-detection event (success) or not (failure)~\cite{Shin2015}.
The probability $p$ of a photon-detection event has a monotonic relationship with the reflectivity of the scene patch, and a monotonic function of an estimate of $p$ becomes the corresponding image pixel value.
For efficiency in acquisition time or illumination energy, we are motivated to form the image from a small number of illumination pulses, under conditions where $p$ is small.\footnote{For applications using time-correlated single photon counting driven by a detector with dead time,
such as a single-photon avalanche diode (SPAD),
it is recommended to keep $p$ below $0.05$ to avoid time skew and missed detections~\cite{tcspcWahl2015}.}
Other types of raster-scanned imaging can be modeled similarly assuming that the dwell time is an integer multiple of some base time interval, during which the observations are binary.

Conventional methods are not adaptive. With a fixed number of trials $n$, the number of successes $K$ is a binomial random variable, and the maximum likelihood (ML) estimate of $p$ is $K/n$.
Though less common in active imaging, a well-known alternative in the statistics literature is to fix a number of successes $\ell$.
Repeating trials until success $\ell$ occurs results in a random number of trials $M$ that is a negative binomial random variable,%
\footnote{Note that the negative binomial distribution is defined inconsistently in the literature, with sometimes the number of failures being fixed rather than the number of successes (reversing the roles of $p$ and $1-p$).
}
and the ML estimate of $p$ is $\ell/M$.
While there may seem to be nothing to design here,
a constraint on the mean number of trials opens up possibilities for data acquisition that results in neither binomial nor negative binomial distributions.
The mean may be
over a multiplicity of (non-random) Bernoulli process parameters to estimate
(such as in active imaging with one parameter per pixel) or
over a prior for a single Bernoulli parameter.
The two cases are formally linked through the relative frequency interpretation of probability,
with the empirical distribution of the multiplicity of deterministic parameters in the former case
playing the role of the prior distribution in the latter case~\cite{Fine2006}.
For multiple deterministic parameters,
we have a resource allocation problem reminiscent of bit allocation in transform coding~\cite{Segall:76,Goyal:01a}.
As we will demonstrate,
in an oracle-aided setting, trials can be allocated to maximize a \emph{trial allocation gain} that is analogous to the coding gain of transform coding.
For a single random parameter, a simple and implementable approach~-- not requiring an oracle~-- asymptotically achieves the optimal trial allocation gain and may perform better than the oracle-aided method for moderate numbers of trials. 

The focus of this paper is on allocating trials in the estimation of a single random parameter through the design of a stopping rule.
A stopping rule may~-- implicitly and stochastically~-- allocate trials differently for different values of $p$, even though $p$ is not known \emph{a priori}.
We show that any optimal stopping rule can be described by a trellis rather than a more complicated graph,
and greedy construction of the trellis is very nearly optimal.
For a rectangular array of Bernoulli processes representing a scene in an imaging problem, applying a good stopping rule allocates more trials to the pixels where they provide the most benefit.
The final image formation may include a method such as
total variation (TV) regularization for exploiting spatial correlations among neighbors.
{Regularized image formation makes it more difficult to optimize the acquisition, but it does not invalidate the advantage from adaptive acquisition.}
In simulations with parameters realistic for active optical imaging, we demonstrate a reduction in mean-squared error (MSE) by a factor of up to 2.73 (4.36~$\mathrm{dB}$) in comparison to the same regularized reconstruction approach applied without adaptation in numbers of trials.
Such gains vary based on image content, and gains without regularization are predictable from the trial allocation gain formulation.

\subsection{Related Work}

\subsubsection{Statistics Literature} 

In statistics, forming a parameter estimate from a number of i.i.d.\ observations that is dependent on the observations themselves is called \emph{sequential estimation}~\cite{Anscombe1953}.
Early interest in sequential estimation of a Bernoulli process parameter was inspired by the high relative error of deterministically stopping after $n$ trials when $p$ is small.
Specifically, the standard error of the ML estimate is $\sqrt{p(1-p)/n}$, which for small $p$ is unfavorable compared to anything proportional to $p$.
This shortcoming manifests, for example, in requiring large $n$ to distinguish between two small possible values for $p$.

Haldane~\cite{Haldane1945} observed that if one stops after $\ell$ successes, the (random) number of trials $M$ is informative about $p$.
Specifically, $(\ell-1)/(M-1)$ is an unbiased estimate of $p$ (provided $\ell \geq 2$), and its standard error is proportional to $p$ (provided $\ell \geq 3$).
(The ML estimate $\ell/M$ is not unbiased, though $M/\ell$ is an unbiased estimate of $1/p$.)
Tweedie~\cite{Tweedie1945} suggested to call this inverse binomial sampling, but the resulting random variable is now commonly known as negative binomial or Pascal distributed.
More recent works have focused on
non-MSE performance metrics~\cite{Cabilio1975,Cabilio1977},
estimation of functions of $p$~\cite{Hubert2000},
estimation from imperfect observations~\cite{Djuric2000},
and composite hypothesis testing~\cite{Ciuonzo2015}.

\subsubsection{Photon-Efficient Imaging and Variable Dwell Time}
First-photon imaging~\cite{FPI2014} introduced sequential estimation to active imaging.
This method uses the number of illumination pulses until the first photon is detected to reveal information about reflectivity, setting $\ell=1$ in the concept of Haldane~\cite{Haldane1945} and thus using geometric sampling as a special case of negative binomial sampling.
A censoring method is used to approximately separate signal and background detections, and
spatial correlations are used to regularize the estimation of the full scene reflectivity image, resulting in good performance from only 1 detected photon per pixel, even when half of the detected photons are attributable to uninformative ambient light.
Subsequent work with binomial sampling (and otherwise identical experimental conditions) resulted in similar performance~\cite{Shin2015}, and
greatly increasing robustness to ambient light is largely attributable to improving the censoring step~\cite{Rapp2017}.
These works leave questions on the importance of negative binomial sampling to first-photon imaging unanswered;
comparing first-photon imaging to photon-efficient methods with deterministic dwell time~\cite{Krichel2010,Morris2015,Shin2015,Shin2015b,Altmann2016,Shin2016camera,Shin2016,Shin2016multidepth,Mertens2017,Rapp2017,Altmann2017,Halimi2017}
was an initial inspiration for the present work.

While recent works have exploited the first-photon idea in imaging techniques such as ghost imaging~\cite{Liu2018,Altmann2018} and \mbox{x-ray} tomography~\cite{Zhu2018},
previous uses of variable dwell time are not closely connected to sequential estimation or the result of optimized resource allocation.
For example, in lidar, varying dwell time to maintain approximately constant signal strength despite varying effective reflectivity (including greater radial fall-off for more distant scene patches) dates back to at least the 1970s~\cite{Lipke:1979}.
He \emph{et al.}~\cite{He2017} closely follow the technique of \cite{Shin2015}, including its background censoring, and vary the dwell time to keep the number of photon detections after censoring (i.e., photon detections attributed to signal rather than background) at each pixel approximately constant.
In scanning electron microscopy, Dahmen \emph{et al.}~\cite{Dahmen2016} increase dwell time where a measure of image detail is large.
To the best of our knowledge, no previous paper has formally optimized dwell time under a Bernoulli process measurement model. 

\subsection{Main Contributions and Preview of Results}

\subsubsection{Framework}
This work discusses a novel framework for depicting and understanding stopping rules for sequential estimation of Bernoulli parameters under number of trials constraints
(Section~\ref{sec:single-bernoulli}).
In this framework, first presented in~\cite{MedinMBG:2018}, each Bernoulli trial corresponds to a transition in a trellis in which each node is identified by the number of trials and number of successes;
it is easily shown that distinct paths to reach a given node need not be distinguished.
A stopping rule is the assignment of probabilities of stopping to each node in the trellis (see Figs.~\ref{fig:GeneralTrellis}--\ref{fig:example-trellis}).
By construction, a stopping rule defined in this way is implementable because it does not depend on knowledge of $p$ or non-causally on the Bernoulli process.
This framework applies equally well under any prior for $p$.

\subsubsection{Stopping Rule Design}
Simple stopping rules lead to binomial
(Fig.~\ref{fig:stopping_rules-bin-negbin}(a))
and negative binomial
(Fig.~\ref{fig:stopping_rules-bin-negbin}(b))
sampling.
Specializing to the Beta family of priors,
which is both convenient and conventional because it is the conjugate prior for the relevant observation distributions,
methods to optimize the stopping rule are presented in decreasing order of computational complexity:
dynamic programming (Section~\ref{ssec:dynamic_prog}),
offline greedy design (Section~\ref{ssec:greedy-offline}), and
online threshold-based termination (Section~\ref{ssec:greedy-online}) first introduced in~\cite{MedinMBG:2018}.
Empirically, all three methods, including the online method requiring no storage of a precomputed stopping rule, provide very similar performance.
Thus, the easily implementable online method provides very nearly optimal performance.

\subsubsection{Analysis in Oracle-Aided Setting}
This paper introduces the concept of oracle-aided trial allocation whereby processes with different parameters are allocated different fractions of an overall trial budget (Section~\ref{sec:oracle-aided}).
This yields a readily-computed \emph{trial allocation gain} that can be arbitrarily large, though it is generally modest
(Section~\ref{sec:trial-allocation-gain}).
Furthermore, we show that under any Beta prior the threshold-based stopping asymptotically allocates trials identically to the oracle-based optimal
(Section~\ref{ssec:comparison_oracle}).

\subsubsection{Evaluation}
In simulations inspired by realistic active imaging scenarios, an MSE improvement factor of up to $4.36\,\mathrm{dB}$ is demonstrated where spatial correlations are exploited through total variation (TV) regularization
(Section~\ref{sec:image_p}).
Without TV regularization, achieved gains are close to the values predicted by the theoretical trial allocation gain.
For example, the theoretical trial allocation gain is 2.29~dB, and a gain of 2.27~dB is realized for the Shepp-Logan phantom.

\subsubsection{Estimating functions of $p$}
Inspired by applications where estimating functions of $p$ is of interest \cite{Hubert2000}, online threshold-based termination is also extended to estimating $\log p$ from Bernoulli observations
(Section~\ref{sec:estimating-functions}).
Experimental results without TV regularization demonstrate improvements of up to $1.86$~dB using the threshold-based stopping rule versus the conventional binomial sampling
(Section~\ref{sec:image_log_p}).

\section{Trial Allocation Across Multiple Processes}
\label{sec:trial-allocation}

Consider the estimation of the parameters $\{p_i\}_{i=1}^r$ of a finite number $r$ of Bernoulli processes with binomial sampling of each process.
When $m_i$ trials of process $i$ are observed,
the MSE of the ML estimate of $p_i$ is $p_i(1-p_i)/m_i$.
Suppose that we are interested in making the average of the MSEs,
\[
  \frac{1}{r}\sum_{i=1}^r \frac{p_i(1-p_i)}{m_i},
\]
small under a constraint on the average of the numbers of trials
$(1/r)\sum_{i=1}^r m_i \leq \trialbudget$.

Since $p_i(1-p_i)$ varies over $[0,\,1/4]$ for $p_i \in [0,1]$, there can be an advantage to varying the $m_i$ values.
However, that allocation of trials depends on parameters that are to be estimated.
In this section, we suspend the need for implementability and instead study the optimal trial allocation as if the parameters were known.
This provides a benchmark for the implementable methods developed in the remainder of the paper,
with $\{p_i\}_{i=1}^r$ playing the role of a discrete prior on $p$.
We also consider $r \rightarrow \infty$ to reach a distributional limit.

\subsection{Oracle-Aided Optimal Allocation}
\label{sec:oracle-aided}

In optimizations such as
\begin{equation}
\min_{m_i, \ i=1,\,2,\,\ldots,\,r} \, \sum_{i=1}^r \frac{p_i(1-p_i)}{m_i} \mathrm{\ \ s.t. \ \ } \sum_{i=1}^{r} m_i \leq r \trialbudget,
\label{eq:opt_oracleaided_bin}
\end{equation}
ignoring that each $m_i$ should be a positive integer,
each MSE vs.\ number of trials trade-off should be at the same slope,
else it would be advantageous to shift trial resources to the process for which the benefit (MSE reduction) per trial is largest.
This is formalized using the method of Lagrange multipliers.
The resulting optimal allocation is
\begin{equation}
m_i^\ast = r \trialbudget \frac{\sqrt{p_i(1-p_i)}}{\sum_{j=1}^r \sqrt{p_j(1-p_j)}},
\qquad
i = 1,\,2\,\ldots,\,r.
\label{eq:oracleaided_solution}
\end{equation}
Since each process has a fixed number of trials $m_i^\ast$, independent of the experimental outcome of each trial, we call using these numbers of trials \textit{oracle-aided binomial sampling}.

\begin{myExample}[Oracle-aided allocations]
\label{ex:allocations}
\begin{enumerate}
\item[(a)] Let $p_1 = \varepsilon$ and $p_2 = 1/2$.  Then the fractional oracle-aided allocations are
\[
  \frac{m_1^*}{2\trialbudget} = \frac{\sqrt{\varepsilon(1-\varepsilon)}}
                             {\sqrt{\varepsilon(1-\varepsilon)}+\Frac{1}{2}},
\quad
  \frac{m_2^*}{2\trialbudget} = \frac{\Frac{1}{2}}
                             {\sqrt{\varepsilon(1-\varepsilon)}+\Frac{1}{2}}.
\]
These are plotted as functions of $\varepsilon$ in Fig.~\ref{fig:oracle_allocations}(a).

\begin{figure}
\centering
\begin{tabular}{@{}cc@{}}
\includegraphics[width=0.48\linewidth]{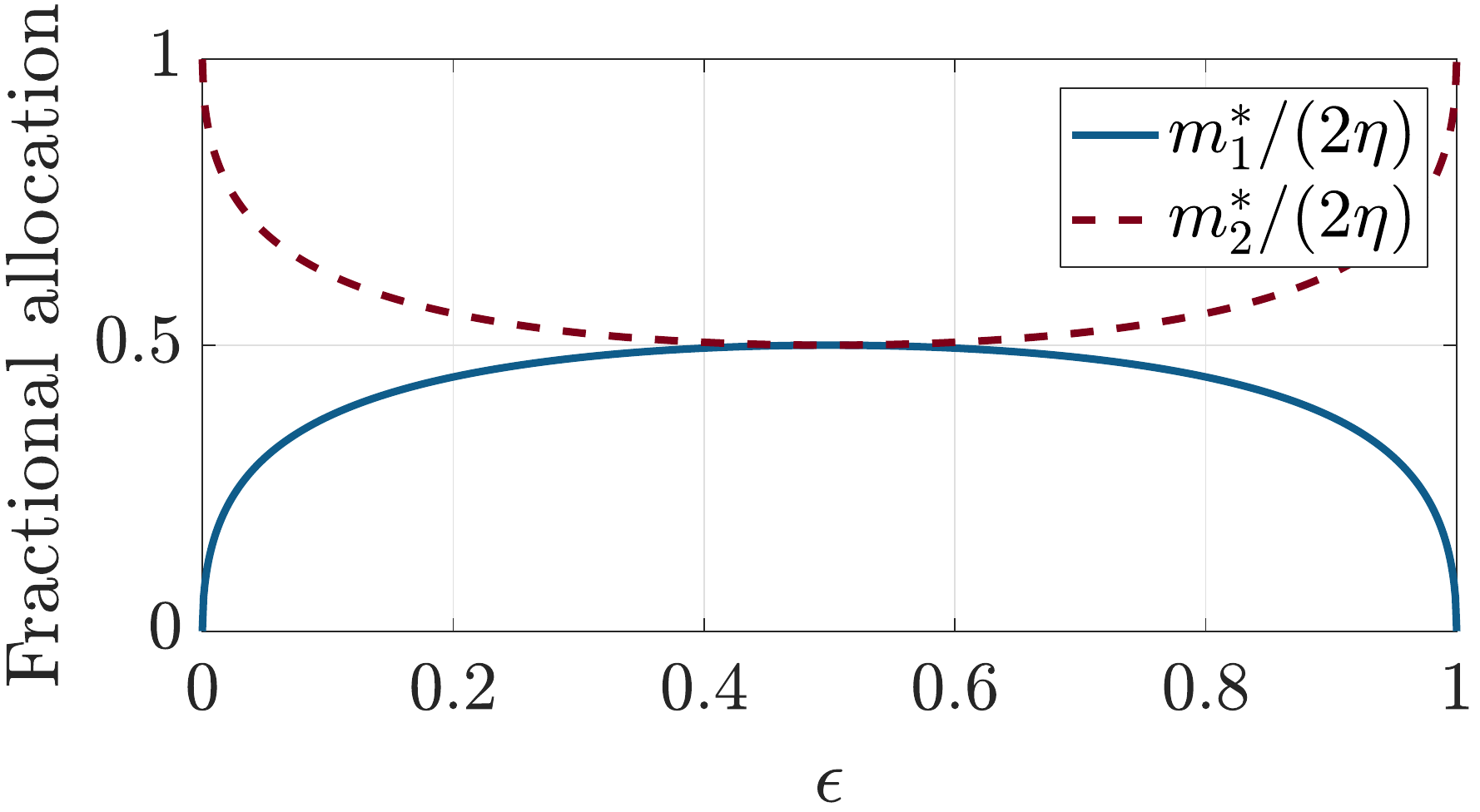} &
\includegraphics[width=0.48\linewidth]{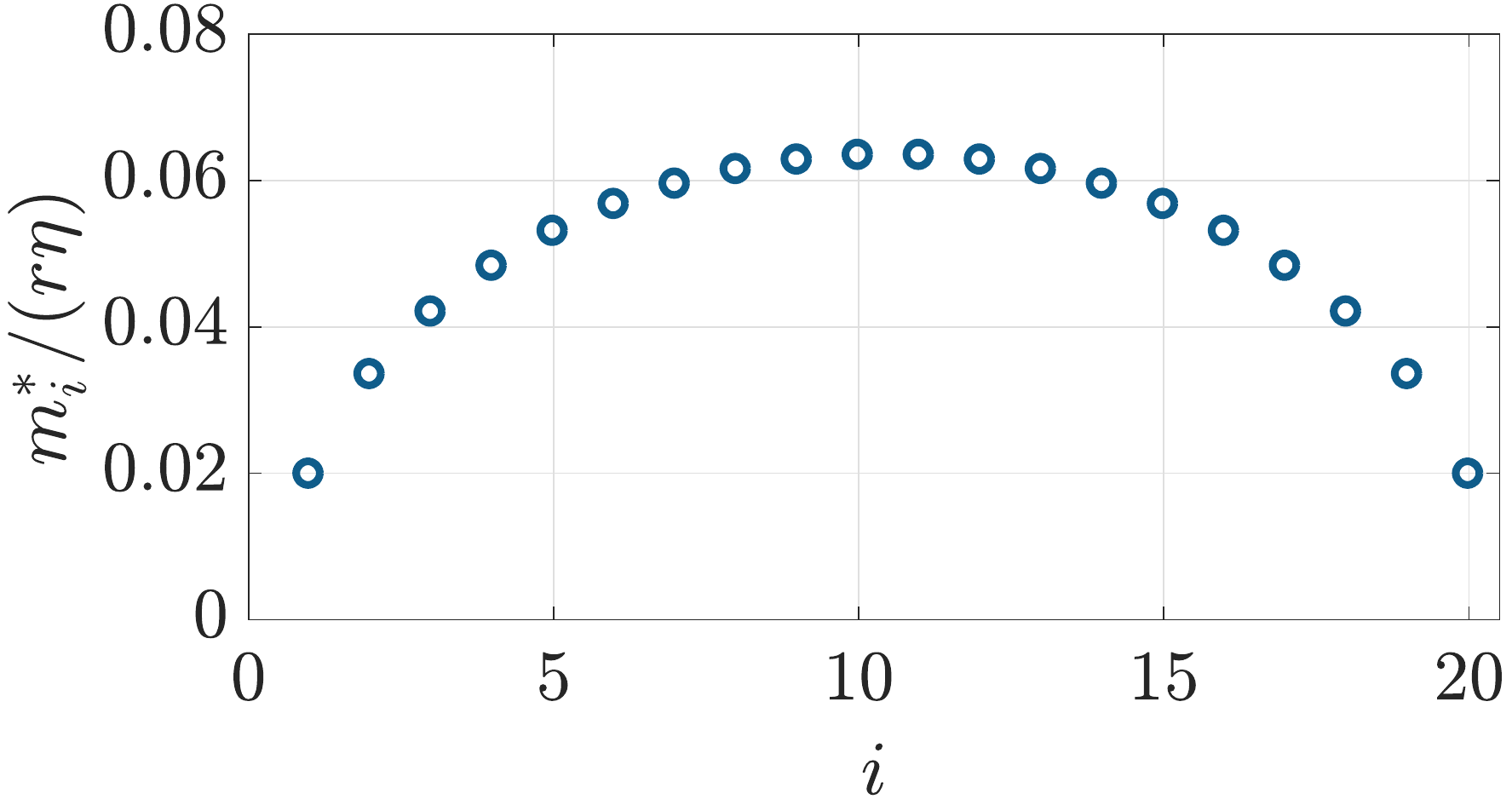} \\
{\small (a) Example~\ref{ex:allocations}(a)} &
{\small (b) Example~\ref{ex:allocations}(c) with $r=20$} \\
\end{tabular}
\caption{Oracle-aided optimal allocations in Examples~\ref{ex:allocations}(a) and~\ref{ex:allocations}(c).}
\label{fig:oracle_allocations}
\end{figure}

\item[(b)] Let $p_1 = p_2 = \cdots = p_{r-1} = \varepsilon$ and $p_r = 1/2$.  Then the fractional oracle-aided allocations are
\begin{eqnarray*}
  \frac{m_i^*}{r\trialbudget} & \!\!\!=\!\!\! & \frac{\sqrt{\varepsilon(1-\varepsilon)}}
                             {(r-1)\sqrt{\varepsilon(1-\varepsilon)}+\Frac{1}{2}},
\quad i {=} 1,\,2,\,\ldots,\,r{-}1, \\
  \frac{m_r^*}{r\trialbudget} & \!\!\!=\!\!\! & \frac{\Frac{1}{2}}
                             {(r-1)\sqrt{\varepsilon(1-\varepsilon)}+\Frac{1}{2}}.
\end{eqnarray*}
\item[(c)] Let $p_i = (2i-1)/(2r)$, $i = 1,\,2,\,\ldots,\,r$.
The fractional oracle-aided allocations $m_i^*/(r \trialbudget)$ are plotted for $r = 20$ in Fig.~\ref{fig:oracle_allocations}(b).
\end{enumerate}
\end{myExample}

\subsection{Trial Allocation Gain}
\label{sec:trial-allocation-gain}

Using the oracle-aided allocations \eqref{eq:oracleaided_solution} reduces the average MSE relative to a constant allocation $m_1 = m_2 = \cdots = m_r = \trialbudget$.
The constant allocation results in the average MSE
\begin{equation}
  \frac{1}{r} \sum_{i=1}^r \frac{p_i(1-p_i)}{\trialbudget},
  \label{eq:mse_constant_allocation}
\end{equation}
whereas using \eqref{eq:oracleaided_solution} yields
\begin{eqnarray}
 \frac{1}{r} \sum_{i=1}^r \frac{p_i(1-p_i)}{m_i^*} 
  & \!\!=\!\! & \frac{1}{r} \sum_{i=1}^r \frac{p_i(1-p_i)}{\sqrt{p_i(1-p_i)}} \frac{1}{r \trialbudget} {\sum_{j=1}^r \sqrt{p_j(1-p_j)}} \nonumber \\
  & \!\!=\!\! & \frac{1}{r^2 \trialbudget} \Bigg( \sum_{j=1}^r \sqrt{p_j(1-p_j)} \Bigg)^2.
  \label{eq:mse_optimal_allocation}
\end{eqnarray}
We define the ratio of \eqref{eq:mse_constant_allocation} and \eqref{eq:mse_optimal_allocation} as the \emph{trial allocation gain}:
\begin{equation}
  \label{eq:allocation_gain_deterministic}
\allocationgain = \frac{r \sum_{i=1}^r p_i(1-p_i)}
                       {\left( \sum_{j=1}^r \sqrt{p_j(1-p_j)} \right)^2}.
\end{equation}
Trial allocation gain is reminiscent of the coding gain in transform coding~\cite{Segall:76,Goyal:01a}.

\begin{figure}
\centering
\begin{tabular}{@{}cc@{}}
\hspace{1mm}
\includegraphics[width=0.40\linewidth]{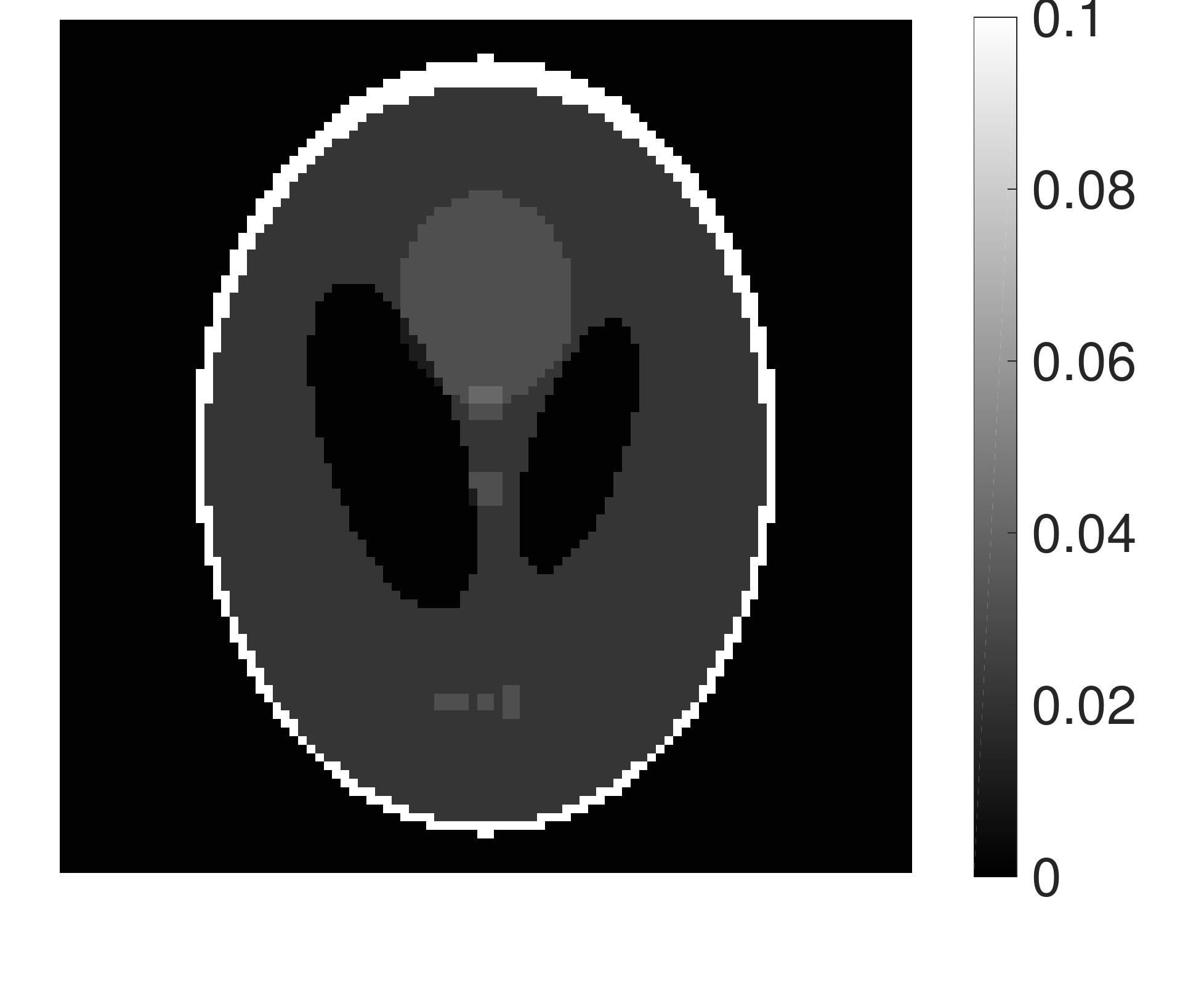}&
\hspace{-6mm}
\includegraphics[width=0.58\linewidth]{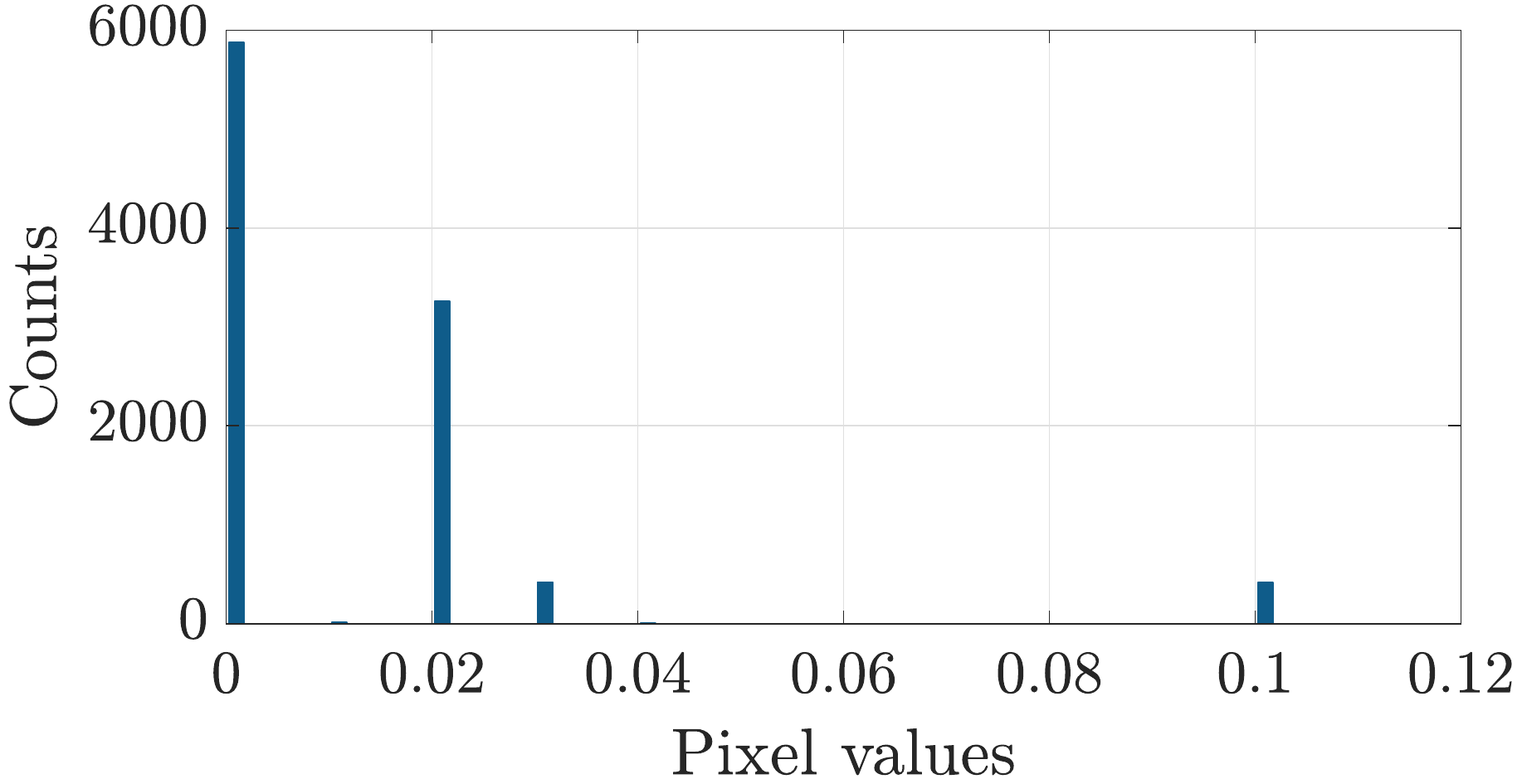} \\
\hspace{-1mm}
{\small (a) Shepp--Logan phantom} &
{\small (b) Histogram of phantom values} \\
\end{tabular}
\caption{Phantom of size $100 \times 100$ with trial allocation gain of 1.686
(Example~\ref{ex:gains}(d)).
}
\label{fig:Shepp-Logan}
\end{figure}

\begin{myExample}[Trial allocation gains]
\label{ex:gains}
\begin{enumerate}
\item[(a)] For the parameters in Example~\ref{ex:allocations}(a),
  the trial allocation gain is plotted as a function of $\varepsilon$ in Fig.~\ref{fig:allocation_gains}.
  Notice that in the limit of $\varepsilon \rightarrow 0$, all the trials are allocated to the nontrivial Bernoulli process,
  doubling its number of trials, which halves the average MSE\@.  Thus $\allocationgain \rightarrow 2$.
  
\begin{figure}
\centering
\begin{tabular}{@{}cc@{}}
\includegraphics[width=0.48\linewidth]{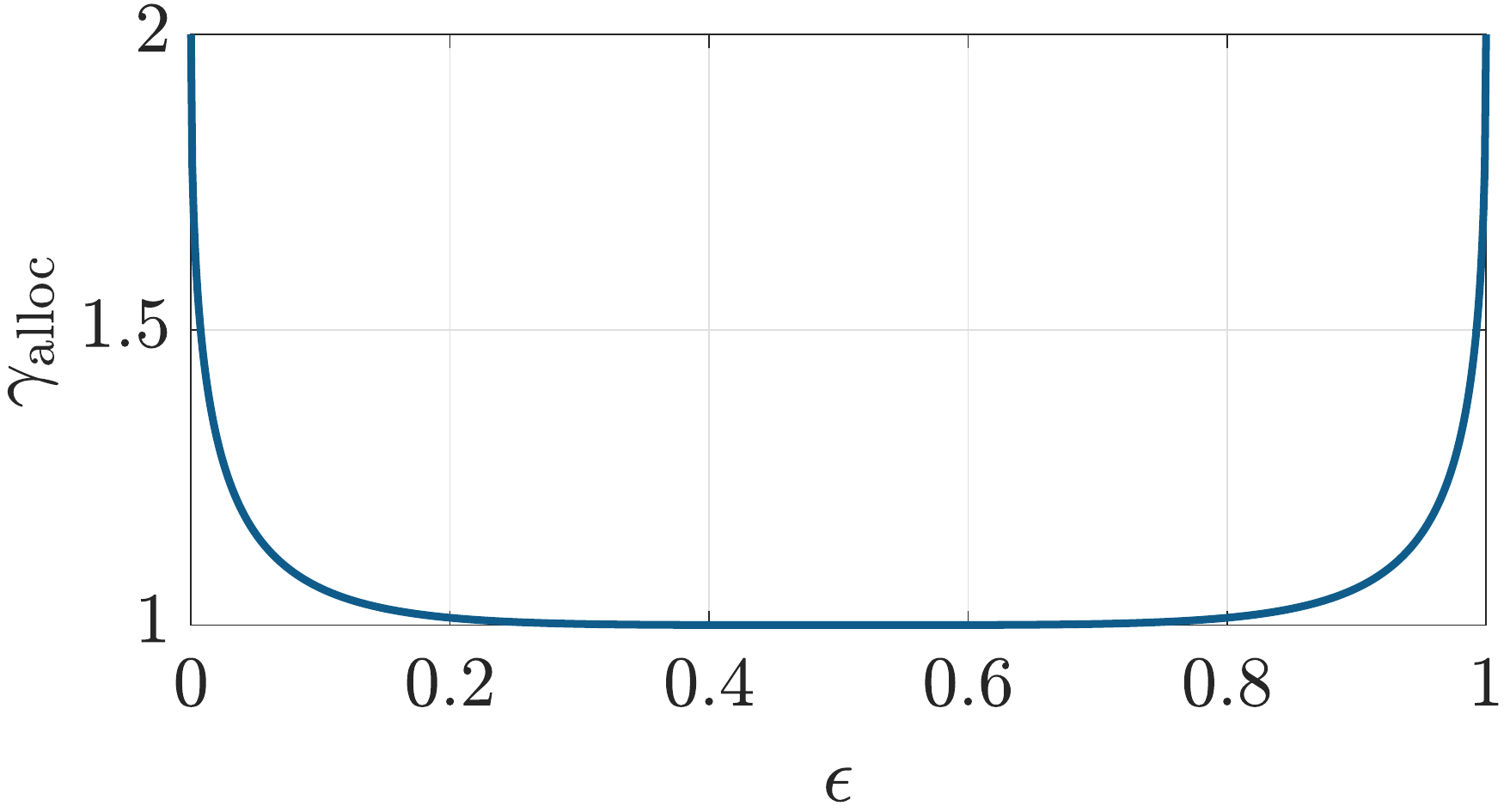} &
\includegraphics[width=0.49\linewidth]{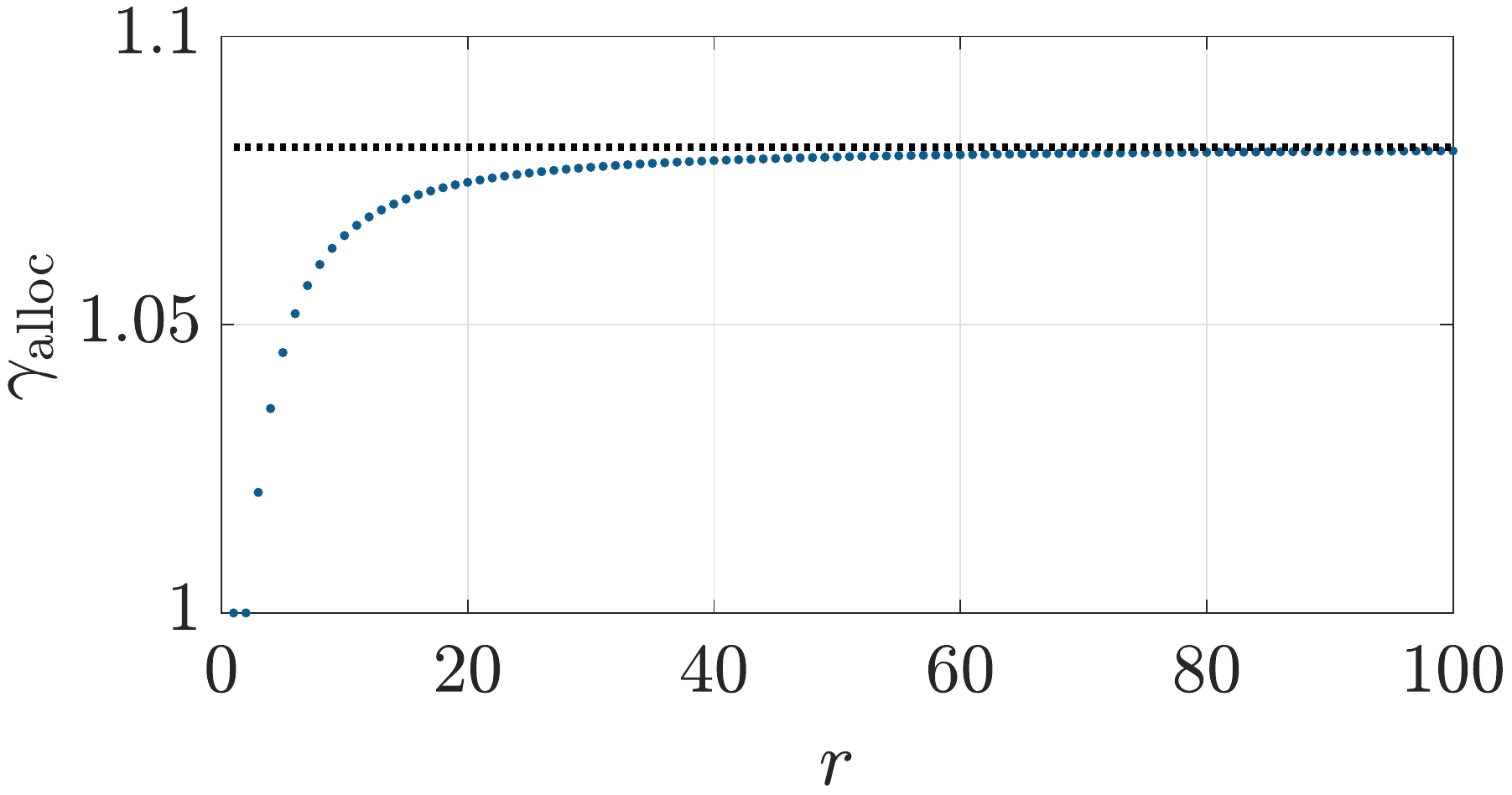} \\
{\small (a) Example~\ref{ex:gains}(a)} &
{\small (b) Example~\ref{ex:gains}(c)} \\
\end{tabular}
\caption{Trial allocation gains in Examples~\ref{ex:gains}(a) and~\ref{ex:gains}(c).
}
\label{fig:allocation_gains}
\end{figure}

\item[(b)] For the parameters in Example~\ref{ex:allocations}(b), $\lim_{\varepsilon \rightarrow 0} \allocationgain = r$.
\item[(c)] For the parameters in Example~\ref{ex:allocations}(c), the trial allocation gain is plotted as a function of $r$ in Fig.~\ref{fig:allocation_gains}(b).
\item[(d)] Fig.~\ref{fig:Shepp-Logan}(a) shows the ``'Modified Shepp--Logan phantom'' provided by the Matlab {\tt phantom} command, at size $100 \times 100$ and scaled to $[0.001,\, 0.101]$.
Fig.~\ref{fig:Shepp-Logan}(b) shows a histogram of the $10^4$ intensity values of the phantom.  Evaluating \eqref{eq:allocation_gain_deterministic} gives 1.6944, or 2.29 dB\@.
\end{enumerate}
\end{myExample}
One can show that $\allocationgain \in [1,\,r]$.
The upper bound is illustrated in part (b) of the example.

\subsection{Distributional Limit}
\label{sec:trial-allocation-distributional}

Suppose now that the Bernoulli process parameter is modeled with random variable $P$ and
the number of trials $M$ is to be assigned by an oracle (i.e., it is allowed to depend on the realization of $P$)
to minimize the MSE of the ML estimate of $P$ under the constraint $\E[M] \leq \trialbudget$.
By analogy to the computations giving \eqref{eq:oracleaided_solution}~--
or formally taking a limit of $r \rightarrow \infty$ with the empirical distribution of $\{p_i\}_{i=1}^r$ converging to the distribution of $P$~--
the number of trials should be assigned based on how large $\sqrt{P(1-P)}$ is relative to
$\E[\sqrt{P(1-P)}]$:
\begin{equation}
  M = \trialbudget \, \frac{\sqrt{P(1-P)}}{\E\!\big[\sqrt{P(1-P)}\big]}.
  \label{eq:trial-allocation-random}
\end{equation}
The resulting trial allocation gain is
\begin{equation}
\allocationgain = \frac{\E[P(1-P)]}{\left(\E\!\big[\sqrt{P(1-P)}\big]\right)^2}.
  \label{eq:allocation-gain-random}
\end{equation}
This can also be written as
\begin{equation}
\allocationgain = 1 + \frac{\var{V}}{\left(\E[V]\right)^2},
  \label{eq:allocation-gain-random-2}
\end{equation}
where $V = \sqrt{P(1-P)}$.
It follows that $\allocationgain \geq 1$, with equality if and only if the random variable $\sqrt{P(1-P)}$ has zero variance.

\begin{myExample}[Trial allocation gains~-- random parameter]
\label{ex:gains-random}
\begin{enumerate}
\item[(a)] Let $P$ have the continuous uniform distribution on $[0,\,1]$.
  Then evaluating \eqref{eq:allocation-gain-random} gives $\allocationgain = 32/(3\pi^2) \approx 1.0808$.
  This value is the asymptote in Fig.~\ref{fig:allocation_gains}(b).
\item[(b)] Let $P$ take two values:  $\frac{1}{2}$ with probability $\delta$ and $0$ with probability $1-\delta$.
  Then $\E[P(1-P)] = \delta/4$ and $\E[\sqrt{P(1-P)}] = \delta/2$.
  Substituting in \eqref{eq:allocation-gain-random} gives $\allocationgain = 1/\delta$.
  We can interpret this with relative frequencies:  Since $p=0$ requires no trials,
  fraction $\delta$ of the time, $p=1/2$ will occur and should be allocated $1/\delta$ times the mean number of trials.
\item[(c)] When $p \ll 1$ holds, $p(1-p) \approx p$.
  Therefore, \eqref{eq:allocation-gain-random-2} becomes approximately invariant to rescaling.
  For example, rescaling the phantom in Example~\ref{ex:gains}(d)
  by a factor of $2$ to $[0.002,\,0.202]$
  gives $\allocationgain \approx 1.6633$,
  and by a factor of $\frac{1}{2}$ to $[0.0005,\,0.0505]$
  gives $\allocationgain \approx 1.7096$;
  these are small changes from the value in Example~\ref{ex:gains}(d).
\end{enumerate}
\end{myExample}
The first two parts of the example show that though an allocation gain may typically be modest, it may also be arbitrarily large.
The third part shows that allocation gain is approximately dependent on the coefficient of variation of the Bernoulli parameter,
provided that the parameter is known to be small.

Having established that varying the numbers of trials can be beneficial,
we now turn our attention to methods that do not depend on an oracle.
We will compare to the oracle-aided allocations in certain asymptotic settings.

\section{Observation of a Single Bernoulli Process}
\label{sec:single-bernoulli}

Let $\{X_n : n = 1,\,2,\,\ldots\}$ be a Bernoulli process with an unknown random parameter $p$,
and let $\trialbudget \in \mathbb{R}^+$ be a \emph{trial budget}.
A \emph{stopping rule} consists of a sequence of
\emph{continuation probability} functions
\begin{equation}
\pi_n: \{0,\,1\}^n \rightarrow [0,\,1], \qquad n = 0,\,1\,\ldots,
\label{eq:pi_t}
\end{equation}
that give the probability of continuing observations after trial $n$~--
based on a biased coin flip independent of the Bernoulli process~--
as a function of $(X_1,\,X_2,\,\ldots,\,X_n)$.
The result is a random number of observed trials $N$.%
\footnote{The time $N$ does not satisfy the standard definition of a stopping time when the stopping rule is randomized because randomness independent of the sequence of outcomes $\{X_n\}$ is allowed to influence the decision of whether or not to continue observations.}
The stopping rule is said to satisfy the trial budget when $\E[N] \leq \trialbudget$.
It is said to be \emph{deterministic} when every $\pi_n$ takes values only in $\{0,\,1\}$ and it is said to be \emph{randomized} otherwise.
A randomized stopping rule can be seen as stochastic multiplexing among some number of deterministic stopping rules.

Our goal is to minimize the MSE in estimation of $p$ through the design
of a stopping rule that satisfies the trial budget and
of an estimator $\phat \, (X_1,\,X_2,\,\ldots,\,X_N)$.
We will first show that the continuation probability functions can be simplified greatly with no loss of optimality.
Then, we will provide results on optimizing the stopping rule under a Beta prior on $p$.

\subsection{Framework for Data-Dependent Stopping}

\begin{figure}
\centering
\includegraphics[width=0.8\linewidth]{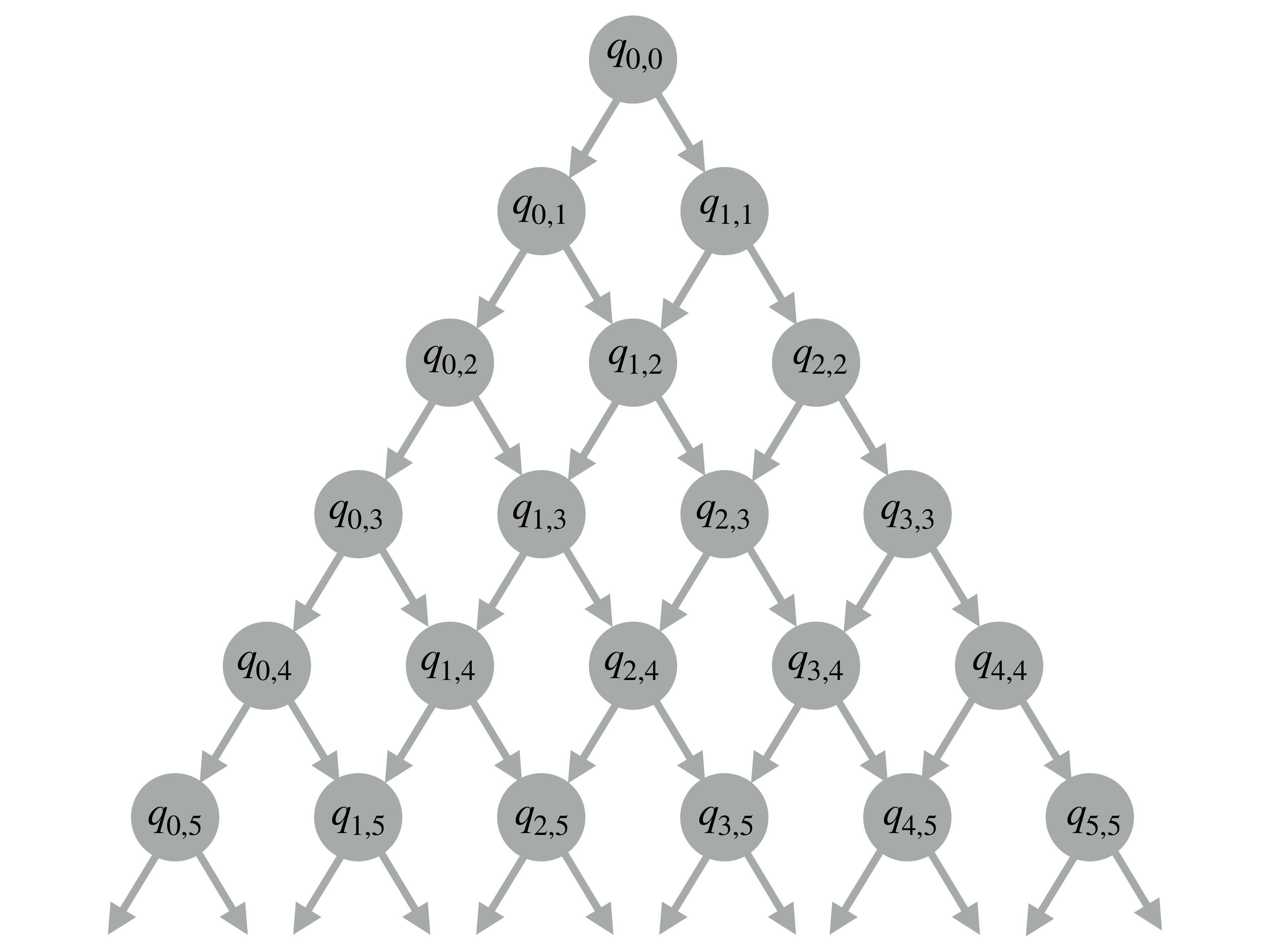}
\caption{A node-labeled trellis showing continuation probabilities for observation sequences up to length 5;
$q_{k,m}$ denotes the probability of continuing after observing $k$ successes in $m$ trials.
}
\label{fig:GeneralTrellis}
\end{figure}

Based on \eqref{eq:pi_t}, a natural representation of a stopping rule is a node-labeled binary tree representing all sample paths of the Bernoulli process,
with a probability of continuation label at each node. 
This representation has
$2^{d+1}-1$
labels for observation sequences up to length
$d$.
However, the tree can be simplified to a trellis without loss of optimality.
Conditioned on observing $k$ successes in $m$ trials,
all $\binom{m}{k}$ sequences of length $m$ with $k$ successes are equally likely.
Thus, 
no improvement can come from having unequal continuation probabilities for the $\binom{m}{k}$ tree nodes
that each represent having $k$ successes in $m$ trials.
Instead, these nodes should be combined, therefore reducing the tree to a trellis.
This representation has
$\frac{1}{2}(d+1)(d+2)$
labels for observation sequences up to length
$d$.
The continuation probability functions are reduced to a set of probabilities
$\{ q_{k,m} : m = 0,\,1\,\,\ldots;\ k = 0,\,1,\,\ldots,\,m \}$
for continuing after $k$ successes in $m$ trials,
as depicted in Fig.~\ref{fig:GeneralTrellis}.

Trellises alone~-- without labels~-- give a simple representation for both data-dependent and data-independent deterministic stopping rules: 
Hence, we begin with some related terminology that will be used throughout this paper.

\begin{myDef}[Complete trellis]
A \emph{complete trellis} of depth $d \in \mathbb{N}$ contains all nodes $\bm{v} = (k,m)$ belonging to the set
\begin{equation}
\mathcal{T}_d = \{ (k,m): k = 0,1,\ldots,m; \ m = 0,1,\ldots,d \}.
\label{eq:trellis_set}
\end{equation}
\end{myDef}

\begin{myDef}[Strategy]
Any $T \in 2^{\mathcal{T}_d}$ is a \emph{strategy} when all nodes in $T$ are connected and $T$ contains the root node $(0,0)$.
\end{myDef}

\smallskip

Henceforth, we restrict our attention to strategies and stochastic multiplexing among strategies.
The stopping rule prescribed by the strategy $T$ is
\begin{equation}
q_{k,m}(T) = 
\begin{cases}
    1, \ \ \bm{v} = (k,m) \in T;  \\ 
    0, \ \ \text{otherwise.}
\end{cases} 
\label{eq:q_km-T}
\end{equation}

\subsection{Standard Sampling Methods and their Representations}
\label{ssec:ExistingProtocols}
\begin{figure}
\begin{tabular}{@{}c@{}c@{}}
\includegraphics[width=0.57\linewidth,trim={2cm 0cm 0.2cm 0cm},clip]{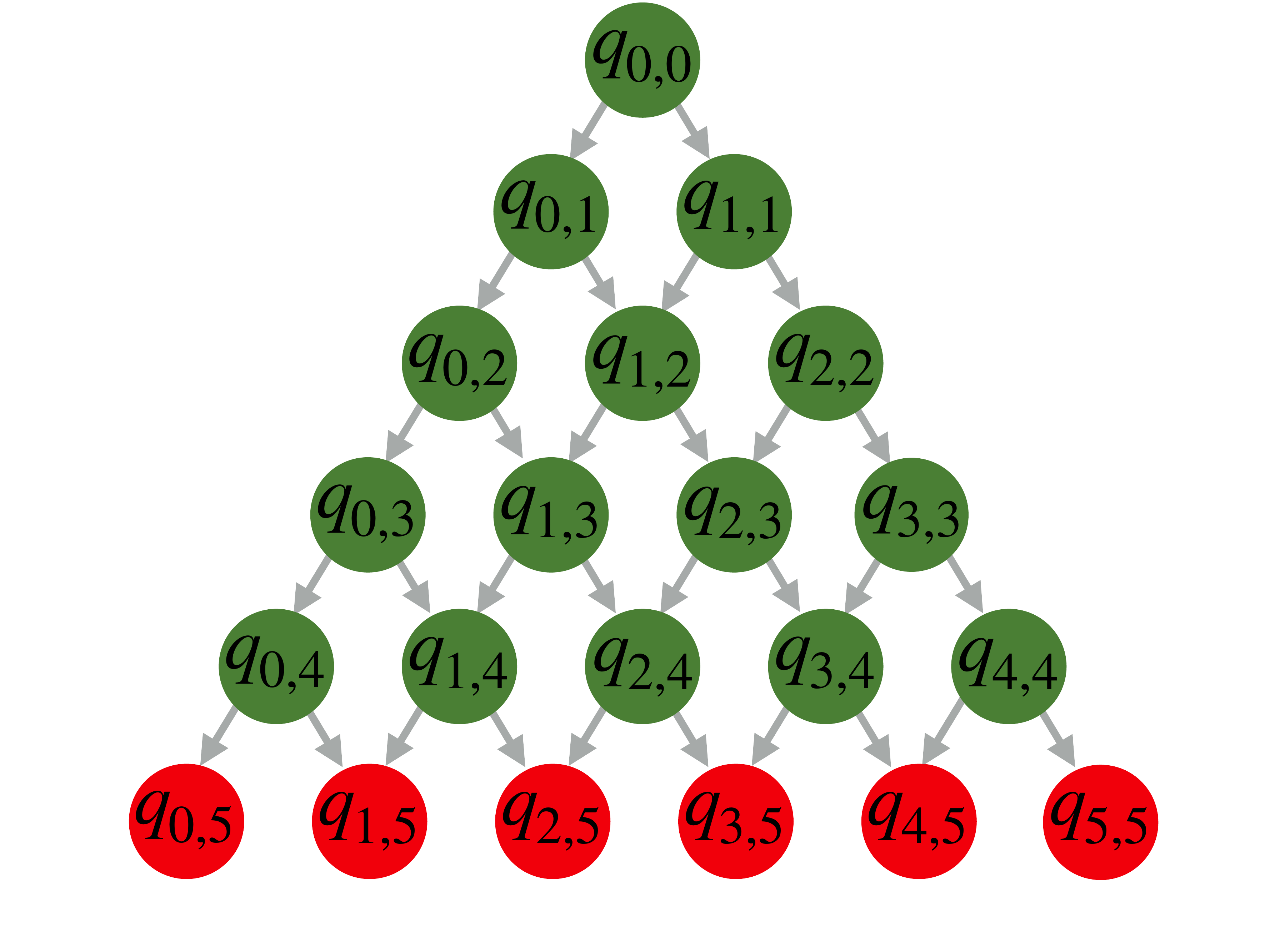}
&
\includegraphics[width=0.41\linewidth,trim={2cm 0cm 9.8cm 0cm},clip]{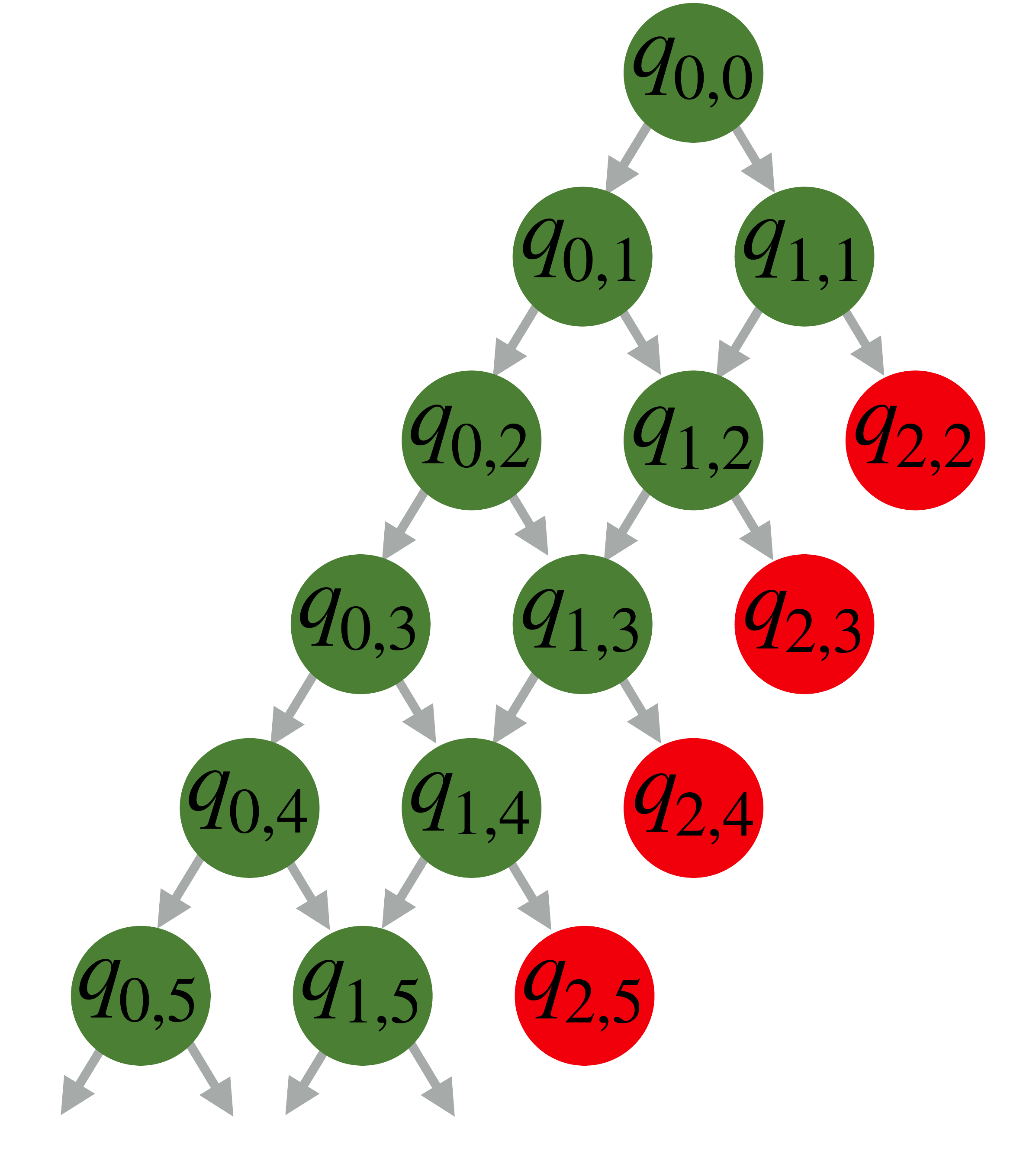}
\\
{\small (a) binomial} & {\small (b) negative binomial}
\end{tabular}
\caption{Green nodes form the trellis representations $T$ of (a) the binomial stopping rule with $n = 5$
and (b) the negative binomial stopping rule with $\ell = 2$;
these are nodes with continuation probability $1$.
Red nodes are not in $T$; these are nodes with continuation probability $0$.}
\label{fig:stopping_rules-bin-negbin}
\end{figure}

The conventional use of a fixed number of trials $n$ corresponds to continuation probabilities
\begin{equation}
 q_{k,m} = 
\begin{cases}
    1, \ \  m < n;  \\ 
    0, \ \ \text{otherwise.}
\end{cases}
\end{equation}
Regardless of the sample path, one
observes exactly $n$ trials,
and the number of successes $K$ is a $\mathrm{Binomial}(n,p)$ random variable.
We refer to this as \emph{binomial sampling} or the \emph{binomial stopping rule}.
An example of the corresponding trellis representation for a fixed number of trials $n=5$ is shown in Fig.~\ref{fig:stopping_rules-bin-negbin}(a).

The technique analyzed by Haldane~\cite{Haldane1945} and employed in first-photon imaging~\cite{FPI2014} with $\ell = 1$ can be expressed with continuation probabilities
\begin{equation}
 q_{k,m} = 
\begin{cases}
    1, \ \  k < \ell;  \\ 
    0, \ \ \text{otherwise.}
\end{cases} 
\end{equation}
Observations cease with $\ell$ successes in $M$ trials, where
$M$ is a $\mathrm{NegativeBinomial}(\ell,p)$ random variable.
We call such a strategy the \emph{negative binomial stopping rule}, or \emph{geometric stopping rule} for the special case where $\ell = 1$.
The trellis representation of the negative binomial stopping rule for $\ell = 2$ is shown in Fig.~\ref{fig:stopping_rules-bin-negbin}(b).

In general, observations cease with $K$ successes in $M$ trials, where $K$ and $M$ are both random variables.
Importantly, the i.i.d.\ nature of a Bernoulli process makes the pair $(K,M)$ contain all the information that is relevant from the sequence of observations.
As noted in the reduction from tree to trellis, conditioned on $(K,M) = (k,m)$, all sequences of length $m$ with $k$ successes are equally likely, so the specific sequence among these is uninformative about $p$.

\subsection{Analysis Under Beta Prior}
\label{ssec:Beta-prior-anal}
Our method for optimizing the design of continuation probabilities is through analyzing mean Bayes risk reduction from continuation.
We define risk function $L$ as \emph{squared error} or \emph{squared loss}
\[
L(p, \phat) = (p-\phat)^2,
\]
where $p$ is the unknown Bernoulli parameter and $\phat$ is the estimate of this parameter.
The Bayes risk $R$ is defined as
\[
R(\phat) = \E\!\left[L(p, \phat)\right] = \E\!\left[(p-\phat)^2\right],
\]
which in this case is the MSE\@.
Using the minimum MSE (MMSE) estimator, for which $\phat = \E[P]$, the Bayes risk
is the \emph{variance} of the posterior distribution.
Thus, key to the optimization is to track posterior variances through the trellis.
For any prior on $p$,
the posterior variance could be computed online or
precomputed for some fixed trellis.
Here we {provide detailed computations} only for the
convenient case of choosing a conjugate prior.

\subsubsection{{Beta Prior}}
The Beta distribution is the conjugate prior for Bernoulli, binomial, and negative binomial distributions:
When $P$ has the $\mathrm{Beta}(a,b)$ distribution with probability density function
\[
f_P(p;a,b) = \frac{\Gamma(a+b)}{\Gamma(a)\Gamma(b)} p^{a-1}(1-p)^{b-1},
\]
where $\Gamma(\cdot) = (\cdot \, - 1)!$ is the gamma function,
the posterior distribution after observing $k$ successes in $m$ trials has the $\mathrm{Beta}(a+k,b+m-k)$ distribution.
The beta distribution $P \sim \mathrm{Beta}(a,b)$
has mean
\begin{equation}
\label{eq:beta_mean}
  \mu_{a,b} = \E[P] = \frac{a}{a+b}
\end{equation}
and variance 
\begin{equation}
  \Rbeta{a}{b} = \var{P} = \frac{ab}{(a+b)^2 (a+b+1)}.
\label{eq:beta_posterior}
\end{equation}

\subsubsection{Expected Number of Trials}
\label{sssec:ExpectedNumTrials}
For the stopping rule represented by the trellis $T$, the expected number of trials is the weighted sum of the depths of all stopping (or leaf) nodes with weights corresponding to probability of reaching that node, under the initial prior.
For a trellis $T \in 2^{\mathcal{T}_d}$ and initial prior $\BetaDist(\alpha,\beta)$, the probability of reaching any node $\bm{v} = (k,m) \in \mathcal{T}_d$
can be expressed recursively using  the probabilities of reaching its parents, $(k-1, m-1)$ and $(k,m-1)$.
Conditioned on reaching $(k-1, m-1)$, the probability of reaching $(k,m)$ is the product of continuation probability $q_{k-1,m-1}(T)$
and success probability
\begin{subequations}
\begin{equation}
  \mu_{\alpha+k-1,\beta+m-k} = \frac{\alpha + k - 1}{\alpha + \beta + m - 1};
\label{eq:prob_recursion_success}
\end{equation}
similarly, conditioned on reaching $(k,m-1)$,
the probability of reaching $(k,m)$ is the product of continuation probability $q_{k,m-1}(T)$ and
failure probability
\begin{equation}
  1 - \mu_{\alpha+k,\beta+m-k-1} = \frac{\beta + m - k - 1}{\alpha + \beta + m - 1}.
\label{eq:prob_recursion_failure}
\end{equation}
\end{subequations}
Hence, we have the recursion
\begin{align}
    u_{k,m}(T) =& u_{k{-}1,m{-}1}(T) q_{k{-}1,m{-}1}(T) \frac{\alpha + k - 1}{\alpha + \beta + m - 1} \nonumber \\
 			    &{+} u_{k,m{-}1}(T) q_{k,m{-}1}(T) \frac{\beta + m - k - 1}{\alpha + \beta + m - 1}
  \label{eq:prob_reaching_km}
\end{align}
for the probability $u_{k,m}(T)$ of reaching node $(k,m)$.
The recursion is initialized with $u_{0,0}(T) = 1$ and $u_{k,m}(T) = 0$ when $k \notin \{0,1,\ldots,m\}$.
Since $T \in 2^{\mathcal{T}_d}$, it suffices to compute up to $m = d+1$.

Using $u_{k,m}(T)$ from \eqref{eq:prob_reaching_km}, the expected number of trials incurred by a strategy $T \in 2^{\mathcal{T}_d}$ starting with a $\BetaDist(\alpha,\beta)$ prior is
\begin{equation}
{h}_{\alpha,\beta}(T) = \sum_{\bm{v} \in \mathcal{T}_{d+1} \setminus T} m \, u_{k,m}(T).
\label{eq:ExpNumTrials}
\end{equation}
The nonzero terms in the sum correspond to the reachable leaf nodes,
which are all contained in $T' = \mathcal{T}_{d+1} \setminus T$.

\subsubsection{Expected Bayes Risk}
\label{sssec:expected_bayes_risk}
Under initial prior $\mathrm{Beta}(\alpha,\beta)$,
the Bayes risk of the estimate of $p$ from observations leading to node $(k,m)$ is given by \eqref{eq:beta_posterior}, with $a = \alpha + k$ and $b = \beta + m - k$.
A strategy $T$ has expected Bayes risk $g_{\alpha,\beta}(T)$ given by the sum of the Bayes risks of nodes with zero continuation probability
weighted by the probabilities of reaching that node:
\begin{align}
g_{\alpha,\beta}&(T)
 = \sum_{\bm{v} \in T'} u_{k,m}(T) \sigma^2_{\alpha+k,\beta+m-k} \nonumber \\
&= \sum_{\bm{v} \in T'} u_{k,m}(T) \frac{ (\alpha+k)(\beta+m-k)}{(\alpha+\beta+m)^2(\alpha+\beta+m+1)}.
\label{eq:EB_risk}
\end{align}

\subsubsection{Optimization Problem Statement}
With the proposed trellis-based framework,
finding an optimal deterministic stopping rule (in the MSE sense) under an average budget constraint becomes a set minimization problem:
\begin{equation}
	\begin{split}
		T^\ast = \, \argmin_{T \in 2^{\mathcal{T}_{d}}} g_{\alpha, \beta} (T) \\
         			\textrm{subject to } h_{\alpha,\beta} \left( T \right) \leq \trialbudget.
	\end{split}	
\label{eq:minimization-problem}
\end{equation}
Implementable solutions to \eqref{eq:minimization-problem}, with varying complexities and deviations from optimality, are presented in the subsequent section.
We seek only solutions on the lower convex hull of the trade-off between $\eta$ and $\min g_{\alpha,\beta}(T)$.
Stochastic multiplexing among these solutions gives optimal randomized stopping rules.

\section{Stopping Rule Design}

\subsection{A Dynamic Programming Solution}
\label{ssec:dynamic_prog}
For a fixed and sufficiently large $d$, total enumeration of the entire solution space is a possible approach for
solving \eqref{eq:minimization-problem} to find
an optimal deterministic stopping rule.
However, the combinatorial structure of the problem means that evaluating the Bayes risks
\eqref{eq:EB_risk}
and expected numbers of trials
\eqref{eq:ExpNumTrials}
for all possible strategies can be computationally prohibitive, even for moderate trial budgets;
this precludes full enumeration.

Conversely, one could start at the leaf nodes of a complete trellis (with depth $d$), traverse the trellis towards its root, whilst deciding whether each visited node merits inclusion in the optimized solution. 
This is the basis of a dynamic programming (DP) solution:
it solves our optimization problem that involves making a sequence of decisions by determining, for each decision, subproblems that can be solved in a similar fashion \cite{bertsekas1996dynamic}.
As such, a solution of the original problem can be found from solutions of subproblems.

Precisely, we first relax \eqref{eq:minimization-problem} by writing its Lagrangian formulation:
\begin{equation}
	\min_{T \in 2^{\mathcal{T}_{d}}} g_{\alpha,\beta}(T) + \lambda h_{\alpha,\beta}(T),
\label{eq:min_problem_relx}
\end{equation}
where $\lambda \in \mathbb{R}_{+}$, can be viewed as the desired MSE reduction per additional trial.
We introduce three compact notations associated with
node $\bm{v} = (k,m)$:
\begin{subequations}
\begin{equation}
\Risk{\,\alpha}{\beta}{k}{m}
  = \sigma^2_{\alpha+k,\beta+m-k} 
  =\frac{ (\alpha+k)(\beta+m-k)}{(\alpha+\beta+m)^2(\alpha+\beta+m+1)}
\label{eq:Rstop_dp}
\end{equation}
is the mean Bayes risk conditioned on stopping at $\bm{v}$,
\begin{equation}
	S_{k,m}^{\,\alpha,\beta} 
  	= \mu_{\alpha+k,\beta+m-k}
    = \frac{\alpha+k}{\alpha+\beta+m}
\label{eq:Skm}
\end{equation}
is the probability of the next trial being a success, and
\begin{equation}
    F_{k,m}^{\,\alpha,\beta}
    = 1 - \mu_{\alpha+k,\beta+m-k} 
    = \frac{\beta+m-k}{\alpha+\beta+m}
\label{eq:Fkm}
\end{equation}
\end{subequations}
is the probability of the next trial being a failure.
The dynamic program summarized in Algorithm~\ref{alg:dynamic-program} iteratively constructs a solution to \eqref{eq:min_problem_relx} by comparing 
$\Risk{\,\alpha}{\beta}{k}{m}$ to the lowest cost achievable from the state that results after a single trial.
{More precisely, at each node $(k,m)$, $ \Risk{\,\alpha}{\beta}{k}{m}$ is compared to the cost of one additional trial plus the expected lowest cost achievable from the subsequent state. We keep track of the lowest of these values in $V_{k,m}$, which is the lowest achievable cost from any node $(k,m)$.}
If $\Risk{\,\alpha}{\beta}{k}{m}$  is lower, then an additional trial is not warranted and the node is eliminated, i.e. $T \leftarrow T\setminus \! \{\bm{v}\}$ and $q_{k,m} = 0$.
Because of the decomposability of the problem, the solutions are optimal.

\begin{algorithm}
  \caption{Dynamic programming algorithm to find optimal deterministic stopping rule for $\BetaDist(\alpha,\beta)$ prior, Lagrange multiplier $\lambda$, and maximum depth $d$}
	\begin{algorithmic}
    	\State \textbf{Input:} $(\alpha, \beta)$, $\lambda \in \mathbb{R}_{+}$, $d \in \mathbb{N}$
        \State \textbf{Output:} $T^\ast \in 2^{\mathcal{T}_{d}}$
        \State \textbf{Initialize:} $T = \mathcal{T}_{d}$ and $[V_{k,m}]_{k,m} = 0$ for all $k,m$
        \For{$k = 1, \ldots, d$}
        	\State Set $V_{k,d} \leftarrow \Risk{\,\alpha}{\beta}{k}{d}$ using \eqref{eq:Rstop_dp}
        \EndFor
   		\For{$m = d-1, ... , 1$}
   		\For{$k = 1, 2, ..., m$}
      		\If{$ \Risk{\,\alpha}{\beta}{k}{m} > \lambda +  S_{k,m}^{\,\alpha,\beta} V_{k+1,m+1} + F_{k,m}^{\,\alpha,\beta} V_{k,m{+}1}$}
     			\State $V_{k,m} \gets \lambda + S_{k,m}^{\,\alpha,\beta} V_{k{+}1,m{+}1} + F_{k,m}^{\,\alpha,\beta} V_{k,m{+}1}$
      		\Else
     			\State $V_{k,m} \gets \Risk{\,\alpha}{\beta}{k}{m}$      
     			\State $T \gets T\setminus \! \{\bm{v}\}$
      		\EndIf
   \EndFor
   \EndFor \\
      \Return $T^\ast \gets T$
  \end{algorithmic}  
  \label{alg:dynamic-program}
\end{algorithm}

\subsection{A Greedy Algorithm}
\label{ssec:greedy-offline}

The DP method (Algorithm~\ref{alg:dynamic-program}) prunes from the complete trellis $\mathcal{T}_d$.
Monotonicity of the objective $g_{\alpha,\beta}(T)$
and cost $h_{\alpha, \beta}(T)$
can be exploited to develop a lower-complexity greedy algorithm that instead builds a trellis starting from just the root node.

The scheme outlined in Algorithm~\ref{alg:greedy-offline} monotonically improves the objective function value for the minimization problem \eqref{eq:minimization-problem} with each iteration.
Specifically, at iteration $i$, the greedy decision is to add to the current trellis $T_i$ a node $\bm{v} \not\in T_i$ that yields the largest reduction in the Bayes risk \textit{per additional trial},
\begin{equation*}
\frac{g_{\alpha, \beta}(T_{i}\cup {\bm{v}})  - g_{\alpha, \beta} (T_{i})}{h_{\alpha, \beta} (T_{i}\cup {\bm{v}}) - h_{\alpha, \beta} (T_{i})},
\end{equation*}
without violating the mean number of trials constraint. The scheme terminates when no such node exists.

\begin{algorithm}
  \caption{Greedy algorithm to find deterministic stopping rule for $\BetaDist(\alpha,\beta)$ prior and trial budget $\eta$}
  \label{alg:greedy-offline}
	\begin{algorithmic}
    	\State \textbf{Input:} $(\alpha, \beta)$, $\trialbudget$
        \State \textbf{Output:} $T^\ast$
        \State \textbf{Initialize:} $i \gets 0, \, T_0 \gets \{\}$
	\Repeat
    	\State $\displaystyle \bm{\tilde{v}} \gets \argmin_{\bm{v} \not\in T_i} \,\frac{g_{\alpha, \beta}(T_{i}\cup {\bm{v}})  - g_{\alpha, \beta} (T_{i})}{h_{\alpha, \beta} (T_{i}\cup {\bm{v}}) - h_{\alpha, \beta} (T_{i})}$
    	\State $T_{i+1} \gets T_{i} \cup {\bm{\tilde{v}}}$
        \State $i \gets i+1$
    \Until{$h_{\alpha,\beta}(T_{i}) > \trialbudget$} \\
    \Return $T^\ast \gets T_{i-1}$
  \end{algorithmic}
\end{algorithm}

\subsection{Online Threshold-Based Termination}
\label{ssec:greedy-online}
Our final method applies a simple rule for termination of trials, depending on the prior parameters $(\alpha,\beta)$ and the $(k,m)$ position in the trellis.
It implies a trellis design, but it does not require storage of a designed trellis.

Suppose a sequence of trials reaches a node corresponding to the posterior distribution $\mathrm{Beta}(a,b)$.
Denote the mean Bayes risk \emph{without} performing an additional trial by
\begin{equation}
	\Rstop(a,b) = \Rbeta{a}{b},
\label{eq:R_stop}
\end{equation}
using the variance given in \eqref{eq:beta_posterior}.
When one additional trial is performed, the posterior distribution is either $\mathrm{Beta}(a+1,b)$ if the outcome of the additional trial is a success,
or $\mathrm{Beta}(a,b+1)$ if the outcome of the additional trial is a failure.
Therefore, the mean Bayes risk resulting from continuing with one additional trial is
\begin{align}
\Rcont(a,b) =&\ \E\!\left[ (1-P) \, \Rbeta{a}{b+1} + P \, \Rbeta{a+1}{b} \right] \nonumber \\
            =&\ \frac{ab}{(a+b)(a+b+1)^2}.
\end{align}
The Bayes risk \emph{reduction} from one additional trial is
\begin{align}
\Delta R(a,b) =&\ \Rstop(a,b) - \Rcont(a,b) \nonumber \\
              =&\ \frac{ab}{(a+b)^2(a+b+1)^2}.
\end{align}

\begin{figure}
\begin{tabular}{@{}c@{}c@{}}
\includegraphics[height=4.2cm,keepaspectratio,trim={2cm 2cm 3cm 0cm},clip]{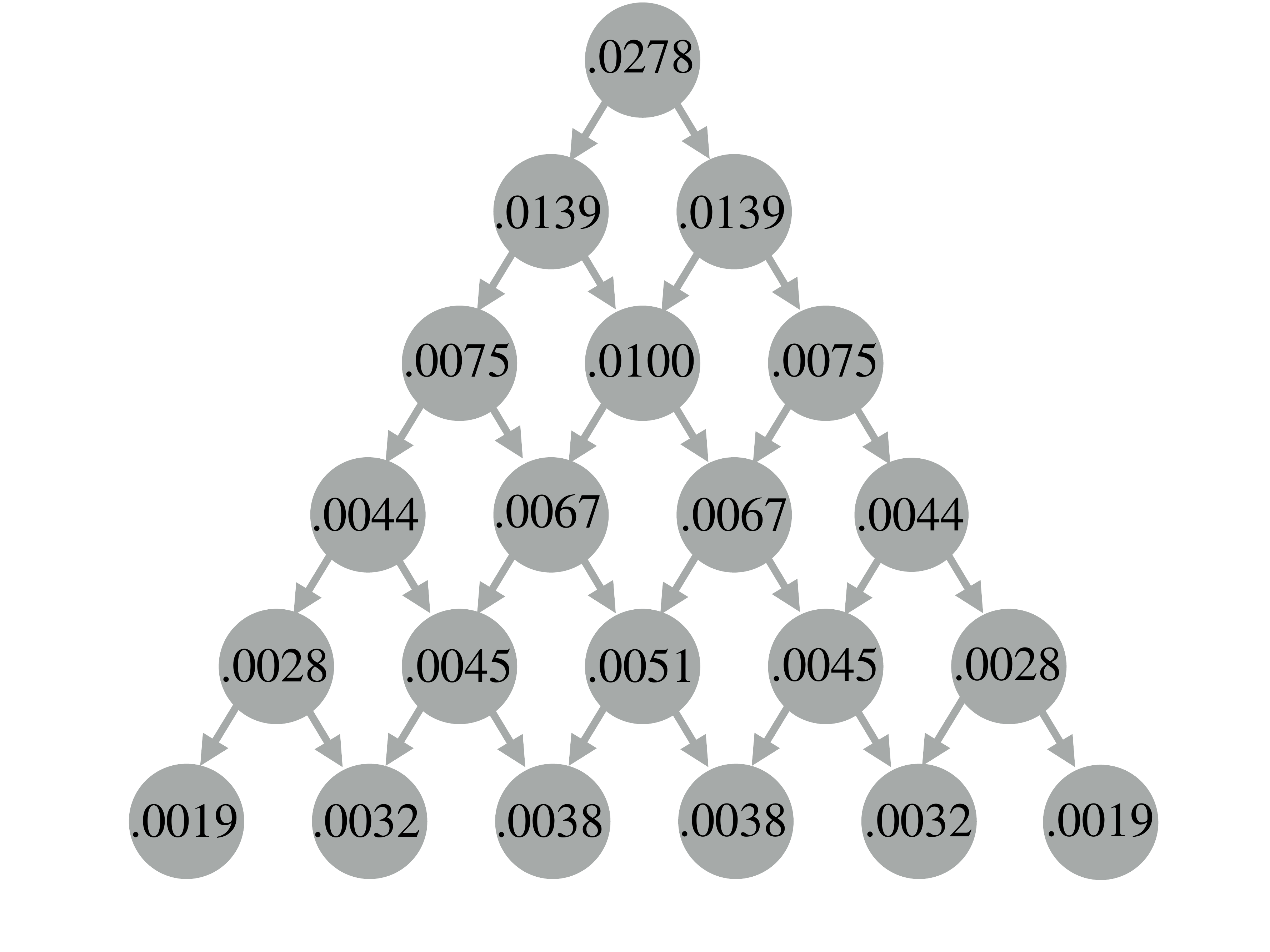} 
&
\includegraphics[height=4.2cm,keepaspectratio, trim={8cm 2cm 8cm 0cm},clip]{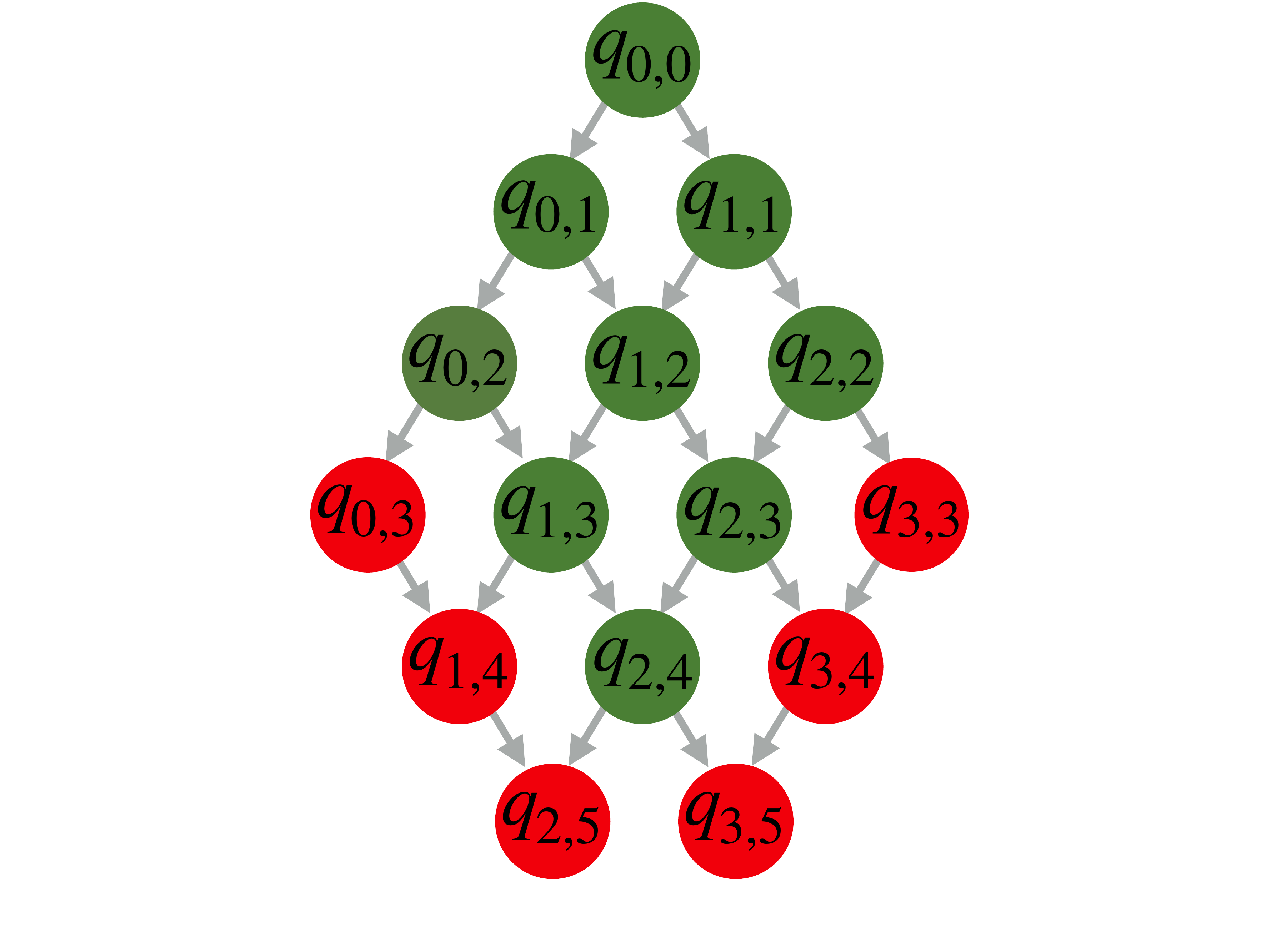} \\
{\small (a)} & {\small (b)}
\end{tabular}
\caption{(a) Bayes risk reductions per additional trial.
(b) Resulting trellis of continuation probabilities for $\DeltaThresh = 0.005$ (right).
$\BetaDist(1,1)$ (i.e., uniform) prior for $P$ has been assumed.}
\label{fig:example-trellis}
\end{figure}

Recall that, starting from a $\BetaDist(\alpha,\beta)$ prior, upon reaching node $(k,m)$, the posterior is $\BetaDist(\alpha+k,\beta+m-k)$.
The Bayes risk reduction from an additional trial,
\begin{equation}
\label{eq:DeltaR-km}
\Delta R(k,m;\alpha,\beta) = \frac{(\alpha+k)(\beta+m-k)}{(\alpha+\beta+m)^2(\alpha+\beta+m+1)^2} ,
\end{equation}
can be the basis of an online stopping rule.
Let $\DeltaThresh > 0$ denote a specified threshold value for the reduction in Bayes risk that justifies an additional trial.
Then stopping based on this threshold induces the probabilities of continuing at each node of the trellis given by
\begin{equation}
 q_{k,m} = 
\begin{cases}
    1, \ \  \Delta R(k,m;\alpha,\beta) \geq \DeltaThresh; \\ 
    0, \ \  \Delta R(k,m;\alpha,\beta)  <   \DeltaThresh.
\end{cases} 
\label{eq:qkm-from-thresholding}
\end{equation}
Fig.~\ref{fig:example-trellis}(a) shows values of $\Delta R(k,m;1,1)$ for $m=0,1,\ldots,5$.
The choice of threshold $\DeltaThresh = 0.005$ results in the trellis shown in
Fig.~\ref{fig:example-trellis}(b).

Notice that for a fixed trellis depth $m$, the denominator of \eqref{eq:DeltaR-km} is fixed, and the numerator of \eqref{eq:DeltaR-km} is a product of factors with fixed sum that is equal to $\alpha + \beta + m$.
Thus, from the arithmetic--geometric mean inequality, $\Delta R(k,m;\alpha,\beta)$ is largest where the posterior distribution is most symmetric.
This is apparent in the example in Fig. \ref{fig:example-trellis}(b);
since we have started with a uniform prior, the center of each row represents a symmetric posterior,
and additional observations are most merited near the center of each row.
Starting with a highly asymmetric prior ($\alpha \ll \beta$ or $\alpha \gg \beta$),
the same principle explains an asymmetry in the greedily optimized trellis of continuation probabilities.

\begin{myExample}[Suboptimality of binomial sampling]
\label{ex:suboptimality-binomial}
Suppose we have a $\BetaDist(1,1)$ (i.e., uniform) prior.
Then \eqref{eq:DeltaR-km} simplifies to
\begin{equation*}
\Delta R(k,m;1,1) = \frac{(k+1)(m-k+1)}{(m+2)^2(m+3)^2} .
\end{equation*}
For the threshold-based termination to induce binomial sampling with $m^*$ trials,
the incremental benefit $\Delta R$ at $(k,m) = (0,m^*)$ must be greater than
$\Delta R$ at $(k,m) = ( \lfloor\half(m^*+1)\rfloor ,\,m^*+1)$:
\begin{align}
& \frac{m^*+1}
       {(m^*+2)^2(m^*+3)^2} \nonumber \\
& \quad \geq
  \frac{( \lfloor\half(m^*+1)\rfloor +1)(m^*- \lfloor\half(m^*+1)\rfloor +2)}
       {(m^*+3)^2(m^*+4)^2}.
\label{eq:binomial-optimality-test}
\end{align}
Since \eqref{eq:binomial-optimality-test} fails to hold for any $m^* > 2$,
threshold-based termination induces binomial sampling only for $1$ and $2$ trials.
This is consistent with Fig.~\ref{fig:example-trellis}.
For such a small trial budget, full enumeration of stopping rules is also feasible,
and one can conclude that binomial sampling is indeed suboptimal for any trial budget
greater than 2\@.
Similar arguments can be made for non-uniform beta priors.
\end{myExample}

\subsection{Comparisons of Designs}

\begin{figure}
\centering
\includegraphics[width=0.8\linewidth]{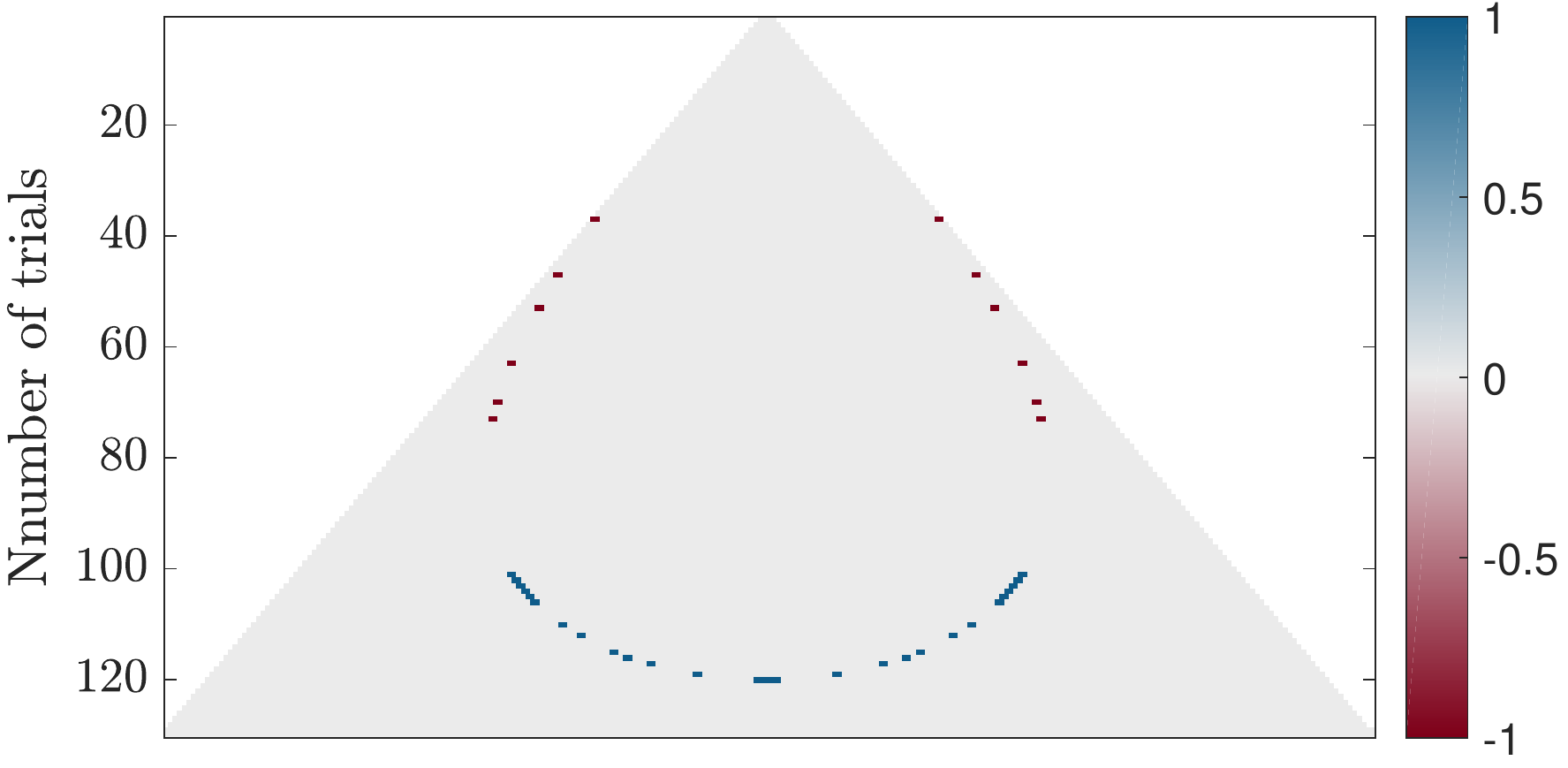}
\caption{Dynamic programming solution minus online threshold-based termination result, both with mean number of trials $\approx 95.36$.
Online threshold-based termination achieves mean Bayes risk of $0.0016037$ whereas DP gives $0.0016036$.
At 24 nodes (red, `-1'), threshold-based rule performs additional trials and DP does not;
at 54 nodes (blue, `+1'), DP performs trials and threshold-based rule does not.
The greedily designed trellis coincides with the DP trellis, hence their difference plot is omitted.}
\label{fig:dp-greedy-diff}
\end{figure}

Sweeping $\DeltaThresh$ in threshold-based termination is very similar to sweeping $\lambda$ in Algorithm~\ref{alg:dynamic-program};
it will achieve certain mean numbers of trials, similar to sweeping $\eta$ in Algorithm~\ref{alg:greedy-offline}.
Intermediate values of the mean number of trials can be achieved by finding $(k^*,m^*)$ such that $\Delta R(k,m;\alpha,\beta)$ is largest among those below $\DeltaThresh$ and varying $q_{k^*,m^*}$ over $(0,1)$.
This idea is used to enforce an equal expected number of trials for trellises optimized with each method, thus allowing a fair comparison of their Bayes risks.
For a mean number of trials $\approx 95.36$, Algorithms~\ref{alg:dynamic-program} and~\ref{alg:greedy-offline} were found to give exactly the same trellis,
while online threshold-based termination gave a slightly different trellis with slightly higher mean Bayes risk.
Fig.~\ref{fig:dp-greedy-diff} illustrates the difference in $q_{k,m}$ values.
It is zero for the vast majority of nodes,
with 24 nodes at which the DP-designed trellis terminates but the threshold-based rule does not (red, $-1$),
and 54 nodes at which the threshold-based rule terminates but the DP-designed trellis does not (blue, $+1$).\footnote{The mean number of trials is equal.  To be convinced that the blue and red nodes can balance, note that while there are more blue nodes, they are for larger values of $m$ and thus have lower probabilities of being reached.}

\begin{figure}
\centering
\includegraphics[width=0.8\linewidth]{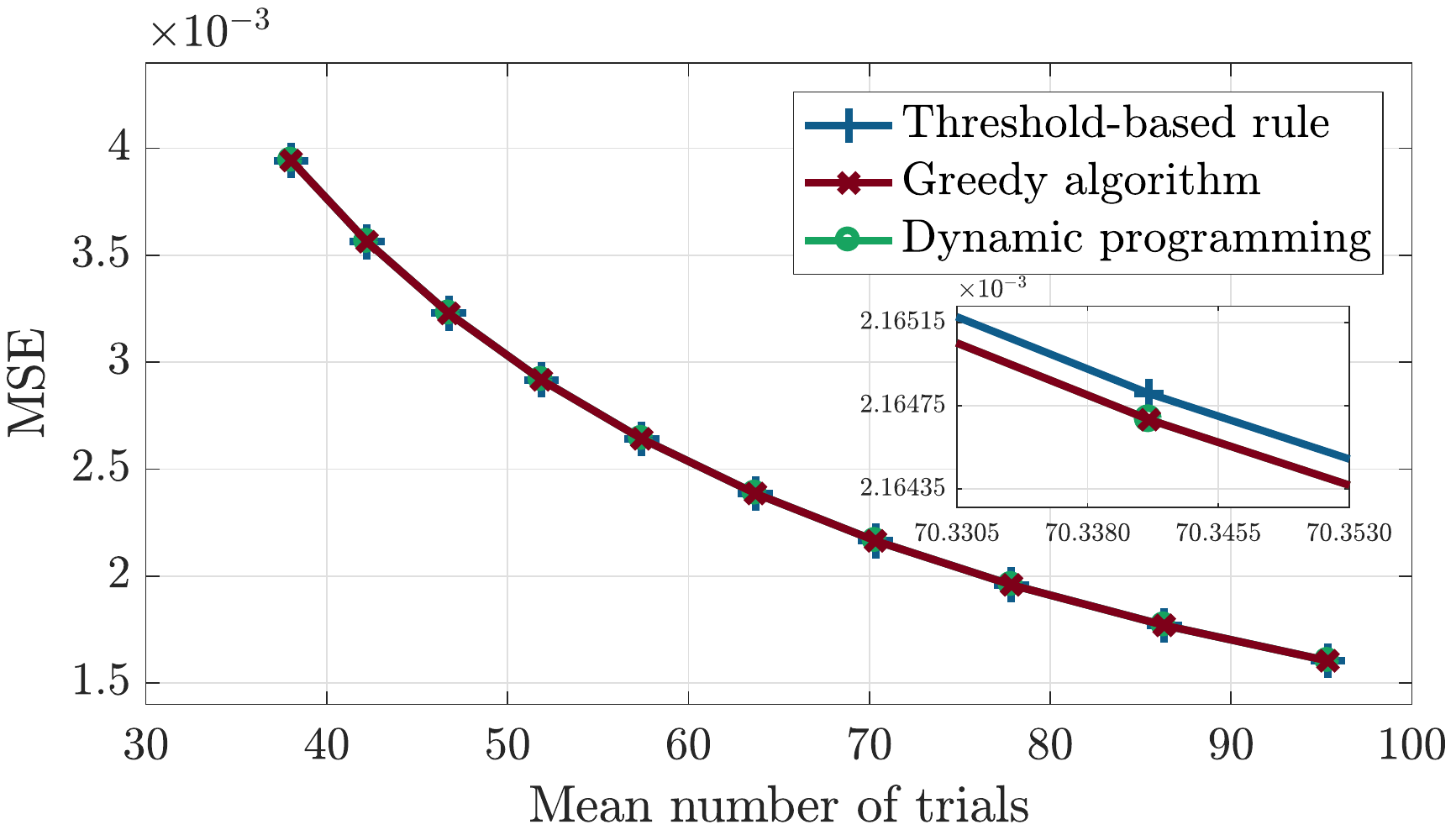}
\caption{Comparison between proposed strategies.
DP and greedy algorithm solutions coincide for all trial budgets,
while threshold-based termination is only very slightly worse by a factor of at most 1.000195.
}
\label{fig:dynamic-vs-greedy}
\end{figure}

Illustrated in Fig.~\ref{fig:dynamic-vs-greedy} is a comparison of our three proposed implementable strategies,
applied for a uniform prior, over a range of trial budgets.
MSEs of DP
(Algorithm~\ref{alg:dynamic-program})
and the greedily optimized trellis
(Algorithm~\ref{alg:greedy-offline})
coincide for all trial budgets
because the trellises are identical~-- though we have not proven that this is guaranteed.
The online threshold-based stopping rule is only very slightly worse by a factor of at most 1.000195 (less than 0.001 dB).

\begin{figure}
\centering
\includegraphics[width=0.8\linewidth]{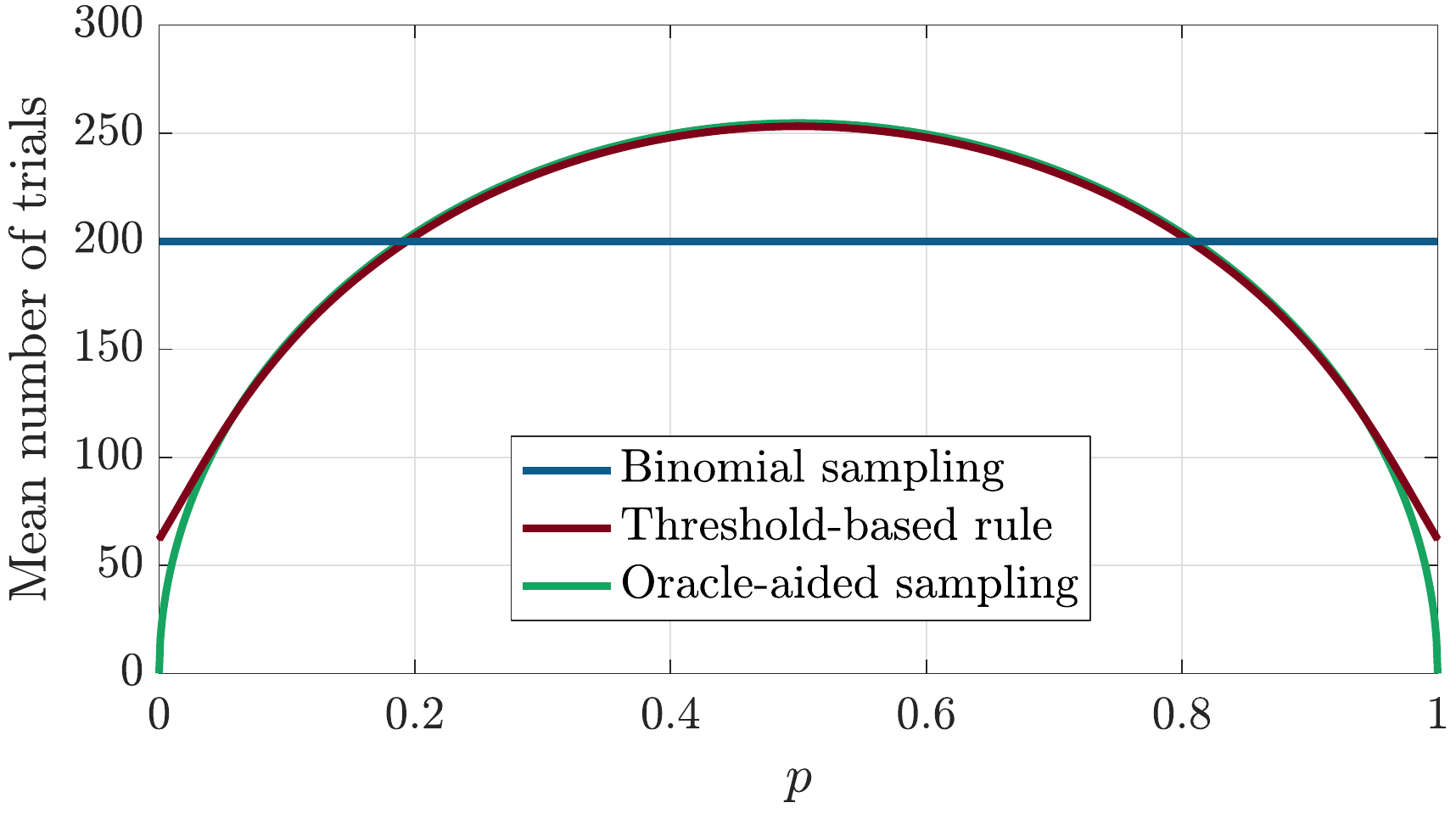} \\
{\small (a)} \\
\includegraphics[width=0.8\linewidth]{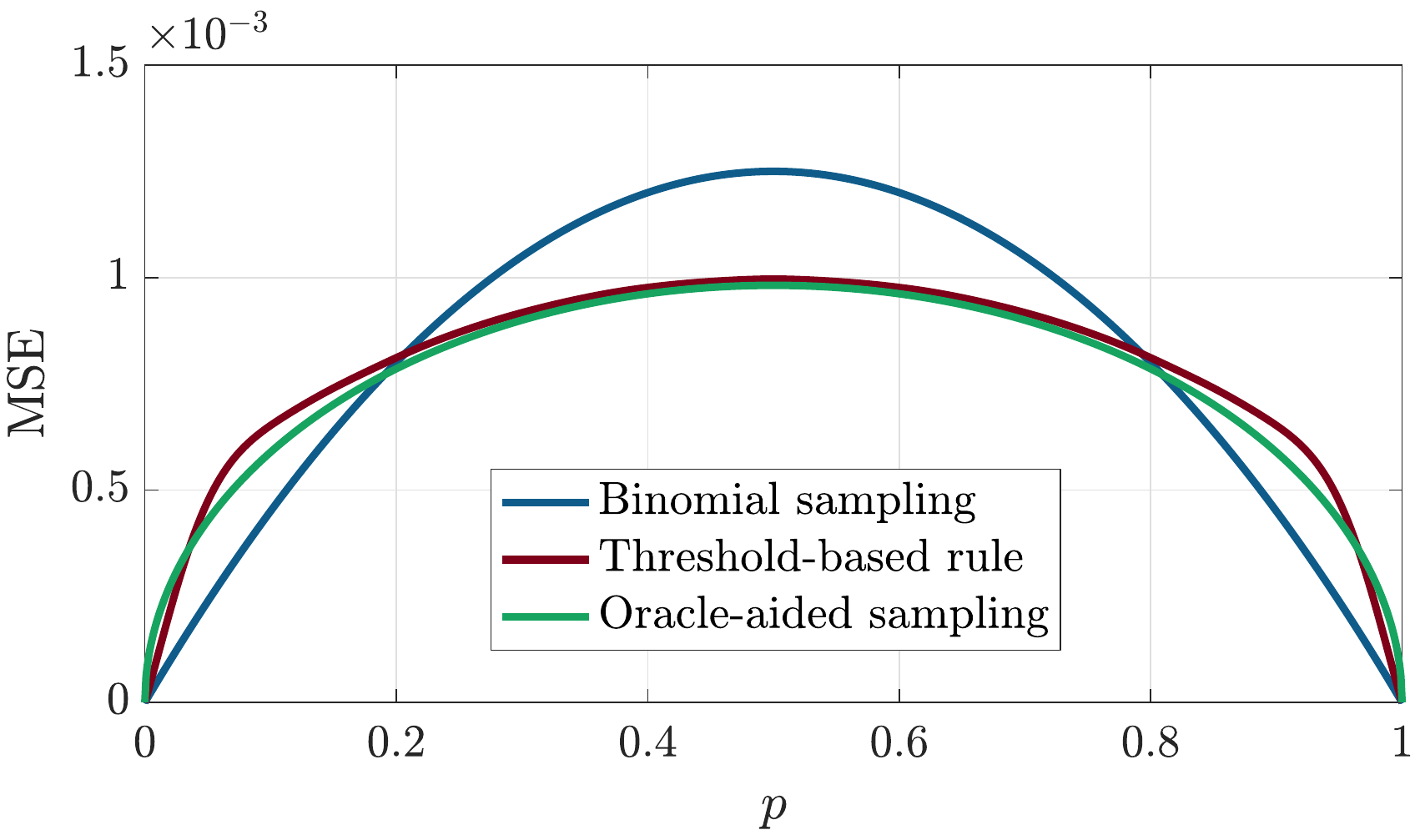} \\
{\small (b)}
\caption{Dependences on $p$ under a uniform prior.
(a) Conditional expectation of the number of trials, conditioned on $p$,
for binomial sampling, online threshold-based termination, and the oracle-aided binomial stopping rule.
Expected number of trials is 200 for all of the three methods.
(b) Dependence of conditional MSEs on the true Bernoulli parameter $p$ under a trial budget of $\trialbudget=200$.
} 
\label{fig:bin_greedy_oracle-EN_ER-comp}
\end{figure}

The phenomenon of more trials being merited when $p$ is near $\frac{1}{2}$ counteracts the MSE of $p(1-p)/n$ being largest for $p$ near $\frac{1}{2}$.
This is illustrated in Fig.~\ref{fig:bin_greedy_oracle-EN_ER-comp}(a), which shows mean numbers of trials allocated as a function of $p$.
We have optimized for MSE averaged over $p$ and, in so doing, obtained a modest improvement factor of $\approx 1.05$ in this average, comparing the online threshold-based termination to conventional binomial sampling.
A more significant reduction in the worst-case MSE is a by-product of the optimization (see Fig.~\ref{fig:bin_greedy_oracle-EN_ER-comp}(b)).

\subsection{Asymptotic Comparison with Oracle-Aided Allocation}
\label{ssec:comparison_oracle}

Considering the non-degenerate cases $p \in (0,1)$, the threshold-based termination is asymptotically equivalent to oracle-aided optimal allocation.
For a large trial budget $\eta$, we will find an approximation for $m^\circ$,
the number of trials at which the online threshold-based rule terminates.
This will match the form of
\eqref{eq:oracleaided_solution}
or \eqref{eq:trial-allocation-random}.

Using \eqref{eq:DeltaR-km}, for an initial $\BetaDist(\alpha,\beta)$ prior,
the online rule continues at node $(k,m)$ if and only if
\begin{equation}
\frac{(\alpha+k)(\beta + m - k)}{(\alpha+\beta+m)^2 (\alpha+\beta+m+1)^2} \leq \DeltaThresh.
\label{eq:BR_reduction_comparison}
\end{equation}
Since the trial budget is large and $p \in (0,1)$,
$k$, $m$, and $m-k$ are all large when nearing termination.
Hence, we approximate the expression in \eqref{eq:BR_reduction_comparison} as
\begin{align}
& \frac{(\alpha+k)(\beta + m - k)}
       {(\alpha+\beta+m)^2 (\alpha+\beta+m+1)^2} \nonumber \\
& \quad
= \frac{m^2 (\alpha/m+k/m) (\beta/m + 1 - k/m)}
       {m^4 (\alpha/m+\beta/m+1)^2 (\alpha/m+\beta/m+1+1/m)^2} \nonumber \\
& \quad
\approx \frac{(k/m) (1 - k/m)}
             {m^2} \nonumber \\
& \quad
= \frac{ \pML (1 - \pML) }
             {m^2},
\label{eq:asymptotic-equivalence-approx}
\end{align}
where $\pML = k/m$ is the ML estimate of $p$.

Substituting \eqref{eq:asymptotic-equivalence-approx}
into \eqref{eq:BR_reduction_comparison},
we obtain
\begin{equation}
m^\circ \approx \sqrt{
 \frac{\pML (1 - \pML)}{\DeltaThresh}
}.
\label{eq:mcirc}
\end{equation}
By the law of large numbers, $\pML \rightarrow p$,
so \eqref{eq:mcirc} shows a match to 
\eqref{eq:oracleaided_solution},
with $\DeltaThresh$ determining the trial budget.
Furthermore, by comparison with \eqref{eq:trial-allocation-random},
we see an equivalence  by choosing
$\DeltaThresh = \big({\E\!\big[\sqrt{P(1-P)}\big]} / \trialbudget\big)^2$.

Fig.~\ref{fig:bin_greedy_oracle-EN_ER-comp}(a) illustrates an example of the approximate match
between threshold-based termination and oracle-aided sampling that is predicted by
the match among
\eqref{eq:oracleaided_solution},
\eqref{eq:trial-allocation-random}, and
\eqref{eq:mcirc}.
Note that convergence is not uniform in $p$;
a larger trial budget is needed to observe approximate equivalence in allocations
for $p$ near $0$ and near $1$.

\begin{figure}
\centering
\vspace{0.04cm}
\includegraphics[width=0.8\linewidth]{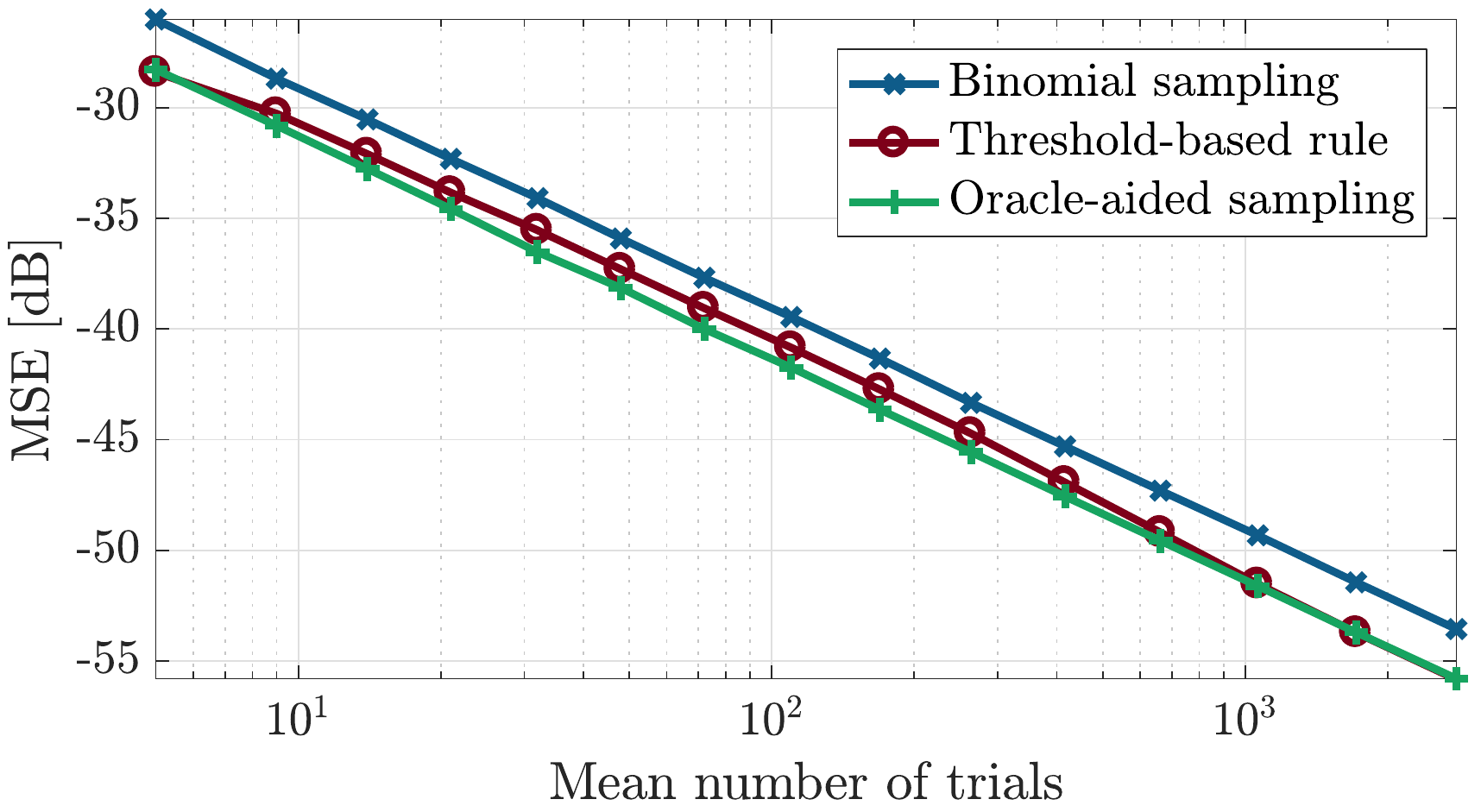} \\[0.03cm]
{\small (a)} \\[0.25cm]
\includegraphics[width=0.8\linewidth]{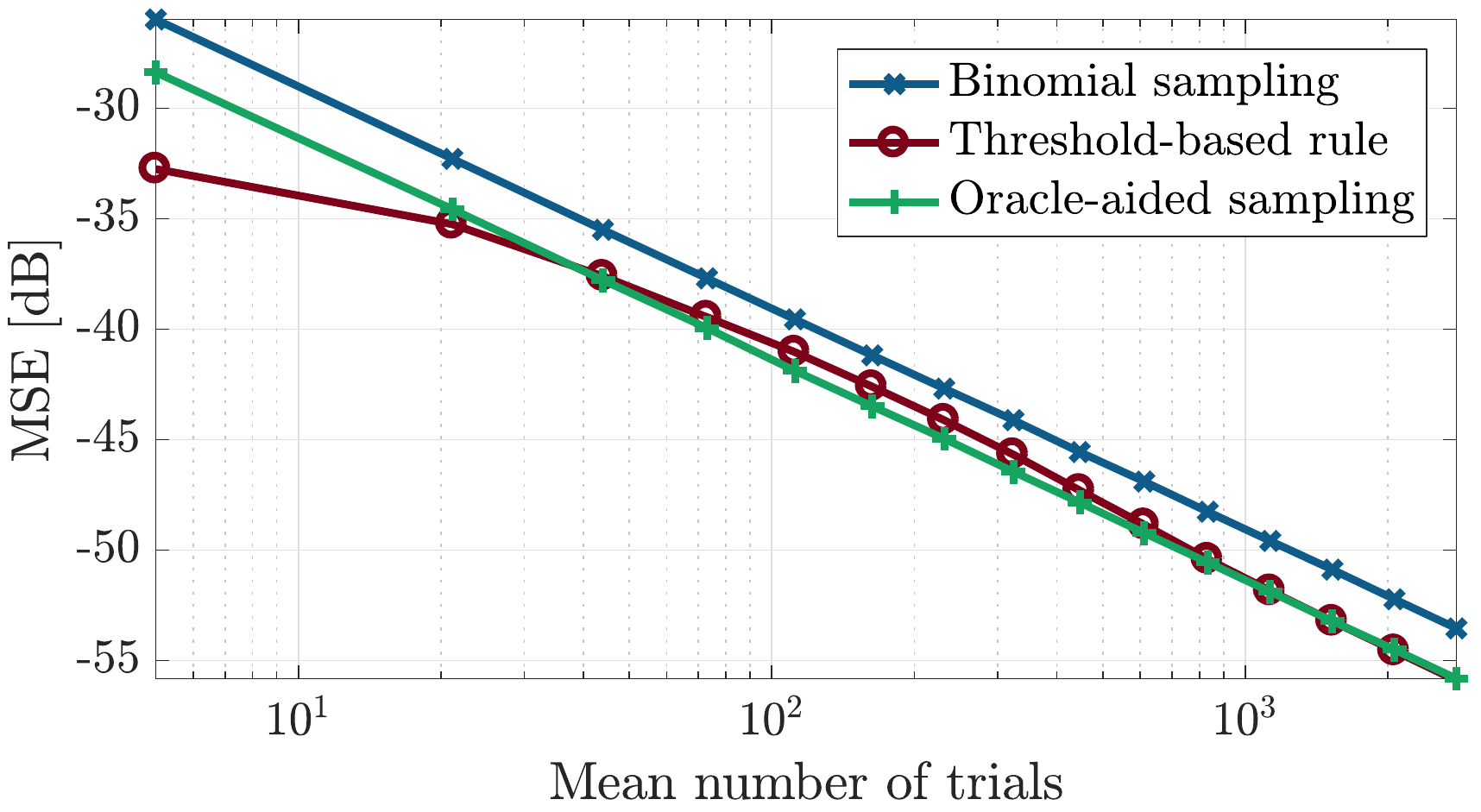} \\
{\small (b)} \vspace{0.05cm}
\caption{Results for the Shepp--Logan Phantom scaled to [0.001,\, 0.101], assuming: (a) $\BetaDist(1,1)$ and (b) $\BetaDist(1,50)$, for the online threshold-based termination.
The MSE has been computed from the average of 20 independent experiments for each mean number of trials.
Improvements are consistent with the trial allocation gain computed in
Example~\ref{ex:gains}(d).
}
\label{fig:coding-gain-Beta-1-x}
\end{figure}

Fig.~\ref{fig:coding-gain-Beta-1-x} shows the variation of the MSEs with mean number of trials budget constraint for conventional binomial sampling,
threshold-based termination,
and oracle-aided allocation.
The results are based on Monte Carlo simulations, with MATLAB, using the phantom image in Fig.~\ref{fig:Shepp-Logan}(a).
As expected the optimized rules consistently achieve MSE improvements over the conventional binomial sampling, for all simulated trial budgets.
In addition, when compared to the unrealizable oracle-aided method, the threshold-based approach only marginally under-performs at moderate mean number of trials budget constraints.
This observation is further underscored in Fig.~\ref{fig:bin_greedy_oracle-EN_ER-comp}, which shows significant overlap between threshold-based termination
and oracle-aided allocation, in terms of both trial allocations and the resulting MSEs. {Using {a} negative binomial sampling strategy yields significantly worse performance than binomial sampling and our proposed rules for estimating $p$; thus, we have omitted it from Figs.~\ref{fig:bin_greedy_oracle-EN_ER-comp} and~\ref{fig:coding-gain-Beta-1-x}, as well as other numerical simulations related to the estimation of $p$.}

In line with the earlier asymptotic analysis, the threshold-based termination and oracle-aided performances coincide for moderate
to high mean numbers of trials, 
independent of the prior.
Under a highly skewed prior consistent with the true distribution of the phantom pixels,
Fig.~\ref{fig:coding-gain-Beta-1-x}(b) demonstrates that
it is possible for online threshold-based termination
to even outperform the oracle-aided binomial method, at low trial budgets.
This phenomenon is attributable to the
online method allocating more trials when the Bernoulli process realization has
a relatively high fraction of successes.
Put simply, it is allocating more trials for ``unlucky'' realizations where the MSE would be higher,
while the oracle-aided binomial method maintains a fixed number of trials.

\section{Estimating Functions of a Bernoulli Parameter}
\label{sec:estimating-functions}
When estimating an arbitrary function $f(p)$ of a Bernoulli parameter is of interest \cite{Hubert2000}, one can derive similar stopping strategies as before. In this section, we concern ourselves only with the estimation of $f(p) = \log p$ due to its prevalence in real-life scenarios. For instance, the subjective brightness perceived by the human vision system is a logarithmic function of the incident light intensity~\cite{Gonzalez2006,Milner2017}.
Also, the common \emph{log odds ratio} $\log( p/(1-p) )$, is approximately equal to $\log p$ when $p \ll 1$.

As before, we begin with a squared error loss
\begin{equation}
L(p,\cdot) = \big(f(p) - \hat{f}(\cdot)\big)^2,
\label{eq:loss-log}
\end{equation}
where $\hat{f}(\cdot)$ is the estimate of $f(p)$.
The expectation of this loss function over $p$ gives the Bayes risk
\begin{equation}
R(\cdot) = \E[L(p,\cdot)] = \E\!\big[\big(f(p) - \hat{f}(\cdot)\big)^2\big].
\label{eq:log-bayes-risk}
\end{equation}

For $f(p) = \log{p}$, suppose a sequence of trials leads to a node in the trellis corresponding to the posterior distribution $\mathrm{Beta}(a,b)$.
Under $P \sim \mathrm{Beta}(a,b)$, the MMSE estimator of $\log p$ is \cite{Karlis2005}
\begin{equation}
\hat{f}(a,b) := \E\!\left[\log P\right] = \psi^{(0)}(a) - \psi^{(0)}(a+b),
\end{equation}
where $\psi^{(m)}$ is the polygamma function of order $m$.
The Bayes risk \eqref{eq:log-bayes-risk} when no additional trial is performed, $R_{\mathrm{stop}}$, becomes the variance of $\log p$ \cite{Dette2017}:
\begin{equation}
R_{\mathrm{stop}} \defeq R(a,b) = \var{\log P} = \psi^{(1)}(a) - \psi^{(1)}(a+b).
\end{equation}
If one additional trial is performed, the Bayes risk reduces to
\begin{align}
\Rcont(a,b) =&\ \E\!\left[ (1-P) \, R(a,b+1) + P \,  R(a+1,b) \right] \nonumber \\
            =&\  \frac{b}{a+b} \, R(a,b+1) + \frac{a}{a+b} \,  R(a+1,b) .
\end{align}
Hence, the Bayes risk reduction from one additional trial is
\begin{equation}
  \Rstop(a,b) - \Rcont(a,b) 
    = \frac{b}{a \, (a+b)^2}.
  \label{eq:log-deltaBR}
\end{equation}
Starting with prior $\BetaDist(\alpha,\beta)$,
the counterpart to \eqref{eq:DeltaR-km} for estimation of $\log p$ is
\begin{equation}
\label{eq:log-DeltaR-km}
\Delta R(k,m;\alpha,\beta)
  = \frac{\beta + m -k }
         {(\alpha + k)(\alpha+\beta+m)^2} .
\end{equation}
As before, this can be used in
\eqref{eq:qkm-from-thresholding}
as an online threshold-based termination method.

\begin{figure}
\centering
\includegraphics[width=0.7\linewidth]{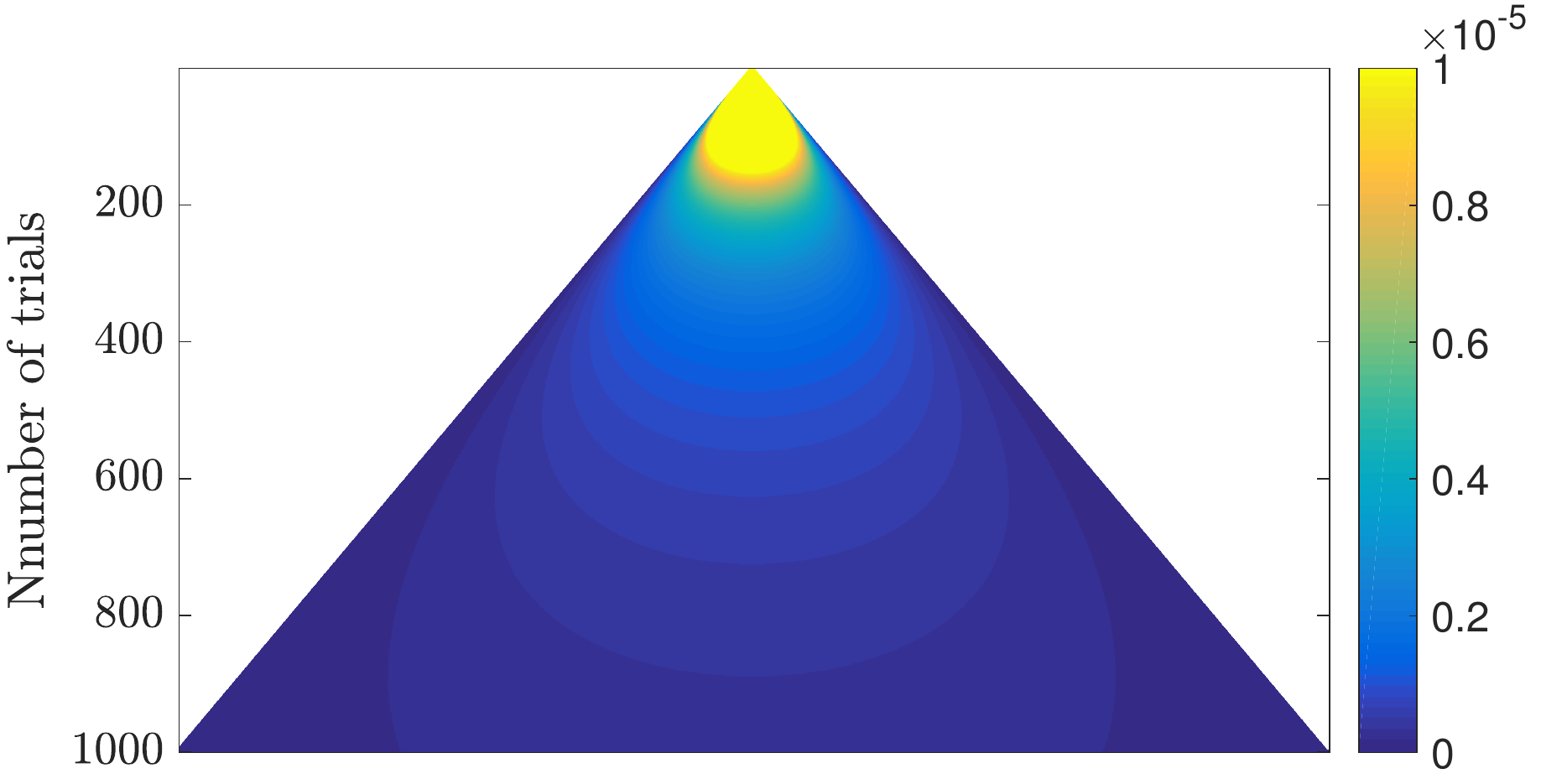} \\
{\small (a)} \\
\includegraphics[width=0.7\linewidth]{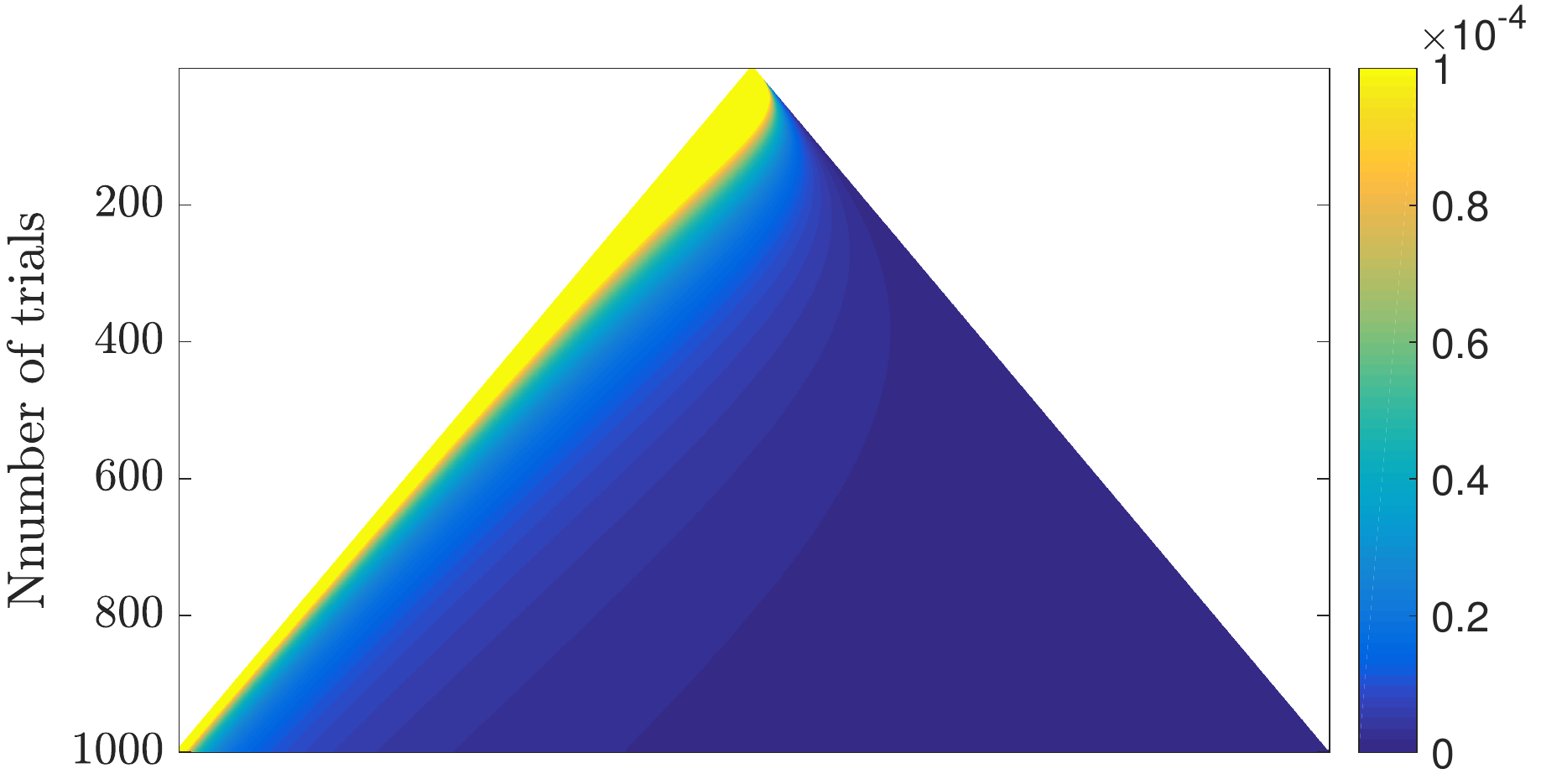} \\
{\small (b)}
\caption{Bayes risk reductions for estimation of (a) $f(p) = p$ and (b) $f(p) = \log{p}$, assuming $\mathrm{Beta}(1,1)$ assumed initially.
Low $p$ values are assigned significantly more trials when $\log{p}$ is estimated.}
\label{fig:p-p-logp-trellises}
\end{figure}

The Bayes risk reductions for both $f(p)=p$
in \eqref{eq:DeltaR-km}
and $f(p) = \log p$
in \eqref{eq:log-DeltaR-km}, starting with a uniform prior,
are shown as heat maps in Fig.~\ref{fig:p-p-logp-trellises}.
When $f(p) = \log p$, the reduction from additional trials after observing sequences with low number of successes is significantly larger.
Thus, the online threshold-based termination of Section~\ref{ssec:greedy-online} is likely to assign more trials for the smaller underlying Bernoulli parameters.
Such a stopping rule is intuitive because a fixed amount of estimation error for $p$ would contribute more to the loss function defined in \eqref{eq:loss-log}, with $f(p) = \log p$, when $p$ is small.
In fact, one can choose a loss function that enforces different penalties for different $p$ values.
An example is the family of weighted mean squared errors,
$\E[w(p)(p-\hat{p})^2]$, where a weighting function $w(p)$ is designed according to the problem.
A special case is relative MSE $\E[(p-\hat{p})^2/p^2]$,
which is approximately the squared loss in \eqref{eq:log-bayes-risk} with $f(p) = \log p$ for estimates sufficiently close to the true value~\cite{Hubert2000}.

\begin{figure}
\centering
\includegraphics[width=0.8\linewidth]{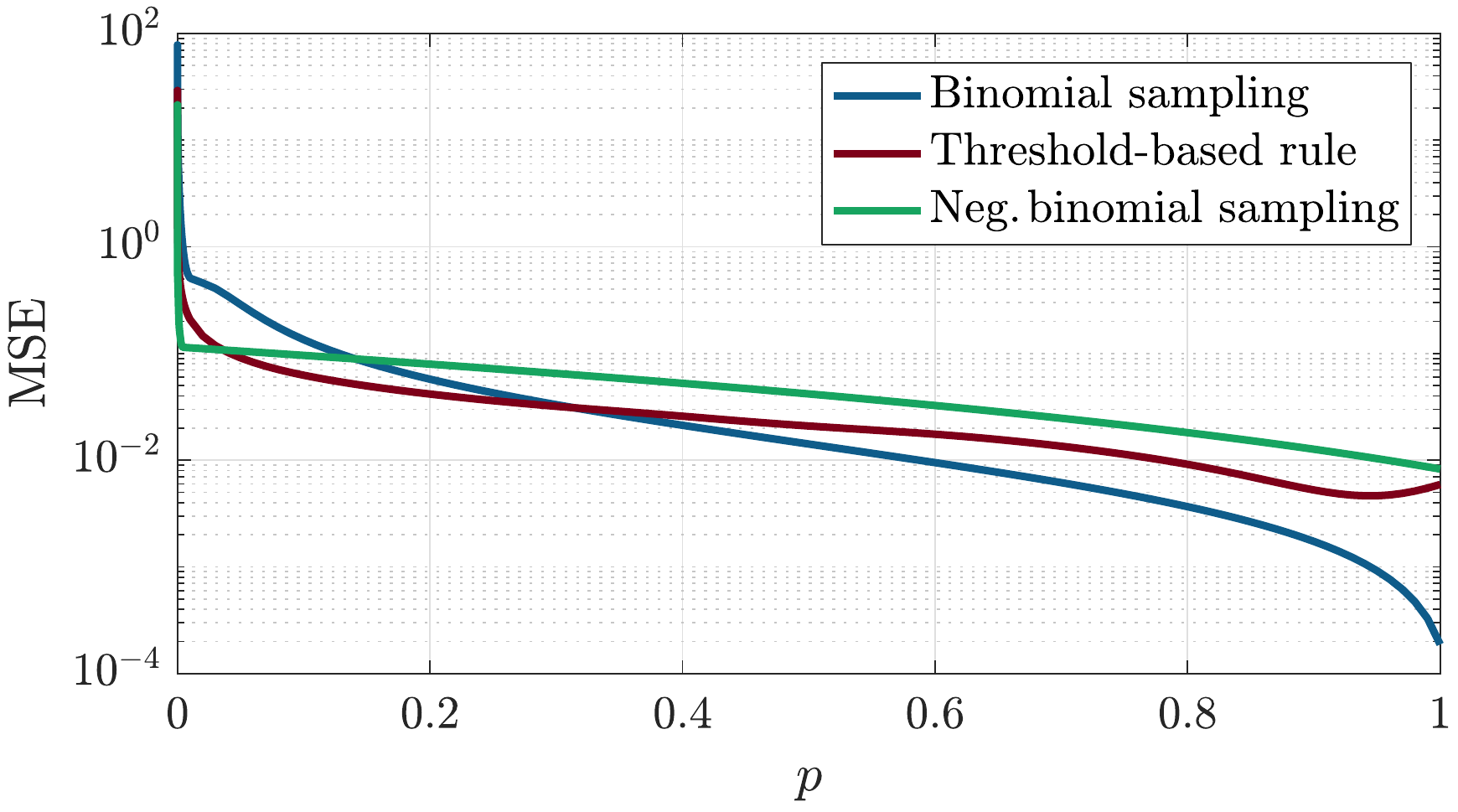}
\caption{Estimation of $\log p$.
Conditional MSE of $\log p$, conditioned on $p$, is shown as a function of $p$ for binomial sampling ($\eta = 72$),
negative binomial sampling ($\ell = 10$, inducing $\eta = 71.66$), and
online threshold-based termination ($\eta = 71.77$).
The threshold-based termination uses a uniform prior,
and averaging over a uniform prior gives
MSE of 0.0668 for binomial sampling,
0.0489 for negative binomial sampling,
and 0.0328 for threshold-based termination\@.
{These values are computed directly from their corresponding trellises (hence exact), not through numerical integration.}
}
\label{fig:log-estimation}
\end{figure}

When smaller $p$ values are of more importance, it makes sense to use a strategy that allocates more trials to these instances.
Negative binomial sampling explained in Section~\ref{ssec:ExistingProtocols} achieves this type of trial allocation.
For the estimation of $f(p) = \log p$, we compare the performances of binomial sampling, negative binomial sampling, and online threshold-based termination in Fig.~\ref{fig:log-estimation}.
Threshold-based termination outperforms both binomial and negative binomial sampling, with improvement factors of $2.037$ and $1.491$, respectively.

\section{Applications to Active Imaging}
\label{sec:applications}

Active imaging systems typically raster scan the scene by probing patch $(i,j)$, $i=1,\ldots,N_i$ and $j=1,\ldots,N_j$, using pulsed illumination.
The measured data~-- used to form an image of the scene~-- are arrays $[k_{i,j}]_{i,j}$ and $[m_{i,j}]_{i,j}$;
i.e., the number of detections (successes) and number of illumination pulses (trials) for each scene patch.
Note that the conventional approach of a fixed number of trials makes $m_{i,j}= \trialbudget$ for all $(i,j)$ and $\{k_{i,j}\}$ random, whereas both $\{k_{i,j}\}$ and $\{m_{i,j}\}$ are random when the proposed approach is applied.

The parameters of the Bernoulli processes generated by probing a scene patch and its neighbors are typically correlated.
This can be exploited in the image formation stage through mechanisms inspired by any of various image compression or denoising methods.
For this initial demonstration of adaptive acquisition, we apply total variation (TV) regularization \cite{louchet2008}.
We present simulation results using the Shepp--Logan phantom in Fig.~\ref{fig:Shepp-Logan}(a),
two lidar {datasets provided by the Alaska Department of Natural Resources~\cite{AlaskaLidar}, and scanning electron microscopy (SEM) images
{\emph{Foraminifera}\footnote{Quanta SEM image of Protozoan group secreting a calcareous shell
by Philippe Crassous,
\url{https://www.fei.com/image-gallery/Foraminifera-Protozoan/}}
and \emph{HairStyle}\footnote{Quanta SEM image of the upper part of the style and stigma from an \emph{Arabidopsis} flower by Guichuan Hou,
\url{https://www.fei.com/uploadedImages/FEISite/Content/Image_Gallery/Images/2013_Image_Contest/FEI/IM_20130718_Hou_18_HairStyle_lg.jpg}}}
taken from ThermoFisher {Scientific}.
All images have been rescaled to take on values in the range $[0.001,0.101]$.}

\subsection{Estimation of $f(p) = p$}
\label{sec:image_p}
We focus here on comparing conventional binomial sampling against online threshold-based termination
\eqref{eq:qkm-from-thresholding}
applied for each pixel.
For $f(p) = p$, the relevant Bayes risk reduction per trial is given by \eqref{eq:DeltaR-km}.

\subsubsection{MMSE Estimation Under I.I.D. Prior}
\label{sec:pixelwise-mmse-images}
When not exploiting any spatial correlations, each pixel estimation is performed separately using the methods of Section~\ref{ssec:greedy-online}.
Under a $\BetaDist(\alpha,\beta)$ prior, the MMSE estimate is
$\phatMMSE{[i,j]} = \Frac{(k_{i,j} + \alpha)}{(m_{i,j} + \alpha + \beta)}$.

\begin{figure}
\centering
\includegraphics[width=0.9\linewidth, trim={8mm 0mm 7mm 2mm},clip]{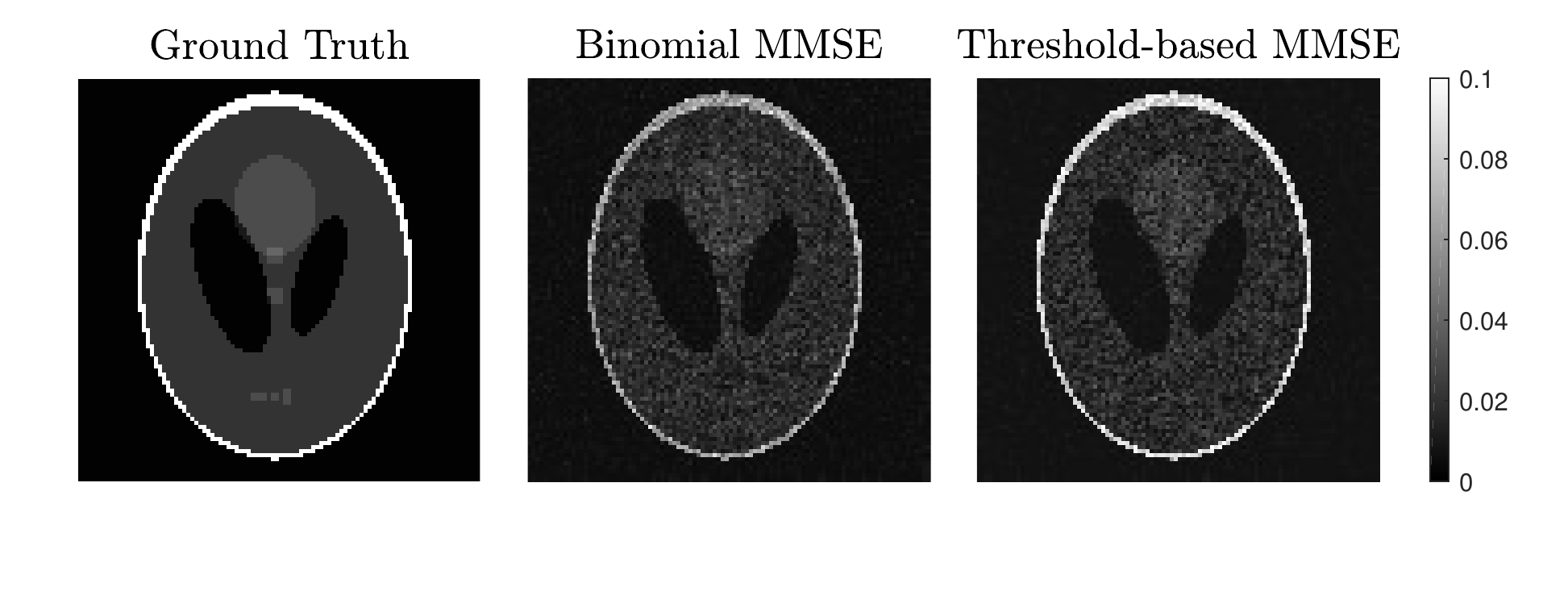} \\[-3mm] \vspace{-0.55cm}
{\footnotesize\begin{flushleft} \hspace*{31.7mm} PSNR 20.0 dB \hspace*{5.5mm} PSNR 22.4 dB \end{flushleft}} \vspace{-0.3cm}
\centerline{\footnotesize(a) Shepp--Logan phantom, $\BetaDist{(2,152)}$ and $\eta = 200$.} \vspace{0.1cm}

\includegraphics[width=0.9\linewidth, trim={8mm 11mm 7mm 8mm},clip]{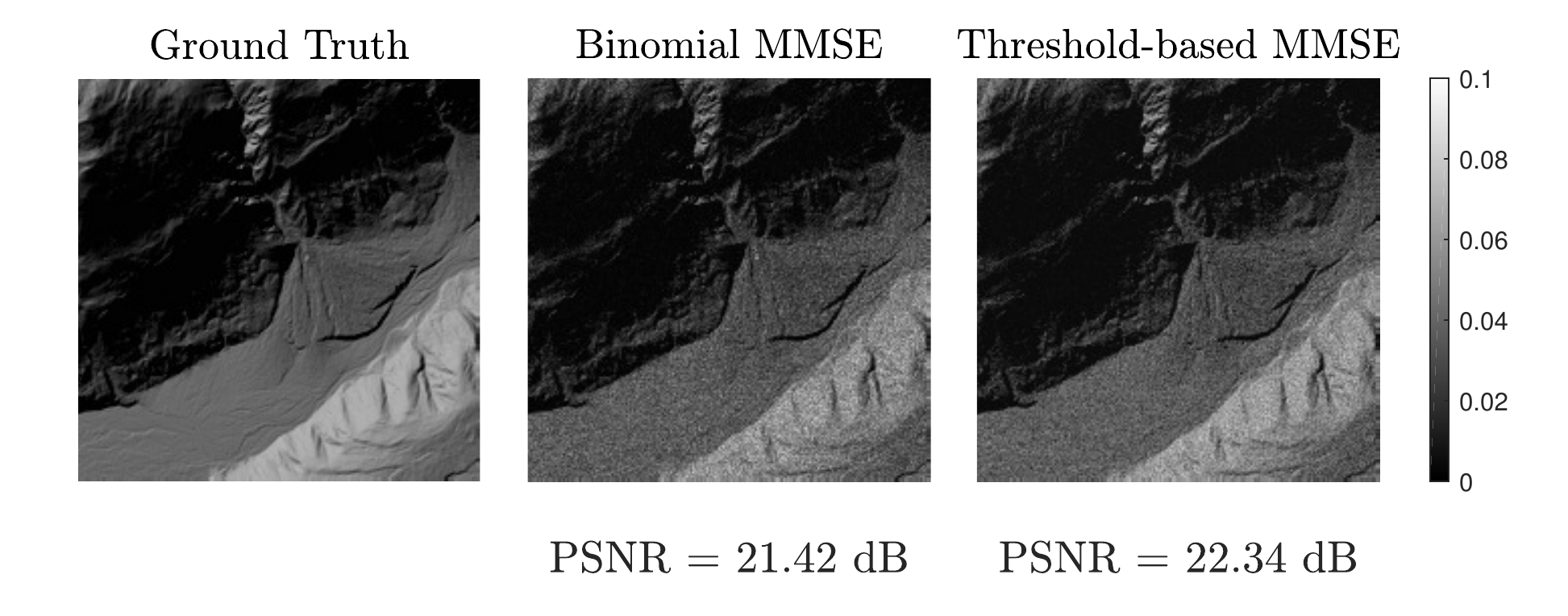}  \\[-3mm] \vspace{-0.05cm}
{\footnotesize\begin{flushleft} \hspace*{31.7mm} PSNR 21.4 dB \hspace*{5.5mm} PSNR 22.3 dB \end{flushleft}} \vspace{-0.3cm}
\centerline{\footnotesize(b) lidar \#1, $\BetaDist{(2,162)}$ and $\eta = 800$.} \vspace{0.1cm}

\includegraphics[width=0.9\linewidth, trim={8mm 11mm 7mm 8mm},clip]{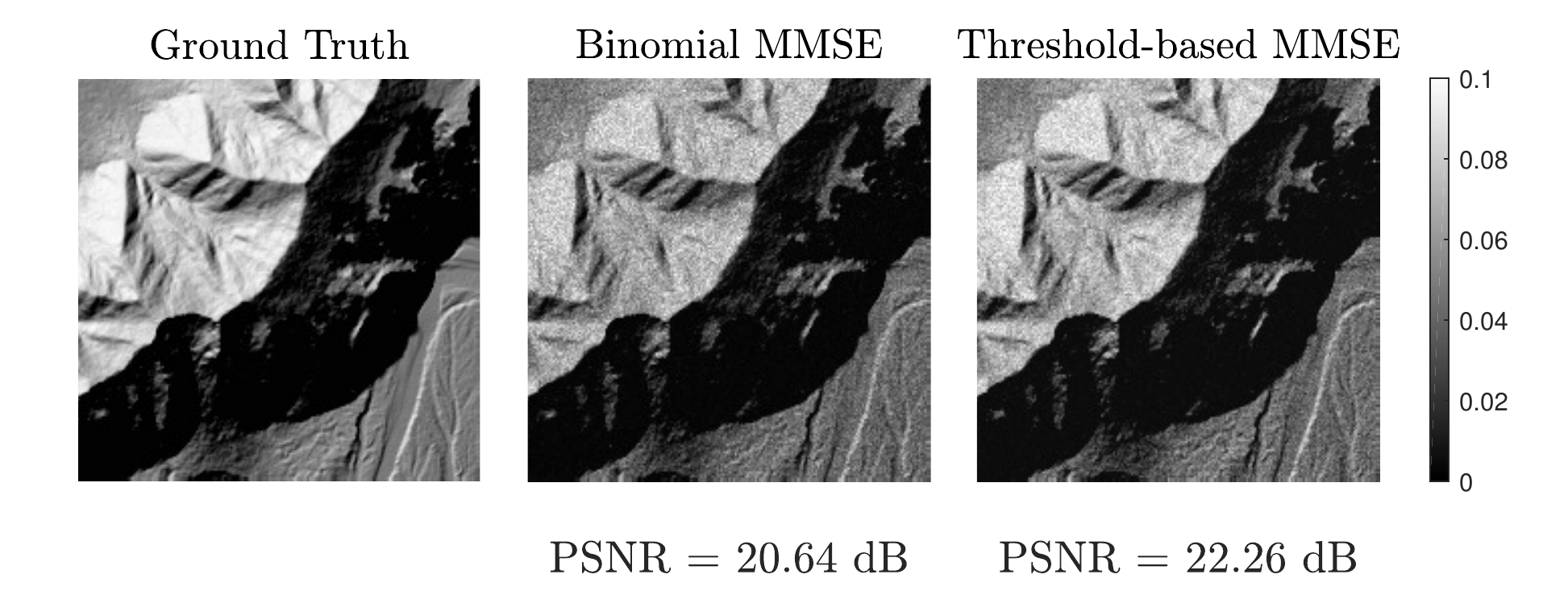}  \\[-3mm] \vspace{-0.05cm}
{\footnotesize\begin{flushleft} \hspace*{31.7mm} PSNR 20.6 dB \hspace*{5.5mm} PSNR 22.3 dB \end{flushleft}} \vspace{-0.3cm}
\centerline{\footnotesize(c) lidar \#2, $\BetaDist{(2,172)}$ and $\eta = 800$.} \vspace{0.1cm}

\includegraphics[width=0.9\linewidth, trim={8mm 11mm 7mm 8mm},clip]{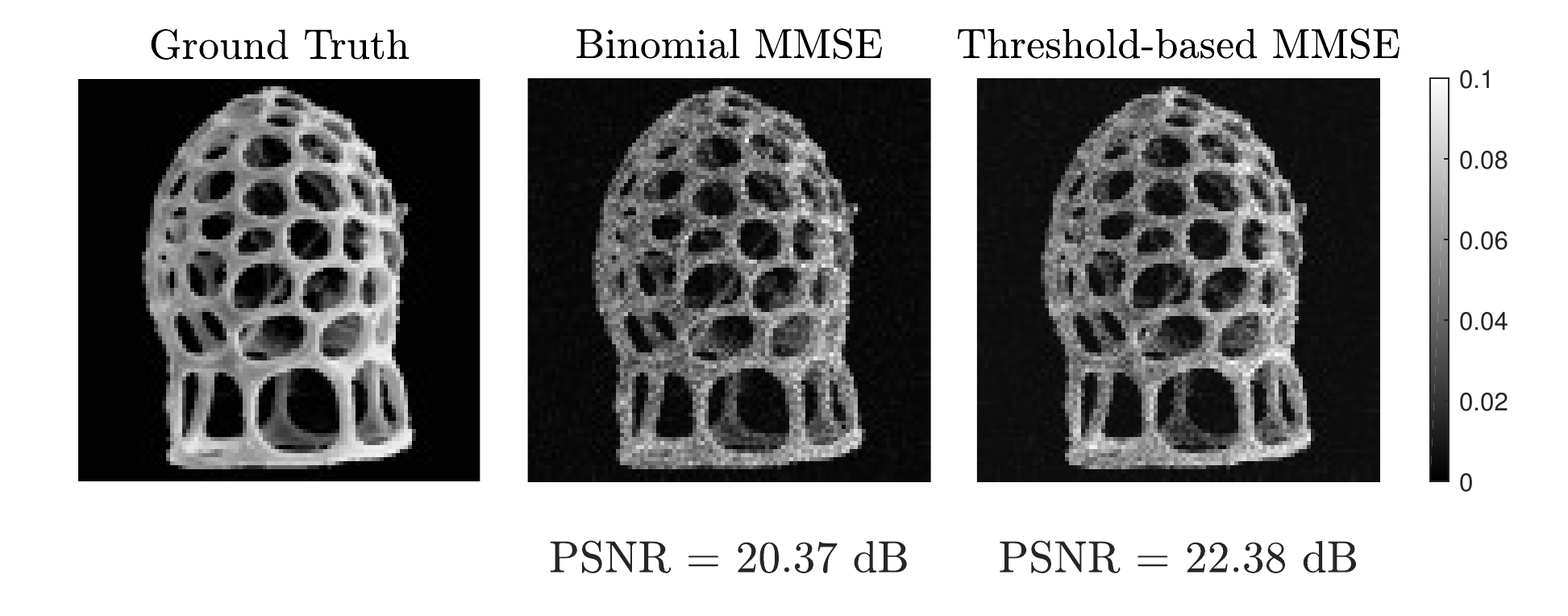}   \\[-3mm] \vspace{-0.05cm}
{\footnotesize\begin{flushleft} \hspace*{31.7mm} PSNR 20.4 dB \hspace*{5.5mm} PSNR 22.4 dB \end{flushleft}} \vspace{-0.3cm}
\centerline{\footnotesize(d) \emph{Foraminifera}, $\BetaDist{(2,162)}$ and $\eta = 500$.} \vspace{0.1cm}

\includegraphics[width=0.9\linewidth, trim={8mm 11mm 7mm 8mm},clip]{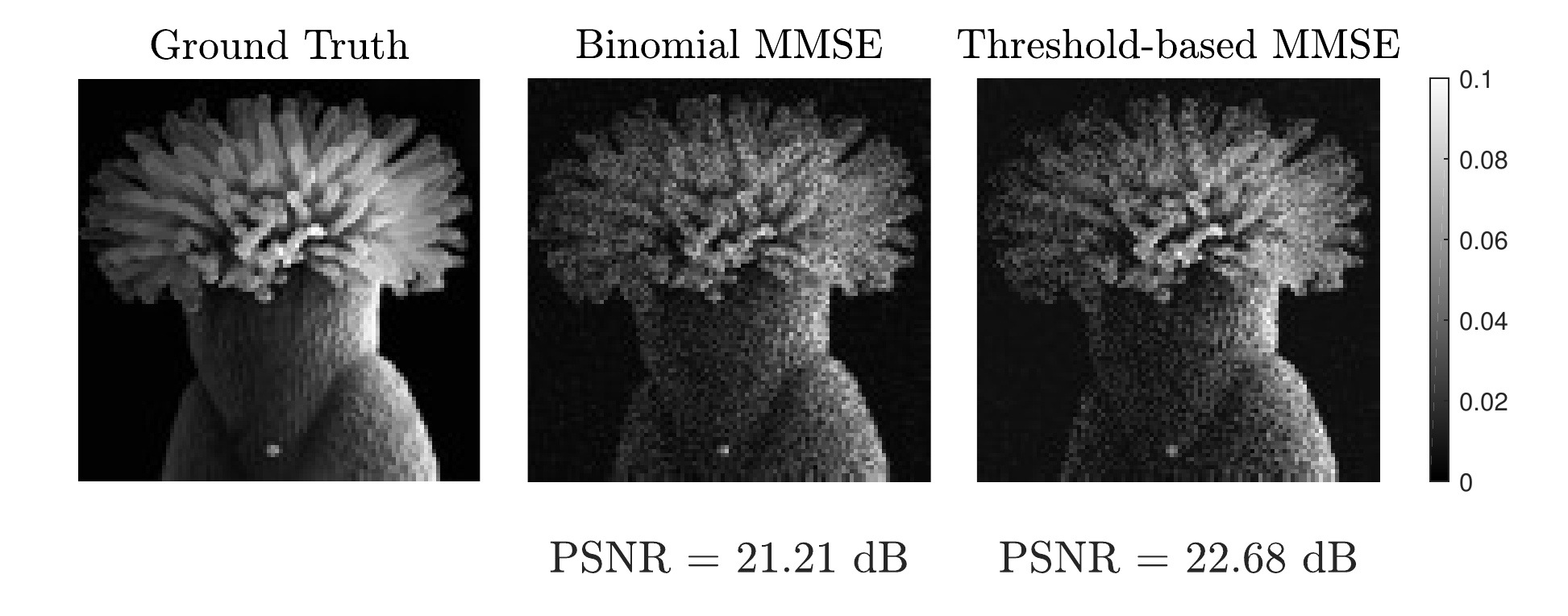}   \\[-3mm] \vspace{-0.05cm}
{\footnotesize\begin{flushleft} \hspace*{31.7mm} PSNR 21.2 dB \hspace*{5.5mm} PSNR 22.7 dB \end{flushleft}} \vspace{-0.3cm}
\centerline{\footnotesize(e) \emph{HairStyle}, $\BetaDist{(2,162)}$ and $\eta = 400$.}
\caption{Images reconstructed through pixelwise MMSE estimation showing MSE improvements of up to $2.42\,\mathrm{dB}$ for online threshold-based termination in place of conventional binomial sampling. Assumed priors and trial budgets are indicated below each test image. All images are scaled to $[0.001,0.101]$.
}
\label{fig:phantom-mmse}
\end{figure}

\begin{figure}
\centering
\includegraphics[width=0.8\linewidth]{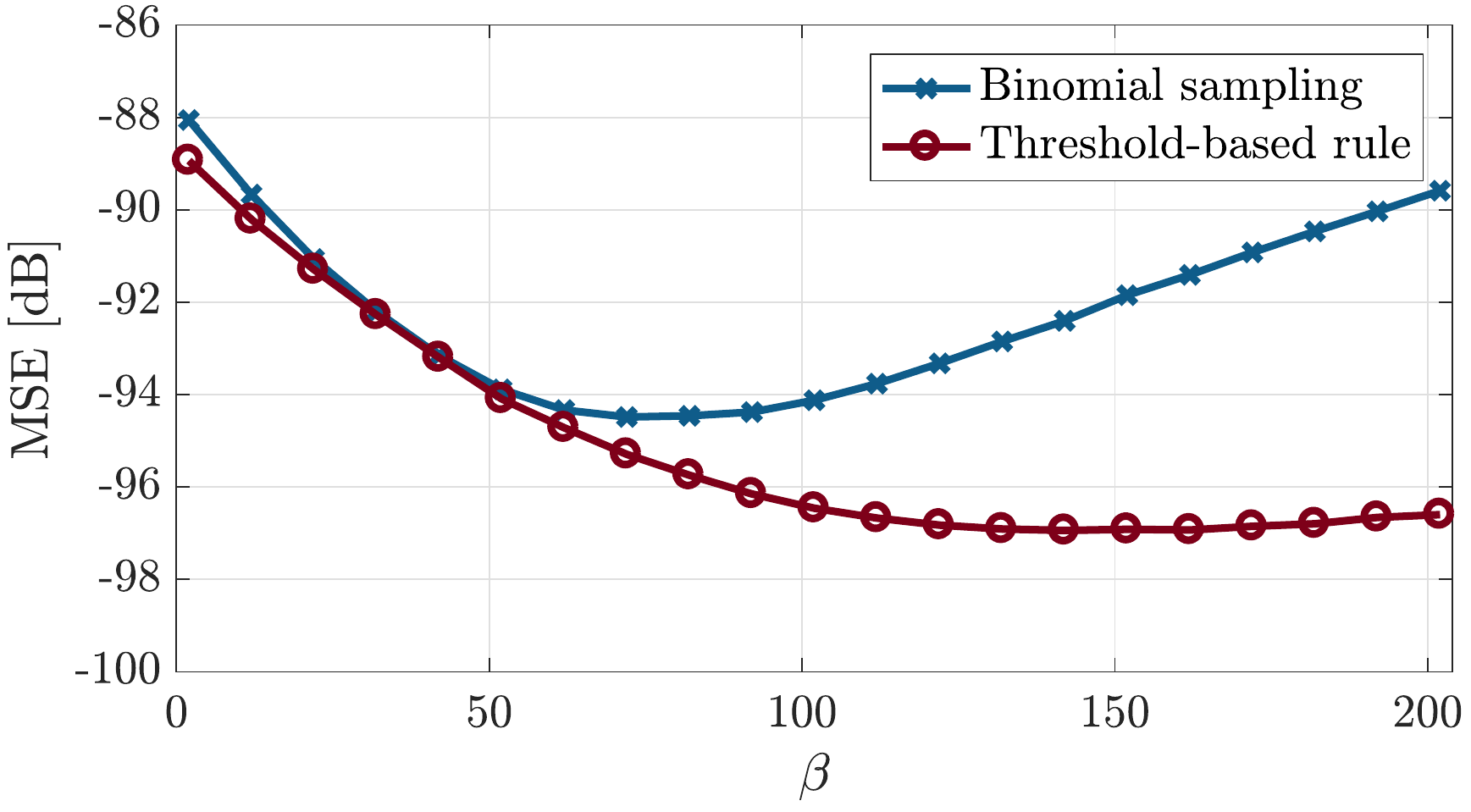}
\centerline{\footnotesize (a) Shepp--Logan phantom, $\eta=200$}

\medskip

\includegraphics[width=0.8\linewidth]{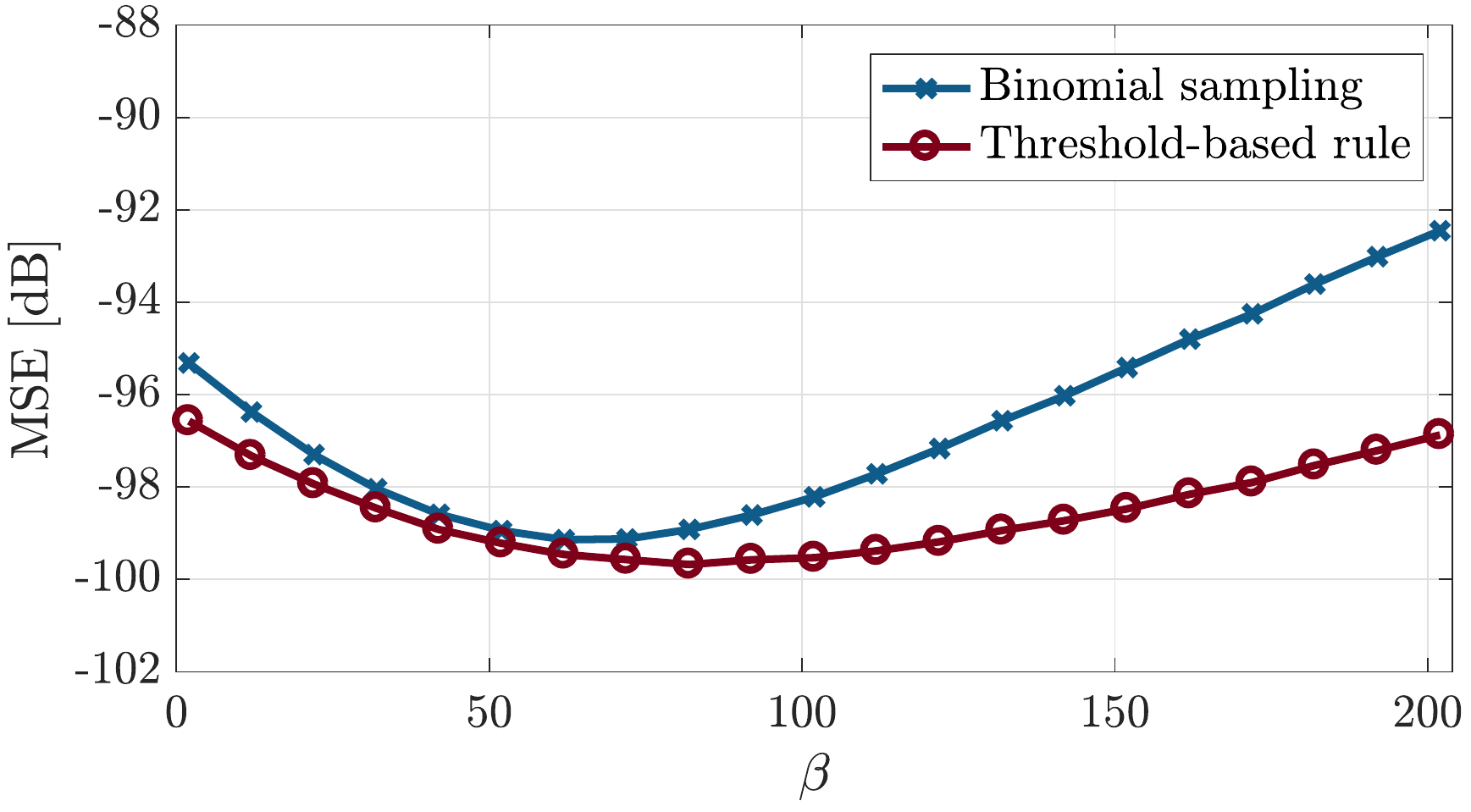}
\centerline{\footnotesize (b) \emph{HairStyle}, $\eta=400$
}
\caption{Dependence of MSE on $\beta$
when $\mathrm{Beta}(2,\beta)$ prior is assumed and pixelwise MMSE estimation is performed; each plotted point is obtained by averaging over $100$ independent experiments.
}
\label{fig:mse-2-beta}
\end{figure}

Fig.~\ref{fig:phantom-mmse}(a) shows that with the choice of the $\BetaDist(2,152)$ prior,
MSE improvement of $2.42 \, \mathrm{dB}$ is attained for trial budget $\eta = 200$ for the Shepp--Logan phantom. In Figs.~\ref{fig:phantom-mmse}(b)--(e), MSE improvements were also demonstrated for the lidar and SEM images under various trial budgets and initial prior parameters. Improvements ranging from $0.92~\mathrm{dB}$ for lidar \#1 to
$2.02~\mathrm{dB}$ for \emph{Foraminifera} were obtained.
For the same corresponding choices of prior,
we also perform $100$ independent experiments
for each test image
at each of
two different trial budgets;
the results
indicated in Table~\ref{tbl:Average_MSE_p} show similar improvement factors.
In particular, the performance gains are close to the prediction from the trial allocation gain
(Example~\ref{ex:gains}(d) for the Shepp--Logan phantom image). The same is observed for other images.
Furthermore, we show in Fig.~\ref{fig:mse-2-beta} that significant MSE improvements can be attained for a large range of Beta priors,
using the Shepp--Logan phantom and \emph{HairStyle} datasets.
We also observe that the performance of binomial sampling
is degraded more by a mismatched prior than the performance of threshold-based termination.

\begin{table}
\centering
\caption{Average reconstruction PSNRs, averaged over 100 experiments,
for conventional binomial sampling and online threshold-based termination, for trial budgets $\eta$.}
  
\label{tbl:Average_MSE_p}
\begin{tabular}{@{}cccccc@{}}
\hline
\multirow{2}{*}{Image}  & \multirow{2}{*}{$\eta$} & \multicolumn{2}{@{}c@{}}{\scriptsize Pixelwise MMSE estimation} 
                        & \multicolumn{2}{@{}c@{}}{\scriptsize TV+ML estimation}
                        \\ \cline{3-6} 
                        &
                        & {\scriptsize Binomial}
                        & {\scriptsize Threshold-based} 
                        & {\scriptsize Binomial}
                        & {\scriptsize Threshold-based} \\ \hline
\multirow{2}{*}{\scriptsize Shepp--Logan}  & $100$       & $17.5$ dB & $\mathbf{20.0}$ $\dB$ & $21.9$ dB & $\mathbf{26.3}$ $\dB$ \\
& $200$      & $19.9$ dB & $\mathbf{22.1}$ $\dB$ & $24.9$ dB & $\mathbf{28.5}$ $\dB$ \\ \hline

\multirow{2}{*}{\scriptsize lidar \#1}  & $400$       & $18.1$ dB & $\mathbf{18.8}$ $\dB$ & $24.0$ dB & $\mathbf{25.9}$ $\dB$ \\
& $800$      & $21.4$ dB & $\mathbf{22.3}$ $\dB$ & $25.2$ dB & $\mathbf{27.7}$ $\dB$ \\ \hline

\multirow{2}{*}{\scriptsize lidar \#2}  & $400$       & $16.9$ dB & $\mathbf{18.3}$ $\dB$ & $26.2$ dB & $\mathbf{26.8}$ $\dB$ \\
& $800$      & $20.7$ dB & $\mathbf{22.3}$ $\dB$ & $27.6$ dB & $\mathbf{28.8}$ $\dB$ \\ \hline

\multirow{2}{*}{\scriptsize \emph{Foraminifera}}  & $400$       & $19.2$ dB & $\mathbf{21.2}$ $\dB$ & $25.5$ dB & $\mathbf{26.7}$ $\dB$ \\
& $500$      & $20.4$ dB & $\mathbf{22.4}$ $\dB$ & $23.5$ dB & $\mathbf{27.5}$ $\dB$ \\ \hline

\multirow{2}{*}{\scriptsize \emph{HairStyle}}  & $400$       & $21.2$ dB & $\mathbf{22.7}$ $\dB$ & $26.8$ dB & $\mathbf{27.8}$ $\dB$ \\
& $500$      & $22.3$ dB & $\mathbf{23.8}$ $\dB$ & $27.8$ dB & $\mathbf{28.5}$ $\dB$ \\ \hline

\end{tabular}
\end{table}

\subsubsection{TV-Regularized ML Estimation}
\label{sec:tv-images}

\begin{figure}
\centering
\includegraphics[width=0.9\linewidth, trim={8mm 0mm 7mm 2mm},clip]{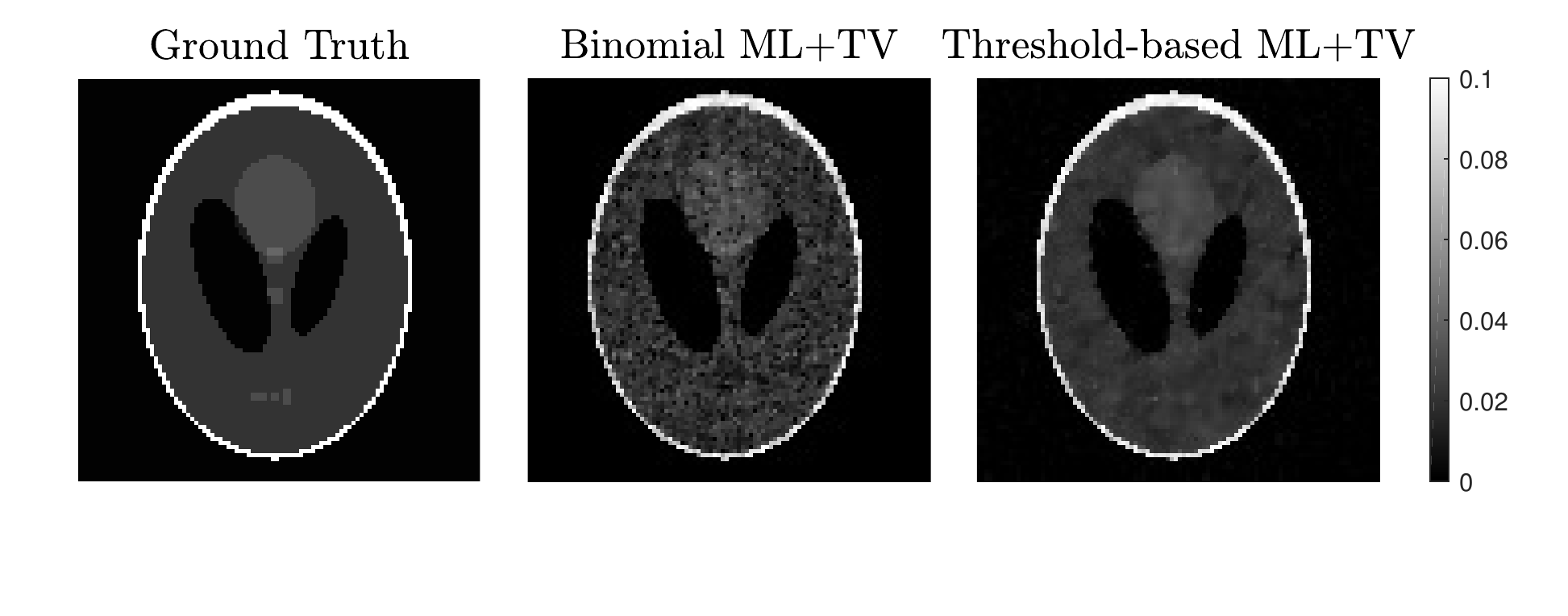} \\[-3mm] \vspace{-0.55cm}
{\footnotesize\begin{flushleft} \hspace*{31.7mm} PSNR 24.7 dB \hspace*{5.5mm} PSNR 29.1 dB \end{flushleft}} \vspace{-0.3cm}
\centerline{\footnotesize(a) $\BetaDist{(2,152)}$ and $\eta = 200$.} \vspace{0.1cm}

\includegraphics[width=0.9\linewidth, trim={8mm 11mm 7mm 8mm},clip]{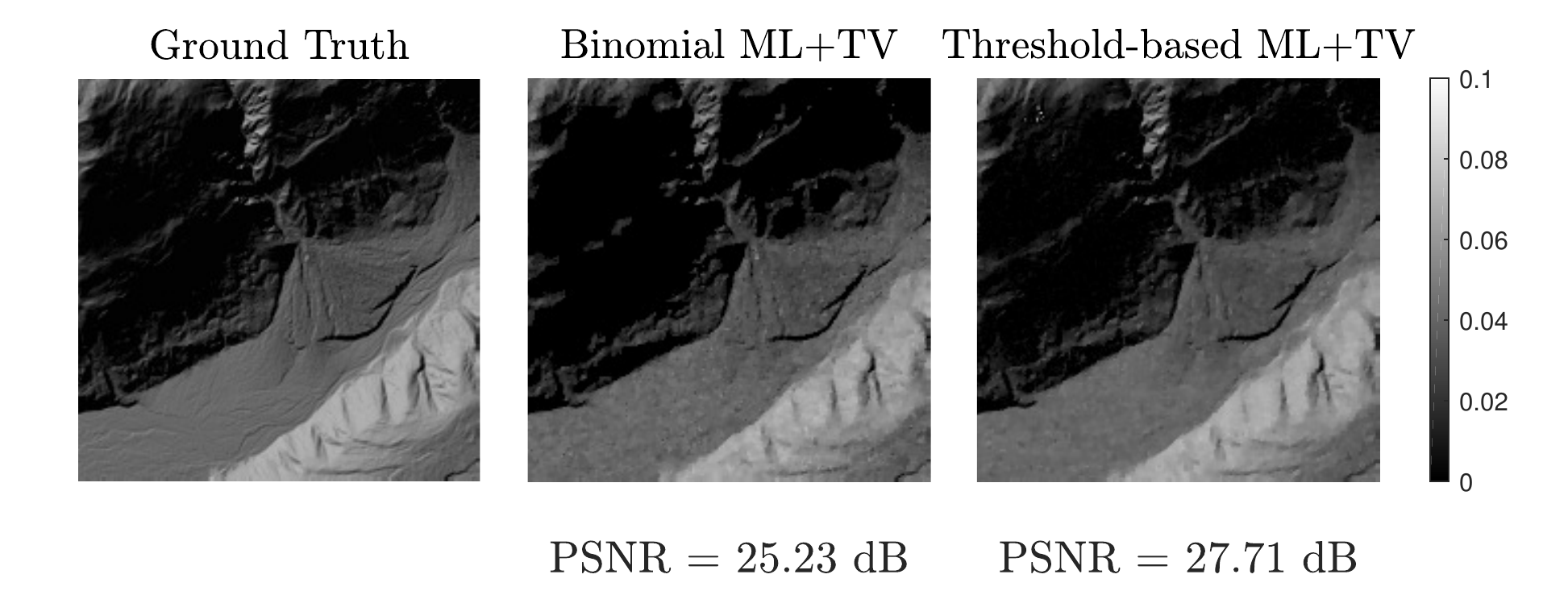}  \\[-3mm] \vspace{-0.05cm}
{\footnotesize\begin{flushleft} \hspace*{31.7mm} PSNR 25.2 dB \hspace*{5.5mm} PSNR 27.7 dB \end{flushleft}} \vspace{-0.3cm}
\centerline{\footnotesize(b) $\BetaDist{(2,162)}$ and $\eta = 800$.} \vspace{0.1cm}

\includegraphics[width=0.9\linewidth, trim={8mm 11mm 7mm 8mm},clip]{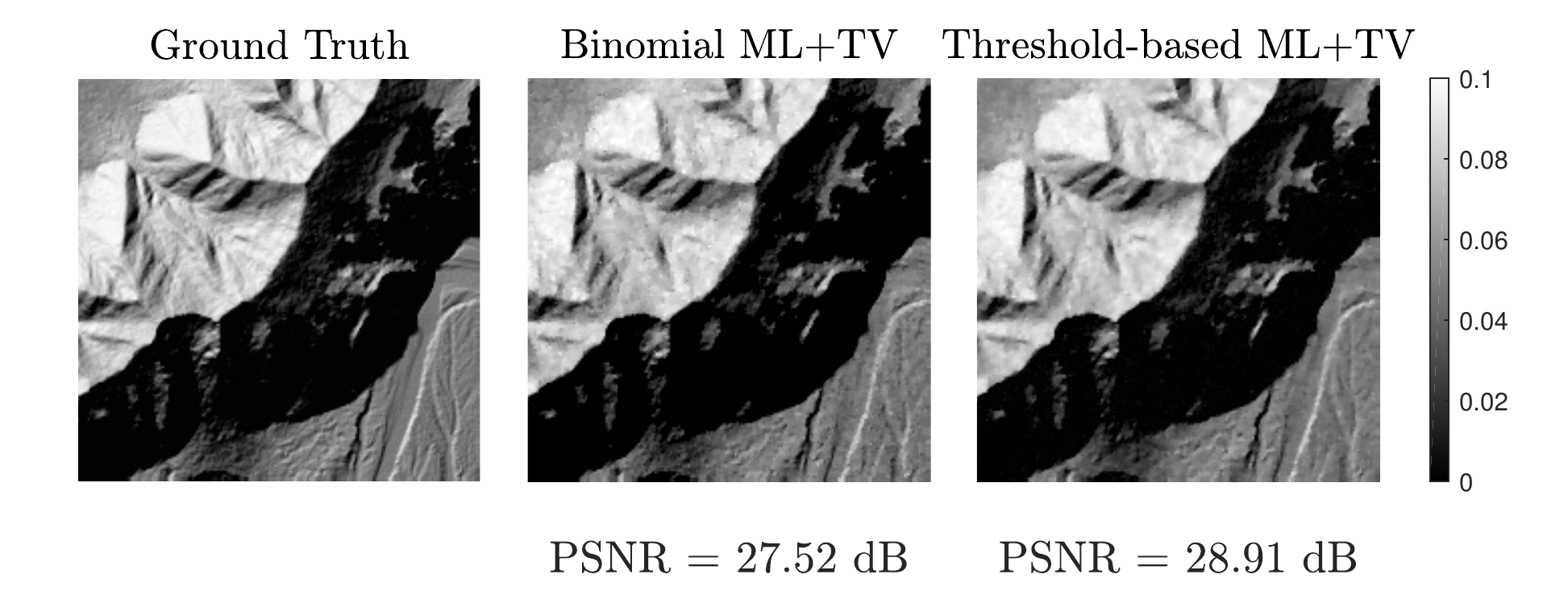}  \\[-3mm] \vspace{-0.05cm}
{\footnotesize\begin{flushleft} \hspace*{31.7mm} PSNR 27.5 dB \hspace*{5.5mm} PSNR 28.9 dB \end{flushleft}} \vspace{-0.3cm}
\centerline{\footnotesize(c) $\BetaDist{(2,172)}$ and $\eta = 800$.} \vspace{0.1cm}

\includegraphics[width=0.9\linewidth, trim={8mm 11mm 7mm 8mm},clip]{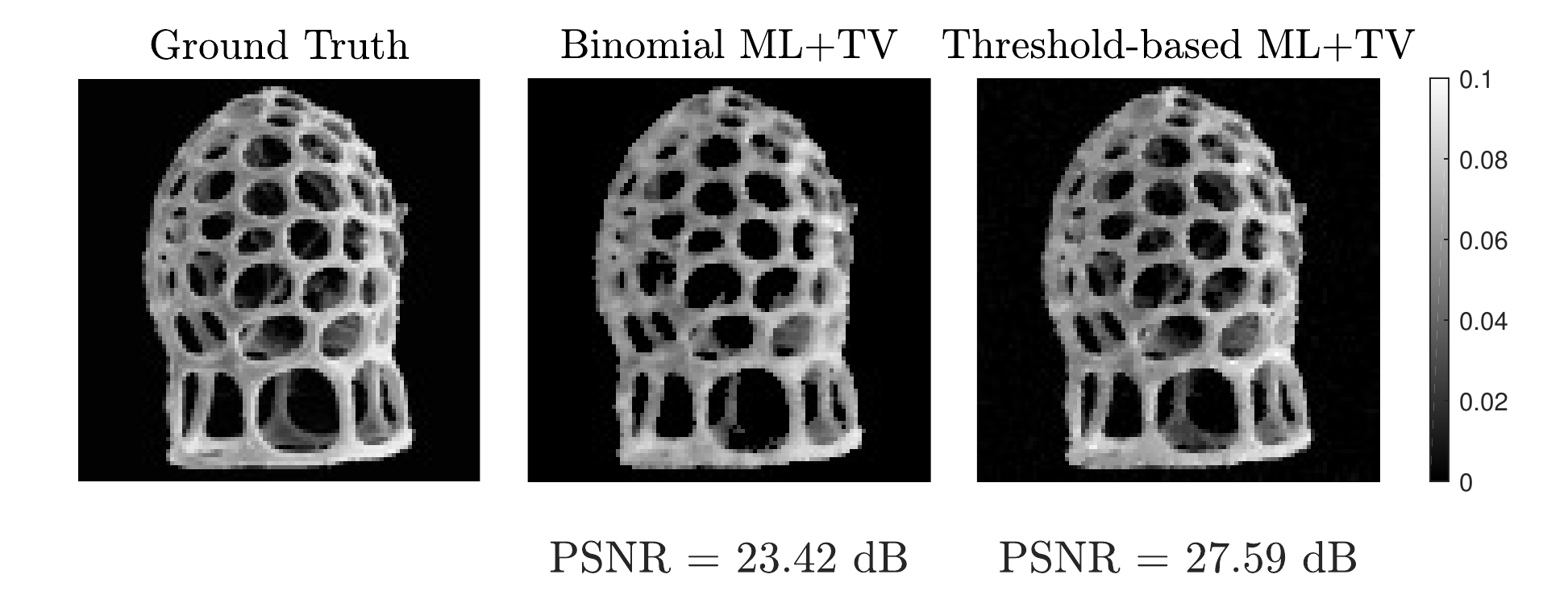}   \\[-3mm] \vspace{-0.05cm}
{\footnotesize\begin{flushleft} \hspace*{31.7mm} PSNR 23.4 dB \hspace*{5.5mm} PSNR 27.6 dB \end{flushleft}} \vspace{-0.3cm}
\centerline{\footnotesize(d) $\BetaDist{(2,162)}$ and $\eta = 500$.}  \vspace{0.1cm}

\includegraphics[width=0.9\linewidth, trim={8mm 11mm 7mm 8mm},clip]{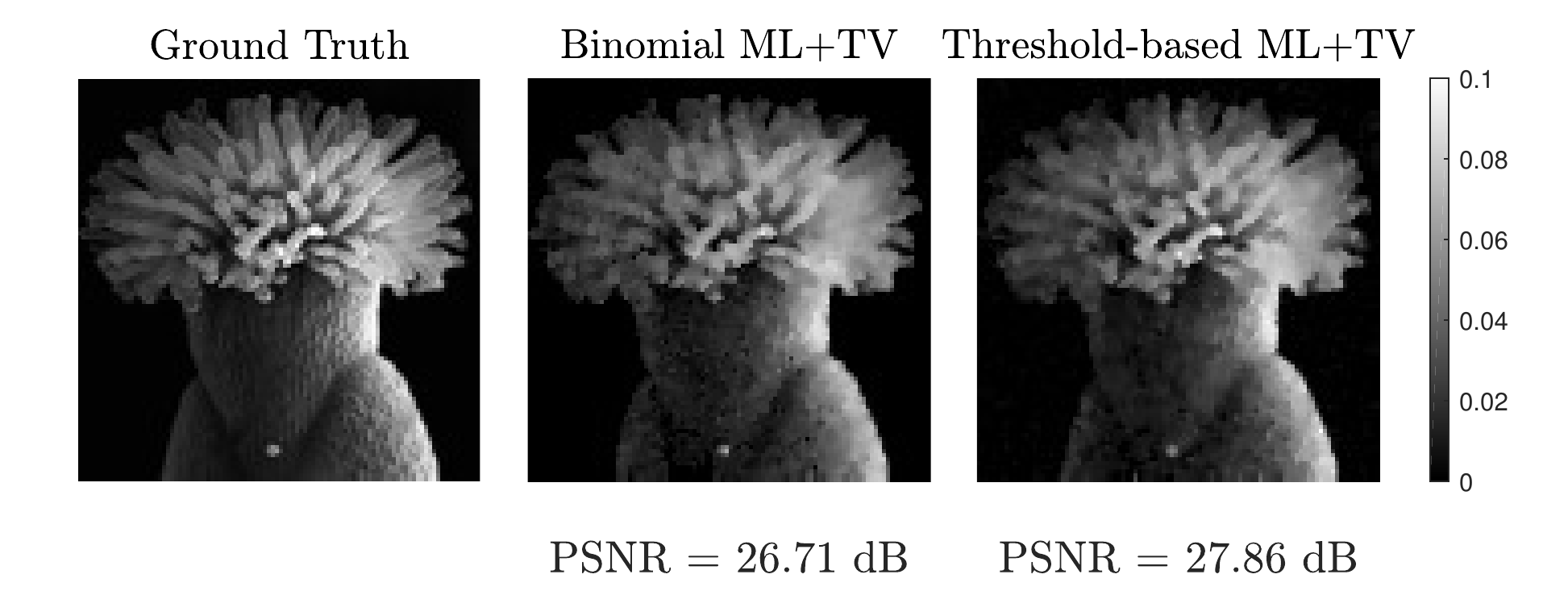}   \\[-3mm] \vspace{-0.05cm}
{\footnotesize\begin{flushleft} \hspace*{31.7mm} PSNR 26.7 dB \hspace*{5.5mm} PSNR 27.9 dB \end{flushleft}} \vspace{-0.3cm}
\centerline{\footnotesize(e) $\BetaDist{(2,162)}$ and $\eta = 400$.}
\caption{Images reconstructed through TV-regularized ML estimation showing MSE improvements of up to $4.36\,\mathrm{dB}$ for online threshold-based termination in place of conventional binomial sampling. Assumed priors and trial budgets are indicated below each test image. All images are scaled to $[0.001,0.101]$.
}
\label{fig:phantom-mmse-TV}
\end{figure}

Reconstruction quality can be improved through the use of TV-regularized ML estimation~\cite{louchet2008,Shin2015}.
In one typical experimental trial shown in Fig.~\ref{fig:phantom-mmse-TV}(a),
the TV-regularized reconstruction from data obtained with online threshold-based termination  outperforms the conventional binomial sampling by $4.36\,\mathrm{dB}$ in MSE for the Shepp--Logan phantom;
the trial budget of $\eta = 200$ and prior of $\BetaDist(2,152)$ are the same as used previously. As anticipated, Figs.~\ref{fig:phantom-mmse-TV}(b)--(e) also demonstrate significant improvements in MSE, ranging from $1.15\,\mathrm{dB}$ to $4.17\,\mathrm{dB}$, for the remaining test images.
Furthermore, keeping the corresponding priors and trial budgets used for each test image in Figs.~\ref{fig:phantom-mmse-TV}, Table~\ref{tbl:Average_MSE_p} also provides results averaged over 100 experiments for statistical significance.
In many cases, imposing TV regularization increases the performance gained from the data-adaptive stopping rule. Most importantly, it does not completely diminish the gain of adapting the acquisition.

\subsection{Estimation of $f(p) = \log{p}$}
\label{sec:image_log_p}
\begin{figure}
\centering
\includegraphics[width=1\linewidth, trim={8mm 18mm 8mm 8mm},clip]{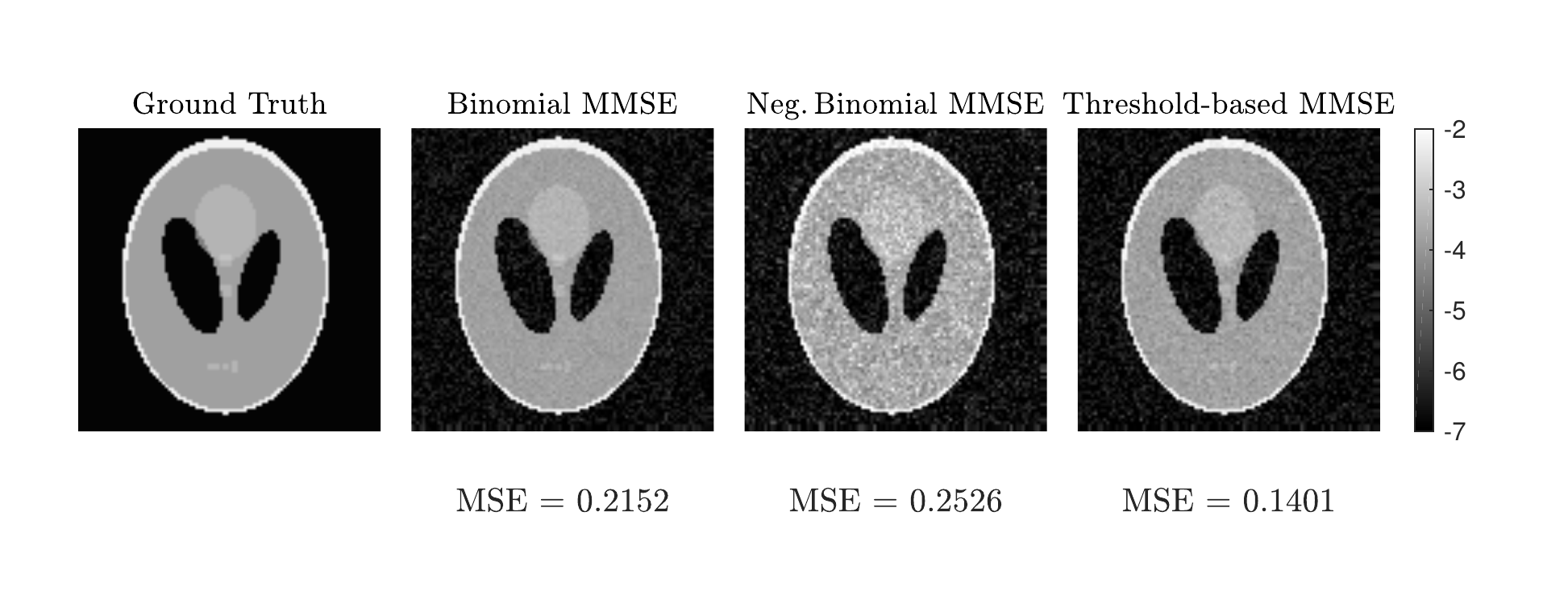} \\[-3mm] \vspace{-0.1cm}
{\footnotesize\begin{flushleft} \hspace*{22.5mm} MSE 0.2152 \hspace*{4.2mm} MSE 0.2526 \hspace*{4.2mm} MSE 0.1401 \end{flushleft}} \vspace{-0.3cm}
\centerline{\footnotesize(a) $\eta = 3000$} \vspace{0.1cm}

\includegraphics[width=1\linewidth, trim={8mm 18mm 8mm 15mm},clip]{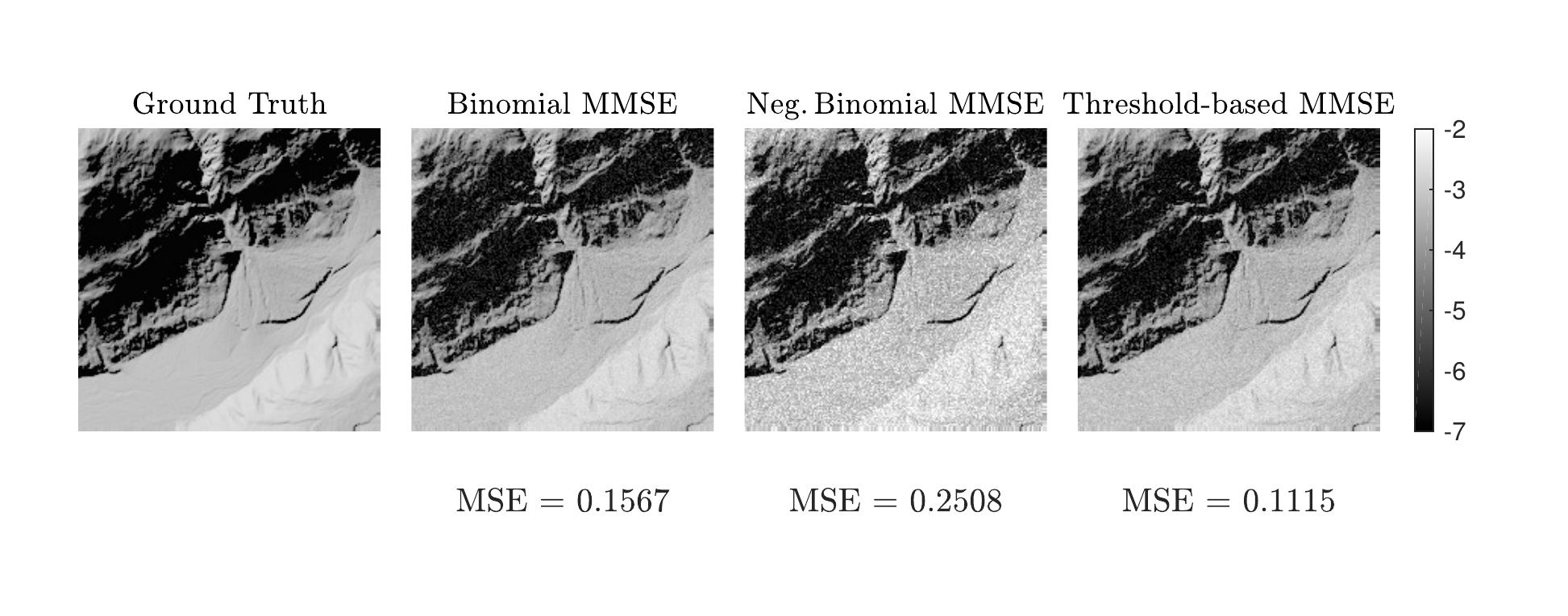}  \\[-3mm] \vspace{-0.1cm}
{\footnotesize\begin{flushleft} \hspace*{22.5mm} MSE 0.1567 \hspace*{4.2mm} MSE 0.2508 \hspace*{4.2mm} MSE 0.1115 \end{flushleft}} \vspace{-0.3cm}
\centerline{\footnotesize(b) $\eta = 1600$} \vspace{0.1cm}

\includegraphics[width=1\linewidth, trim={8mm 18mm 8mm 15mm},clip]{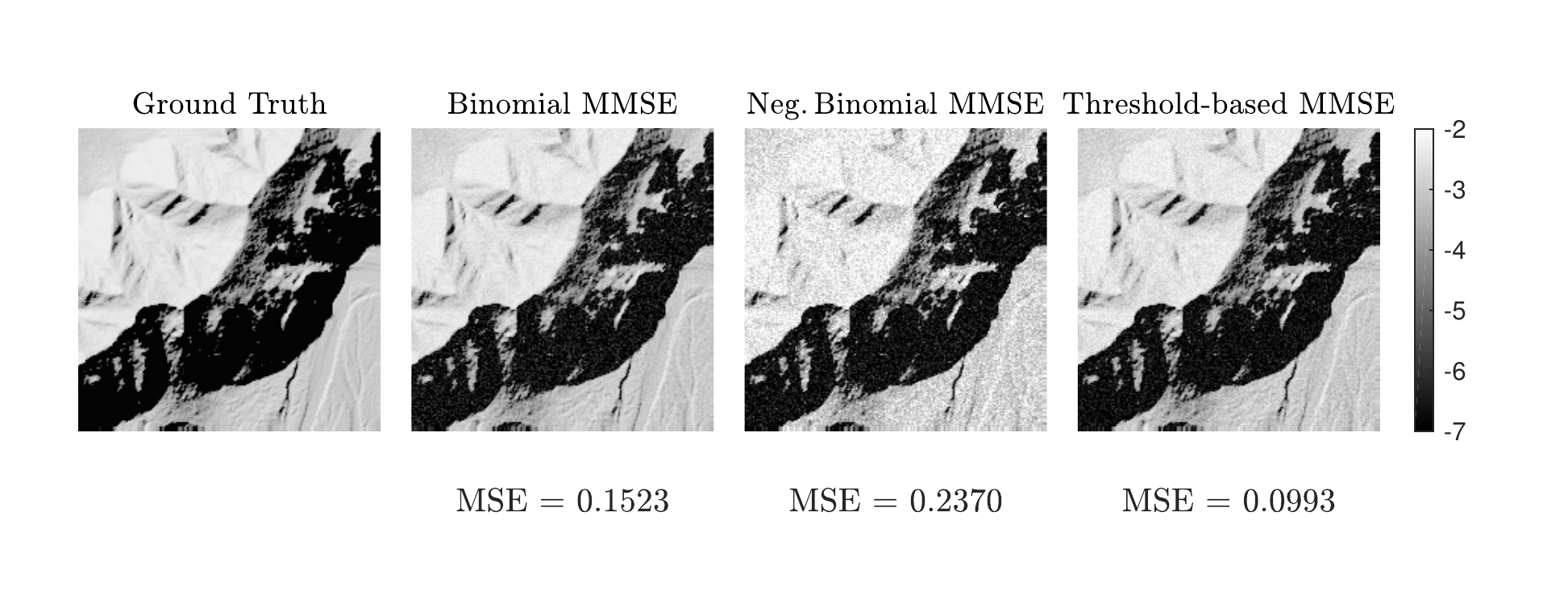}  \\[-3mm] \vspace{-0.1cm}
{\footnotesize\begin{flushleft} \hspace*{22.5mm} MSE 0.1523 \hspace*{4.2mm} MSE 0.2370 \hspace*{4.2mm} MSE 0.0993 \end{flushleft}} \vspace{-0.3cm}
\centerline{\footnotesize(c) $\eta = 1700$} \vspace{0.1cm}

\includegraphics[width=1\linewidth, trim={8mm 18mm 8mm 15mm},clip]{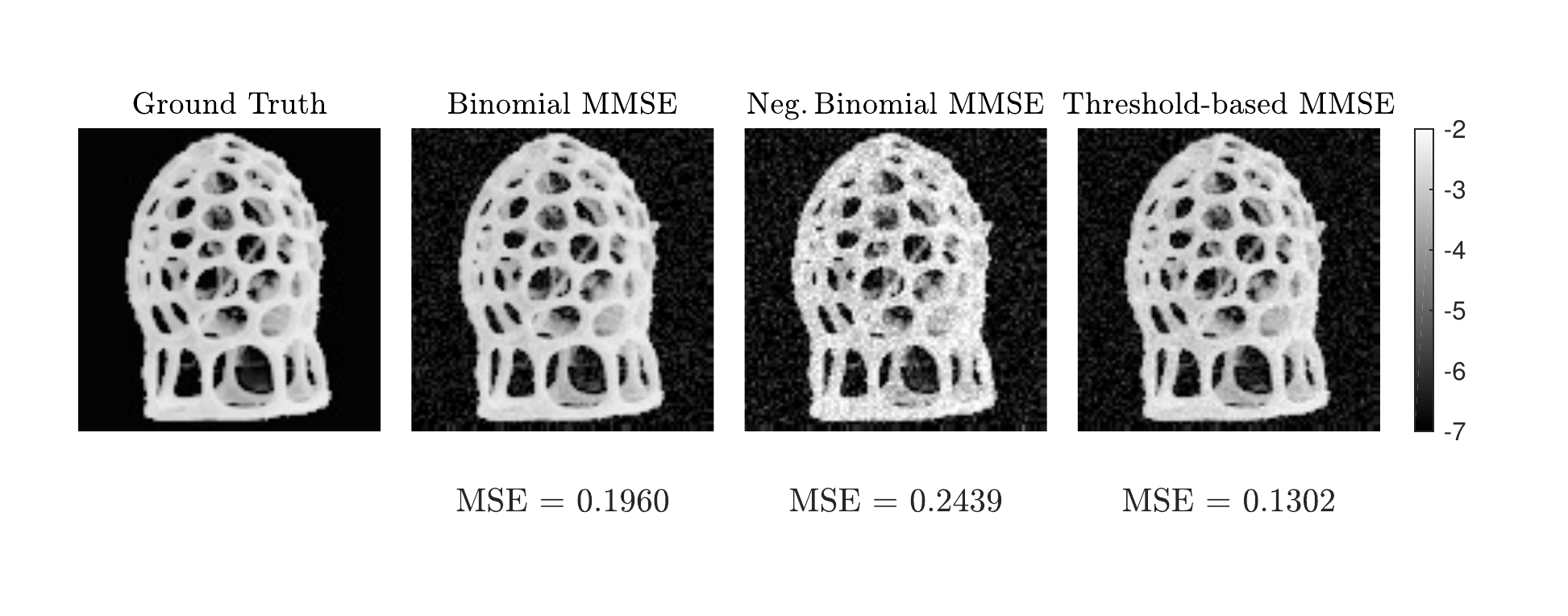}   \\[-3mm] \vspace{-0.1cm}
{\footnotesize\begin{flushleft} \hspace*{22.5mm} MSE 0.1960 \hspace*{4.2mm} MSE 0.2439 \hspace*{4.2mm} MSE 0.1302 \end{flushleft}} \vspace{-0.3cm}

\centerline{\footnotesize(d) $\eta = 2700$} \vspace{0.1cm}
\includegraphics[width=1\linewidth, trim={8mm 18mm 8mm 15mm},clip]{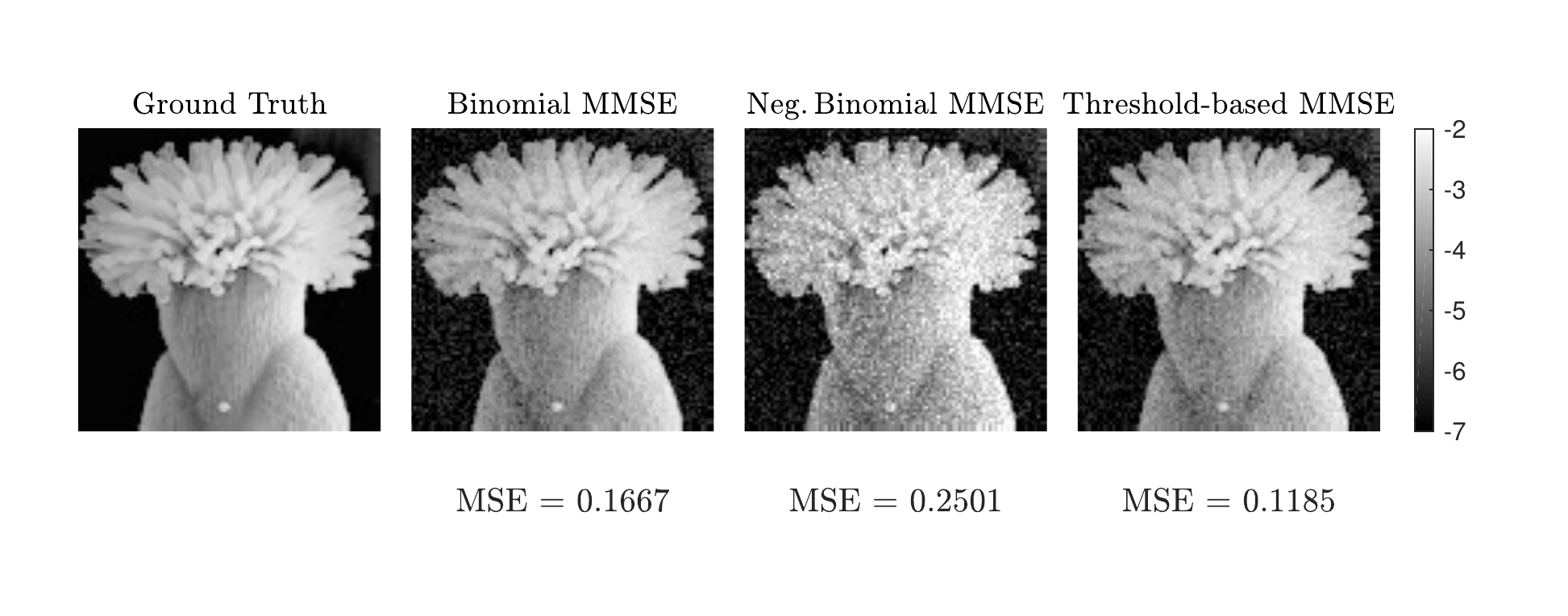}   \\[-3mm] \vspace{-0.1cm}
{\footnotesize\begin{flushleft} \hspace*{22.5mm} MSE 0.1667 \hspace*{4.2mm} MSE 0.2501 \hspace*{4.2mm} MSE 0.1185 \end{flushleft}} \vspace{-0.3cm}
\centerline{\footnotesize(e) $\eta = 1800$}
\caption{Estimation of $\log p$.
Images reconstructed through pixelwise MMSE estimation.
All images are scaled to $[0.001,0.101]$ and uniform prior is assumed.
MSE improvements for online threshold-based termination in place of conventional binomial and negative binomial sampling respectively, are:
(a) $1.86\,\mathrm{dB}$ and $2.56\,\mathrm{dB}$.
(b) $1.48\,\mathrm{dB}$ and $3.52\,\mathrm{dB}$.
(c) $1.86\,\mathrm{dB}$ and $3.78\,\mathrm{dB}$.
(d) $1.78\,\mathrm{dB}$ and $2.73\,\mathrm{dB}$.
(e) $1.48\,\mathrm{dB}$ and $3.24\,\mathrm{dB}$.
The negative binomial results have been obtained with $\ell = 5$.}
\label{fig:phantom-mmse-log}
\end{figure}

Now we present simulation results for the estimation of the logarithm of the previous test images.
For $f(p) = \log p$, the Bayes risk reduction per trial to use in online threshold-based termination
\eqref{eq:qkm-from-thresholding}
is given by
\eqref{eq:log-DeltaR-km}.
Fig.~\ref{fig:phantom-mmse-log} shows simulation results wherein
improvement factors of 
$1.48\,\mathrm{dB}$ to $1.86\,\mathrm{dB}$ are observed using threshold-based termination compared with binomial sampling,
and $2.56\,\mathrm{dB}$ to $3.78\,\mathrm{dB}$ when using threshold-based termination over negative binomial.
Note that since the contributions to the error from the lower pixel values are higher,
the comparison is provided at much higher trial budgets (i.e., $\eta = 3000$, {$1600$, $1700$}, $2800$, and $1800$ for Figs.~\ref{fig:phantom-mmse-log}(a)--(e), respectively) than in Section~\ref{sec:image_p} to obtain meaningful results.
Dark regions having values of $0.001$, for instance, require $1000$ trials on average to observe a success.
The effect of increasing number of trials is apparent in Table~\ref{tbl:Average_MSE_log}.
An increase in the improvement factor obtained by using the threshold-based rule, compared to both binomial and negative binomial stopping is observed, as the trial budget is increased.
For example, for {the} Shepp--Logan phantom, an increase in the improvement factor from $0.77\,\mathrm{dB}$ to $1.75\,\mathrm{dB}$ is observed when increasing $\eta$ from $1800$ to $3000$, if the threshold-based rule is used over conventional binomial stopping. An even larger increase is obtained when the threshold-based rule is compared against negative binomial stopping. The trend of increasing MSE improvements with increased trial budgets is persistent for other test images too.

\begin{table}
\centering
\caption{Estimation of $f(p) = \log(p)$.
Average reconstruction MSEs, averaged over 100 experiments,
for conventional binomial sampling, negative binomial sampling and threshold-based termination,
for trial budgets $\eta$.
}
\label{tbl:Average_MSE_log}
\begin{tabular}{ccccc}
\hline
\multirow{2}{*}{Image}  & \multirow{2}{*}{$\eta$} & \multicolumn{3}{c}{\scriptsize Method}   \\ 
\cline{3-5} 
& & \multicolumn{1}{c}{\scriptsize Binomial} & \multicolumn{1}{c}{\scriptsize Neg. Binomial} & \multicolumn{1}{c}{\scriptsize Threshold-based} \\  \hline
  \multirow{2}{*}{Shepp--Logan}   & $1800$ &      $0.271$     & $0.481$                       & $\mathbf{0.227}$                          \\& $3000$ &    $0.208$         & $0.253$                        & $\mathbf{0.139}$                      \\
     \hline
  
  \multirow{2}{*}{lidar \#1}   & $1600$ &     $0.156$       & $0.246$                       & $\mathbf{0.113}$                          \\ 
    & $2200$ &   $0.133$      & $0.165$                        & $\mathbf{0.080}$                      \\
    \hline  
    
    \multirow{2}{*}{lidar \#2}   & $1700$ &     $0.154$       & $0.237$                       & $\mathbf{0.099}$                          \\ 
    & $2300$ &   $0.135$      & $0.161$                        & $\mathbf{0.073}$                      \\
    \hline  

  \multirow{2}{*}{\emph{Foraminifera}}   & $1700$ &     $0.253$       & $0.462$                       & $\mathbf{0.208}$                          \\ 
  & $2700$ &      $0.206$      & $0.245$                        & $\mathbf{0.131}$                      \\
    \hline
  \multirow{2}{*}{\emph{HairStyle}}   & $1800$ &   $0.166$           & $0.247$                       & $\mathbf{0.118}$                          \\ 
  & $2500$ &    $0.143$        & $0.166$                        & $\mathbf{0.087}$                      \\
    \hline            
\end{tabular}
\end{table}

\section{Conclusion}
We established a novel framework for estimating Bernoulli parameters where we represent each Bernoulli process with a simple trellis graph.
By exploiting the mathematical convenience that comes from assuming Beta priors, we propose and study three stopping strategies with varying complexities but yielding very nearly equal performances.
All strategies give significant performance improvements over the conventional binomial and negative binomial stopping rules in simulated active imaging applications.

The simple online threshold-based termination was shown to asymptotically allocate trials in the same manner as an
oracle-aided solution that assumes the Bernoulli parameters are \emph{a priori} known.
Whilst we only study herein oracle-aided binomial sampling, similar analyses are possible for functions of Bernoulli parameters or oracle-aided negative binomial sampling;
these are omitted here because they do not yield clean expressions like the binomial case.

Finally, the proposed online threshold-based termination is extended to the estimation of $\log{p}$.
Other functions of $p$, such as $f(p) = 1/p$, can prove useful for scenarios wherein distinguishing between small parameters is paramount.

In the formulation of optimizing a trellis to minimize MSE,
a beta prior is convenient but not at all fundamental.
The reduction of the design problem from a general tree to a trellis holds for any prior,
and one may in principle compute Bayes risk reduction per trial for any trellis node
and any prior.
Developing an analogous theory for minimax estimation is also of interest
but is less clear because of a lack of additivity of the cost function.
This is especially intriguing because the optimization for MSE incidentally reduces the maximum over $p$ of the risk (see Fig.~\ref{fig:bin_greedy_oracle-EN_ER-comp}(b)).

\section*{Acknowledgment}
The authors thank Charles Saunders for assistance with implementation of regularized
estimators and Joshua Rapp for discussions on influence in spatial neighborhoods.
\balance

\bibliographystyle{ieeetr}
\bibliography{IEEEabrv,references,bibl}

\newcommand{\SortNoop}[1]{}
\begin{thebibliography}{10}

\bibitem{Shin2015}
D.~Shin, A.~Kirmani, V.~K. Goyal, and J.~H. Shapiro, ``Photon-efficient
  computational 3{D} and reflectivity imaging with single-photon detectors,''
  {\em IEEE Trans. Comput. Imaging}, vol.~1, pp.~112--125, June 2015.

\bibitem{tcspcWahl2015}
M.~Wahl, ``Time-correlated single photon counting ({TCSPC}),'' tech. rep.,
  PicoQuant, Berlin, Germany, 2014.

\bibitem{Fine2006}
T.~L. Fine, {\em Probability and Probabilistic Reasoning for Electrical
  Engineering}.
\newblock Upper Saddle River, NJ: Pearson Prentice Hall, 2006.

\bibitem{Segall:76}
A.~Segall, ``Bit allocation and encoding for vector sources,'' {\em IEEE Trans.
  Inform. Theory}, vol.~IT-22, pp.~162--169, Mar. 1976.

\bibitem{Goyal:01a}
V.~K. Goyal, ``Theoretical foundations of transform coding,'' {\em IEEE Signal
  Process. Mag.}, vol.~18, pp.~9--21, Sept. 2001.

\bibitem{Anscombe1953}
F.~J. Anscombe, ``Sequential estimation,'' {\em J. Roy. Statist. Soc. Ser. B},
  vol.~15, no.~1, pp.~1--29, 1953.

\bibitem{Haldane1945}
J.~B.~S. Haldane, ``On a method of estimating frequencies,'' {\em Biometrika},
  vol.~33, pp.~222--225, Nov. 1945.

\bibitem{Tweedie1945}
M.~C.~K. Tweedie, ``Inverse statistical variates,'' {\em Nature}, vol.~155,
  p.~453, Apr. 14, 1945.

\bibitem{Cabilio1975}
P.~Cabilio and H.~Robbins, ``Sequential estimation of $p$ with squared relative
  error loss,'' {\em Proc. Nat. Acad. Sci. USA}, vol.~72, pp.~191--193, Jan.
  1975.

\bibitem{Cabilio1977}
P.~Cabilio, ``Sequential estimation in {B}ernoulli trials,'' {\em Ann.
  Statist.}, vol.~5, pp.~342--356, Mar. 1977.

\bibitem{Hubert2000}
S.~L. Hubert and R.~Pyke, ``Sequential estimation of functions of $p$ for
  {B}ernoulli trials,'' in {\em Game Theory, Optimal Stopping, Probability and
  Statistics}, vol.~35 of {\em Lecture Notes-Monograph Series}, pp.~263--294,
  Institute of Mathematical Statistics, 2000.

\bibitem{Djuric2000}
P.~M. Djuri{\'c} and Y.~Huang, ``Estimation of a {B}ernoulli parameter $p$ from
  imperfect trials,'' {\em IEEE Signal Process. Lett.}, vol.~7, pp.~160--163,
  June 2000.

\bibitem{Ciuonzo2015}
D.~Ciuonzo, A.~{De Maio}, and P.~{Salvo Rossi}, ``A systematic framework for
  composite hypothesis testing of independent {B}ernoulli trials,'' {\em IEEE
  Signal Process. Lett.}, vol.~22, pp.~1249--1253, Sept. 2015.

\bibitem{FPI2014}
A.~Kirmani, D.~Venkatraman, D.~Shin, A.~Cola{\c c}o, F.~N.~C. Wong, J.~H.
  Shapiro, and V.~K. Goyal, ``First-photon imaging,'' {\em Science}, vol.~343,
  no.~6166, pp.~58--61, 2014.

\bibitem{Rapp2017}
J.~Rapp and V.~K. Goyal, ``A few photons among many: Unmixing signal and noise
  for photon-efficient active imaging,'' {\em IEEE Trans. Comput. Imaging},
  vol.~3, pp.~445--459, Sept. 2017.

\bibitem{Krichel2010}
N.~J. Krichel, A.~McCarthy, and G.~S. Buller, ``Resolving range ambiguity in a
  photon counting depth imager operating at kilometer distances,'' {\em Opt.
  Express}, vol.~18, no.~9, pp.~9192--9206, 2010.

\bibitem{Morris2015}
P.~A. Morris, R.~S. Aspden, J.~E.~C. Bell, R.~W. Boyd, and M.~J. Padgett,
  ``Imaging with a small number of photons,'' {\em Nat. Commun.}, vol.~6, Jan.
  5, 2015.
\newblock doi: 10.1038/ncomms6913.

\bibitem{Shin2015b}
D.~Shin, J.~H. Shapiro, and V.~K. Goyal, ``Single-photon depth imaging using a
  union-of-subspaces model,'' {\em IEEE Signal Process. Lett.}, vol.~22,
  pp.~2254--2258, Dec. 2015.

\bibitem{Altmann2016}
Y.~Altmann, X.~Ren, A.~McCarthy, G.~S. Buller, and S.~McLaughlin, ``Lidar
  waveform-based analysis of depth images constructed using sparse
  single-photon data,'' {\em IEEE Trans. Image Process.}, vol.~25,
  pp.~1935--1946, May 2016.

\bibitem{Shin2016camera}
D.~Shin, F.~Xu, D.~Venkatraman, R.~Lussana, F.~Villa, F.~Zappa, V.~K. Goyal,
  F.~N.~C. Wong, and J.~H. Shapiro, ``Photon-efficient imaging with a
  single-photon camera,'' {\em Nat. Commun.}, vol.~7, June 24, 2016.
\newblock doi: 10.1038/ncomms12046.

\bibitem{Shin2016}
D.~Shin, J.~H. Shapiro, and V.~K. Goyal, ``Performance analysis of low-flux
  least-squares single-pixel imaging,'' {\em IEEE Signal Process. Lett.},
  vol.~23, pp.~1756--1760, Dec. 2016.

\bibitem{Shin2016multidepth}
D.~Shin, F.~Xu, F.~N.~C. Wong, J.~H. Shapiro, and V.~K. Goyal, ``Computational
  multi-depth single-photon imaging,'' {\em Opt. Express}, vol.~24,
  pp.~1873--1888, Feb. 2016.

\bibitem{Mertens2017}
L.~Mertens, M.~Sonnleitner, J.~Leach, M.~Agnew, and M.~J. Padgett, ``Image
  reconstruction from photon sparse data,'' {\em Sci. Rep.}, vol.~7, Feb. 7,
  2017.
\newblock doi: 10.1038/srep42164.

\bibitem{Altmann2017}
Y.~Altmann, R.~Aspden, M.~Padgett, and S.~McLaughlin, ``A {B}ayesian approach
  to denoising of single-photon binary images,'' {\em IEEE Trans. Comput.
  Imaging}, vol.~3, pp.~460--471, Sept. 2017.

\bibitem{Halimi2017}
A.~Halimi, A.~Maccarone, A.~McCarthy, S.~McLaughlin, and G.~S. Buller, ``Object
  depth profile and reflectivity restoration from sparse single-photon data
  acquired in underwater environments,'' {\em IEEE Trans. Comput. Imaging},
  vol.~3, pp.~472--484, Sept. 2017.

\bibitem{Liu2018}
X.~Liu, J.~Shi, X.~Wu, and G.~Zeng, ``Fast first-photon ghost imaging,'' {\em
  Sci. Rep.}, vol.~8, Mar. 22, 2018.
\newblock doi: 10.1038/s41598-018-23363-w.

\bibitem{Altmann2018}
Y.~Altmann, S.~McLaughlin, M.~J. Padgett, V.~K. Goyal, A.~O. Hero, and
  D.~Faccio, ``Quantum-inspired computational imaging,'' {\em Science},
  vol.~361, p.~660, Aug. 2018.

\bibitem{Zhu2018}
Z.~Zhu and S.~Pang, ``Few-photon computed x-ray imaging,'' {\em Appl. Phys.
  Lett.}, vol.~113, p.~231109, Dec. 2018.

\bibitem{Lipke:1979}
D.~L. Lipke, ``Active imaging system using variable gate width time programmed
  dwell.'' U.S. Patent 4,151,415, Apr. 1979.

\bibitem{He2017}
W.~He, Z.~Feng, J.~Lin, S.~Shen, Q.~Chen, G.~Gu, B.~Zhou, and P.~Zhang,
  ``Adaptive depth imaging with single-photon detectors,'' {\em IEEE Photon.
  J.}, vol.~9, Apr. 2017.

\bibitem{Dahmen2016}
T.~Dahmen, M.~Engstler, C.~Pauly, P.~Trampert, N.~{de Jonge}, F.~M{\"u}cklich,
  and P.~Slusallek, ``Feature adaptive sampling for scanning electron
  microscopy,'' {\em Sci. Rep.}, vol.~6, May 6, 2016.
\newblock doi: 10.1038/srep25350.

\bibitem{MedinMBG:2018}
S.~C. Medin, J.~Murray-Bruce, and V.~K. Goyal, ``Optimal stopping times for
  estimating {B}ernoulli parameters with applications to active imaging,'' in
  {\em Proc. IEEE Int. Conf. Acoust., Speech, and Signal Process.}, (Calgary,
  AB, Canada), pp.~4429--4433, May 2018.

\bibitem{bertsekas1996dynamic}
D.~P. Bertsekas, {\em Dynamic Programming and Optimal Control}, vol.~1.
\newblock Belmont, MA, USA: Athena Scientific, 1996.

\bibitem{Gonzalez2006}
R.~C. Gonzalez and R.~E. Woods, {\em Digital Image Processing (3rd Edition)}.
\newblock Upper Saddle River, NJ, USA: Prentice-Hall, Inc., 2006.

\bibitem{Milner2017}
E.~S. Milner and M.~T.~H. Do, ``A population representation of absolute light
  intensity in the mammalian retina,'' {\em Cell}, vol.~171, pp.~865--876.e16,
  2018/08/19 2017.

\bibitem{Karlis2005}
D.~Karlis, ``{EM} algorithm for mixed {P}oisson and other discrete
  distributions,'' {\em ASTIN Bulletin}, vol.~35, no.~1, p.~3–24, 2005.

\bibitem{Dette2017}
H.~Dette and D.~Tomecki, ``Hankel determinants of random moment sequences,''
  {\em J. Theoretical Probability}, vol.~30, pp.~1539--1564, Dec 2017.

\bibitem{louchet2008}
C.~Louchet and L.~Moisan, ``Total variation denoising using posterior
  expectation,'' in {\em Proc. 16th European Signal Process. Conf.}, pp.~1--5,
  Aug. 2008.

\bibitem{AlaskaLidar}
T.~D. Hubbard, M.~L. Braun, R.~E. Westbrook, and P.~E. Gallagher,
  ``High-resolution lidar data for infrastructure corridors, {H}ealy
  {Q}uadrangle, {A}laska.'' Alaska Division of Geological \& Geophysical
  Surveys, DOI: 10.14509/23163, Dec. 2011.

\end{thebibliography}

\balance

\end{document}